\documentclass[journal]{IEEEtran}
\usepackage[tbtags]{amsmath}
\usepackage{amsfonts}
\usepackage{amssymb}
\usepackage{subfigure}
\usepackage{bm}
\usepackage{cite}
\usepackage{calc}
\usepackage{color}
\usepackage{epsfig}
\usepackage{setspace}

\newtheorem{defn}{Definition}
\newtheorem{thm}{Theorem}[section]
\newtheorem{cor}[thm]{Corollary}
\newtheorem{prop}{Proposition}
\newtheorem{lem}[thm]{Lemma}
\newtheorem{conj}[thm]{Conjecture}
\newtheorem{constr}[thm]{Construction}
\newtheorem{note}{Remark}
\newcommand{\bit}{\begin{itemize}}
\newcommand{\eit}{\end{itemize}}
\newcommand{\bcor}{\begin{cor}}
\newcommand{\ecor}{\end{cor}}
\newcommand{\beq}{\begin{equation}}
\newcommand{\eeq}{\end{equation}}
\newcommand{\beqn}{\begin{equation*}}
\newcommand{\eeqn}{\end{equation*}}
\newcommand{\beqa}{\begin{eqnarray}}
\newcommand{\eeqa}{\end{eqnarray}}
\newcommand{\beqan}{\begin{eqnarray*}}
\newcommand{\eeqan}{\end{eqnarray*}}
\newcommand{\ben}{\begin{enumerate}}
\newcommand{\een}{\end{enumerate}}
\newcommand{\bdefn}{\begin{defn}}
\newcommand{\edefn}{\end{defn}}
\newcommand{\bnote}{\begin{note}}
\newcommand{\enote}{\end{note}}
\newcommand{\bprop}{\begin{prop}}
\newcommand{\eprop}{\end{prop}}
\newcommand{\blem}{\begin{lem}}
\newcommand{\elem}{\end{lem}}
\newcommand{\bthm}{\begin{thm}}
\newcommand{\ethm}{\end{thm}}
\newcommand{\bconj}{\begin{conj}}
\newcommand{\econj}{\end{conj}}
\newcommand{\bconstr}{\begin{constr}}
\newcommand{\econstr}{\end{constr}}
\newcommand{\bpf}{\begin{proof}}
\newcommand{\epf}{\end{proof}}

\begin{document}

\title{DMT of Multi-hop Cooperative Networks - Part II: Half-Duplex Networks with Full-Duplex Performance}

\author{\authorblockN{K. Sreeram, S. Birenjith, and P. Vijay Kumar} \\
\thanks{The authors are with the Department of Electrical Communication Engineering, Indian Institute of Science, Bangalore, India
(Email: sreeramkannan@ece.iisc.ernet.in, biren@ece.iisc.ernet.in,
vijayk@usc.edu). P.~Vijay Kumar is on leave of absence from the
University of Southern California, Los Angeles, USA. }
\thanks{This work was supported in part by NSF-ITR Grant CCR-0326628, in part by the DRDO-IISc Program on Advanced
Mathematical Engineering and in part by Motorola's University
Research Partnership Program with IISc.}
\thanks{The material in this paper was presented in part at the 10th International
Symposium on Wireless Personal Multimedia Communications, Jaipur,
Dec. 2007, the Information Theory and Applications Workshop, San
Diego, Jan. 2008, and at the IEEE International Symposium on
Information Theory, Toronto, July 2008.} }

\date{}
\maketitle
\begin{abstract}

We consider single-source single-sink (ss-ss) multi-hop relay
networks, with slow-fading links and single-antenna half-duplex
relay nodes. While two-hop cooperative relay networks have been
studied in great detail in terms of the diversity-multiplexing
gain tradeoff (DMT), few results are available for more general
networks. In a companion paper, we characterized end points of DMT
of arbitrary networks, and established some basic results which
laid the foundation for the results presented here. In the present
paper, we identify two families of networks that are multi-hop
generalizations of the two-hop network: $K$-Parallel-Path (KPP)
networks and layered networks.

KPP networks may be viewed as the union of $K$ node-disjoint
parallel relaying paths. Generalizations of these networks include
KPP(I) networks, which permit interference between paths and
KPP(D) networks, which possess a direct link between source and
sink. We characterize the DMT of these families of networks
completely for $K > 3$ and show that they can achieve the cut-set
bound, thus proving that the DMT performance of full-duplex
networks can be obtained even in the presence of the half-duplex
constraint. We then consider layered networks, which are comprised
of layers of relays, and prove that a linear DMT between the
maximum diversity $d_{\max}$ and the maximum multiplexing gain of
$1$ is achievable for single-antenna fully-connected(fc) layered
networks. This is shown to be equal to the cut-set bound on DMT if
the number of relaying layers is less than $4$, thus
characterizing the DMT of this family of networks completely. For
multiple-antenna KPP and layered networks, we provide lower bounds
on DMT, that are significantly better than the best-known bounds.

All protocols in this paper are explicit and use only
amplify-and-forward (AF) relaying. We also construct codes with
short block-lengths based on cyclic division algebras that achieve
the optimal DMT for all the proposed schemes. In addition, it is
shown that codes achieving full diversity on a MIMO Rayleigh
channel achieve full diversity on arbitrary fading channels as
well.

Two key implications of the results in the paper are that the
half-duplex constraint does not entail any rate loss for a large
class of cooperative networks and that simple AF protocols are
often sufficient to attain the optimal DMT.

\end{abstract}

\section{Introduction\label{sec:introduction}}

In fading relay networks, cooperative diversity provides a method
of efficient operation of networks. While much of work in the
literature on cooperative diversity is based on two-hop networks,
we focus our attention on multi-hop networks. For a review of
related literature, see Section I-A in the companion paper
\cite{Part1}. In the companion paper, we derived results
pertaining to the DMT of arbitrary full-duplex networks.

 In the present paper, we deal with half-duplex networks, for which
specification of explicit schedules requires some structure in the
network. Therefore, we focus on specific classes of half-duplex
networks in the present paper. For half-duplex networks without a
direct link, even the achievability of a maximum multiplexing gain
of $1$, in the case of single antenna networks, is not clear. We
will show that for a large family of networks, this maximum
multiplexing gain can be achieved by appropriately scheduling the
links. In fact, we will show that the cut-set upper bound on DMT
for many of these networks can be achieved, thus demonstrating
that the half-duplex operation does not entail any loss in DMT
performance as compared to full-duplex operation for these
families of networks.

\subsection{Classification of Networks \label{sec:network_taxonomy}}

\begin{figure}[h]
  \centering
  \subfigure[General two-hop relay network]{\label{fig:classical_relay}\includegraphics[height=30mm]{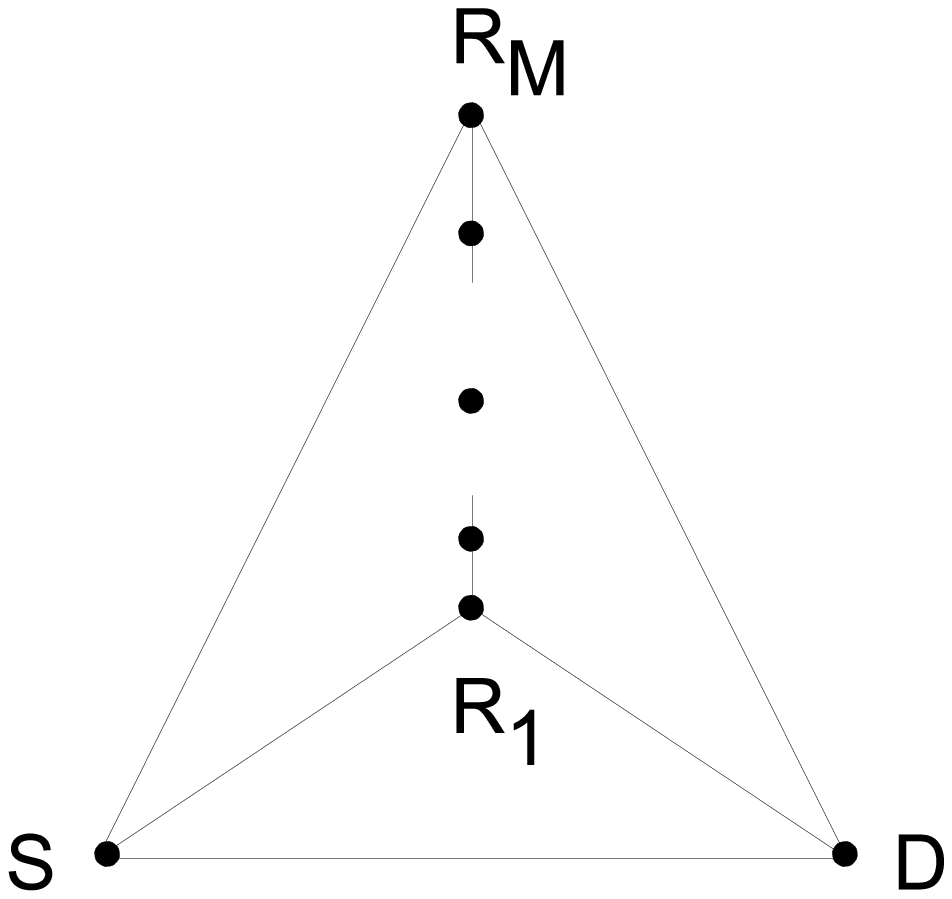}}
  \subfigure[Single relay network]{\label{fig:one_relay}\includegraphics[width=40mm]{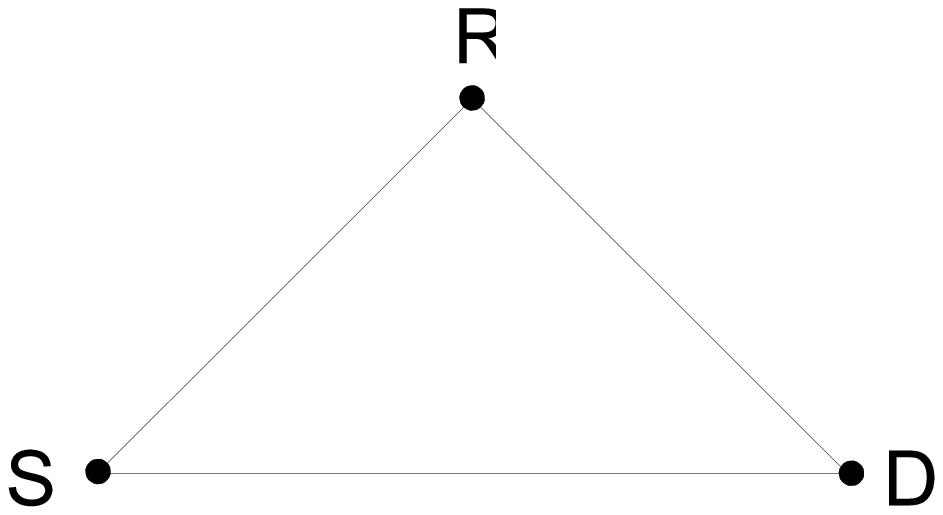}}
  \caption{Two-hop cooperative relay networks}
  \label{fig:two_hop_relay}
\end{figure}

In this section, we define the classes of networks under
consideration. The well-studied two-hop network with direct link
is shown in Fig.~\ref{fig:two_hop_relay}. We will consider two
multi-hop generalizations of two-hop networks in this paper: KPP
and layered networks. Unless otherwise stated, all networks
considered possess a single source and a single sink and we will
apply the abbreviation ss-ss to these networks.

A cooperative wireless network can be built out of a collection of
spatially distributed nodes in many ways. For instance, we can
identify paths connecting source to the sink through a series of
nodes in such a manner that any two adjacent nodes fall within the
Rayleigh zone\cite{YukErk}. This process can be continued barring
those nodes which are already chosen. Such a construction will
result in a set of paths from the source to the sink. In the
simplest model, we can further impose the constraint that these
paths do not interfere each other, see
Fig.\ref{fig:classical_relay}, thus motivating the study of a
class of multi-hop network which we shall refer to as the set of
K-Parallel-Path (KPP) networks.

\begin{figure}[h!]
\centering
\includegraphics[height=60mm]{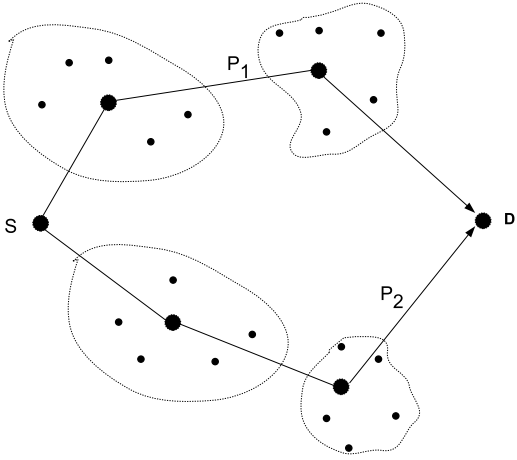}
\caption{Motivation for the KPP networks
\label{fig:kpp_network_motivation}}
\end{figure}

Alternatively, layers of relays can be identified from a
collection of nodes between the source and the sink. This will
result in a layered network model.

\subsubsection{Network Representation by Graph} Any wireless network can be associated with
a directed graph, with vertices representing nodes in the network
and edges representing connectivity between nodes. If an edge is
bidirectional, we will represent it by two edges one pointing in
either direction. An edge in a directed graph is said to be
\emph{live} at a particular time instant if the node at the head
of the edge is transmitting at that instant. An edge in a directed
graph is said to be \emph{active} at a particular time instant if
the node at the head of the edge is transmitting and the tail of
the edge is receiving at that instant. Since most networks
considered in this paper have bidirectional links, we will
represent a bidirectional link by an undirected edge. Therefore,
undirected edges must be interpreted as two directed edges, with
one edge pointing in either direction.

A wireless network is characterized by broadcast and interference
constraints. Under the \emph{broadcast} constraint, all edges
connected to a transmitting node are simultaneously live and
transmit the same information. Under the \emph{interference}
constraint, the symbol received by a receiving end is equal to the
sum of the symbols transmitted on all incoming live edges. We say
a protocol avoids interference if only one incoming edge is live
for all receiving nodes.

In wireless networks, the relay nodes operate in either half or
full-duplex mode. In case of half-duplex operation, a node cannot
simultaneously listen and transmit, i.e., an incoming edge and an
outgoing edge of a node cannot be simultaneously active.

\subsubsection{K-Parallel-Path Networks}

\begin{figure}[h!]
\centering
\includegraphics[width=80mm]{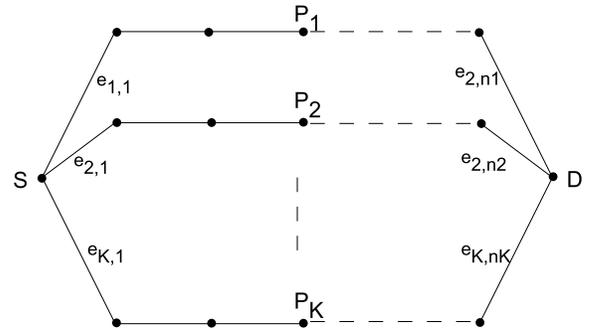}
\caption{The KPP network \label{fig:kpp_network}}
\end{figure}

One way of generalizing the two-hop relay network is to consider
this network as a collection of $K$ parallel, relaying paths from
the source to the sink, each of length greater than $1$. This
immediately leads to a more general network that is comprised of
$K$ parallel paths of varying lengths, linking source and sink.
More formally:

\bdefn \label{defn:kpp_defn} A set of edges $(v_1,v_2)$,
$(v_2,v_3)$, $\ldots$, $(v_{n-1},v_n)$ connecting the vertices
$v_1$ to $v_n$ is called a path. The length of a path is the
number of edges in the path. The K-parallel path (KPP) network  is
defined as a ss-ss network that can be expressed as the union of
$K$ node-disjoint paths, each of length greater than one,
connecting the source to the sink. Each of the node-disjoint paths
is called a relaying path. All edges in a KPP network are
bidirectional (see Fig.~\ref{fig:kpp_network}). \edefn

The communication between the source and the sink takes place
along $K$ parallel paths, labeled with the indices $P_1$, $P_2$,
$\ldots$, $P_K$. Along path $P_i$, the information is transmitted
from source to sink through multiple hops with the aid of $n_i-1$
intermediate relay nodes $\{R_{ij}\}_{j=1}^{n_i-1}$.

\bnote A network similar to the KPP network in
Definition~\ref{defn:kpp_defn} is considered in \cite{RibCaiGia},
albeit from a symbol error probability perspective. \enote

Definition~\ref{defn:kpp_defn} of KPP networks precludes the
possibility of either having a direct link between the source and
the sink, or of having links connecting nodes lying on different
node-disjoint paths. We now extend the definition of KPP networks
to include these possibilities.

\bdefn \label{defn:kpp_general} If a given network is a union of a
KPP network and a direct link between the source and sink, then
the network is called a KPP network with direct link, denoted by
KPP(D). If a given network is a union of a KPP network and links
interconnecting relays in various paths, then the network is
called a KPP network with interference, denoted by KPP(I). If a
given network is a union of a KPP network, a direct link and links
interconnecting relays in various paths, then the network is
called a KPP network with interference and direct path, denoted by
KPP(I, D). \edefn

\begin{figure}[h]
  \centering
  \subfigure[A KPP network with $K=3$]{\label{fig:kpp_eg}\includegraphics[width=40mm]{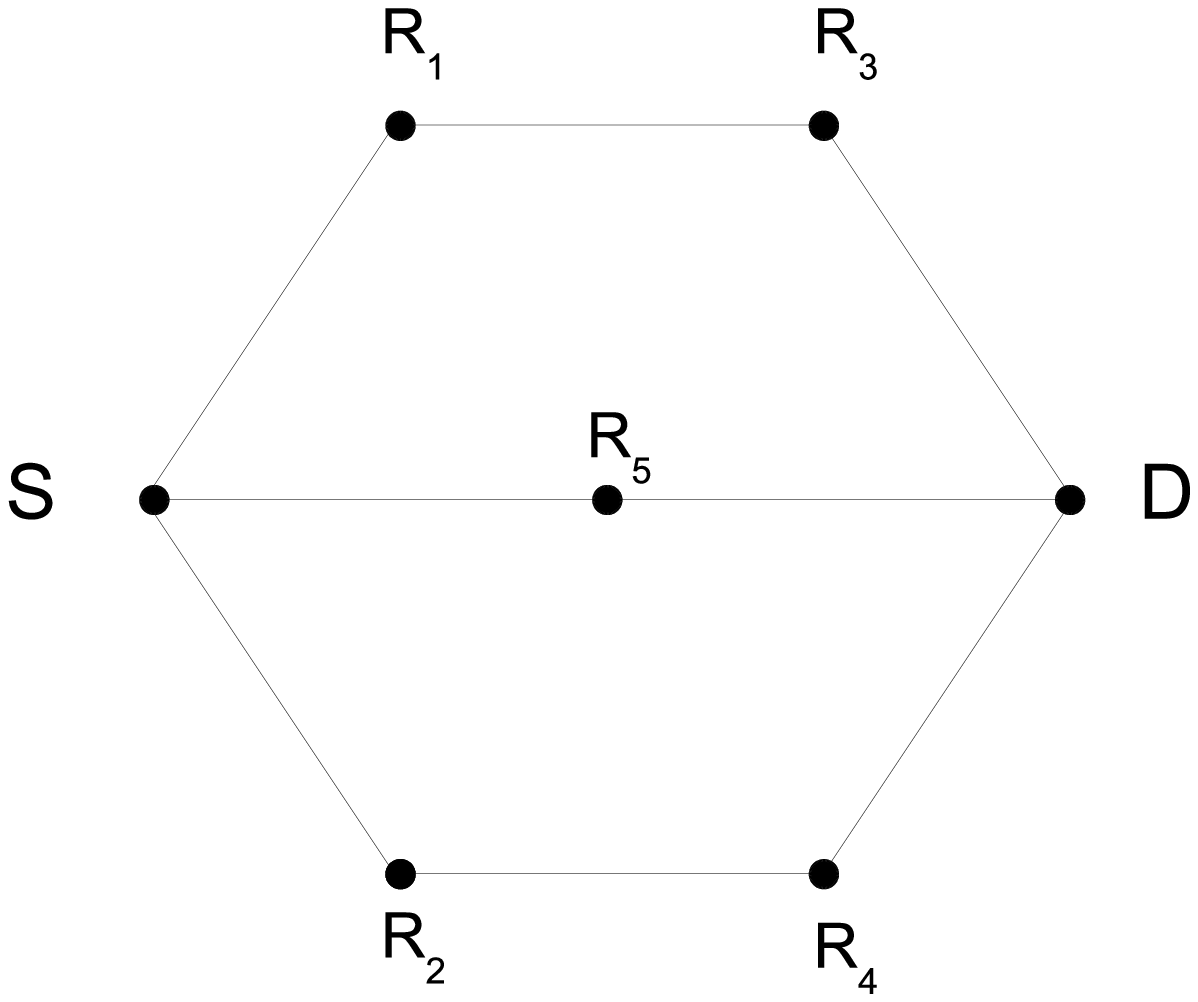}}
  \subfigure[A KPP(I) network with $K=3$]{\label{fig:kppi_eg}\includegraphics[width=40mm]{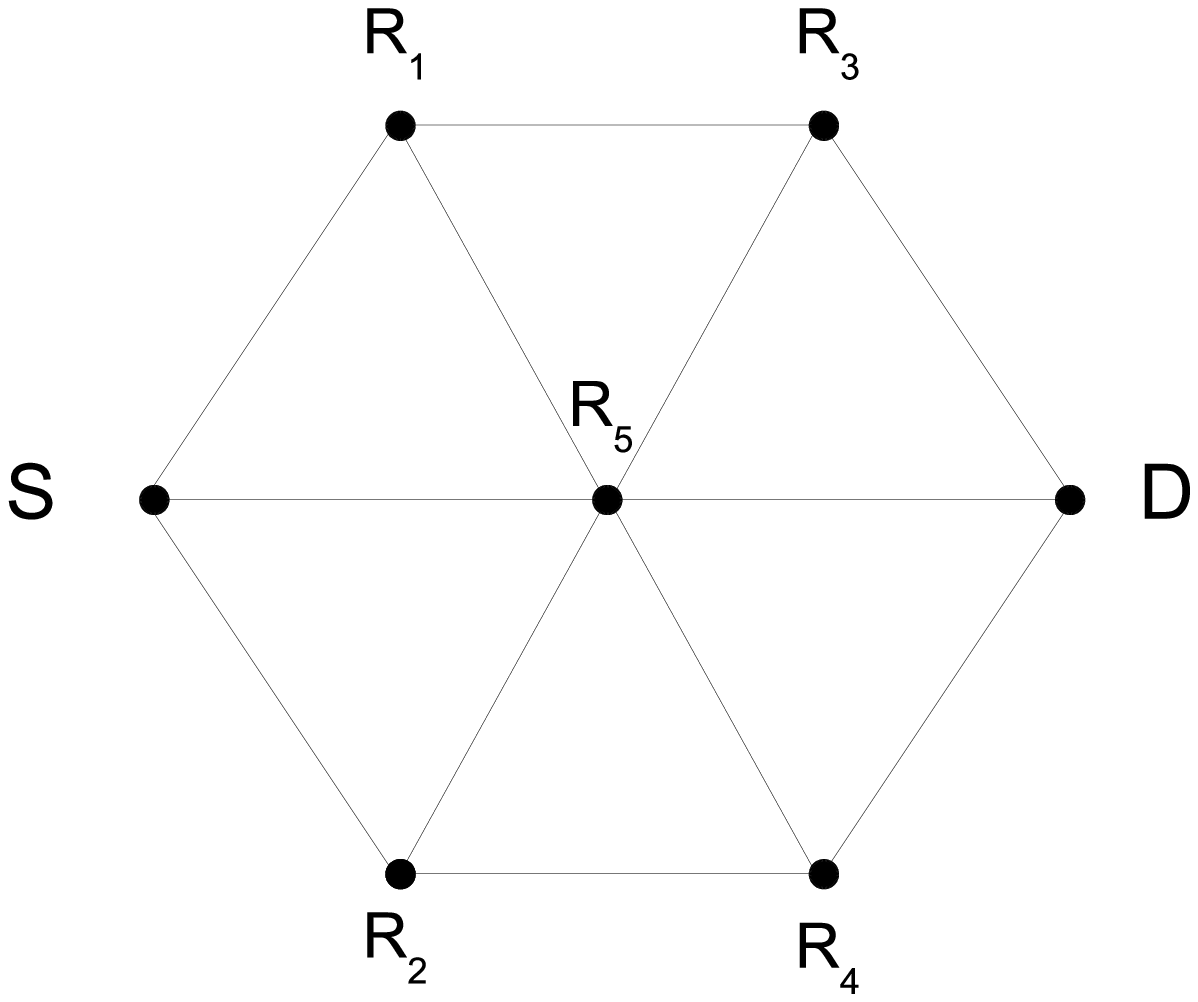}}
  \subfigure[A KPP(D) network with $K=2$]{\label{fig:kppd_eg}\includegraphics[width=40mm]{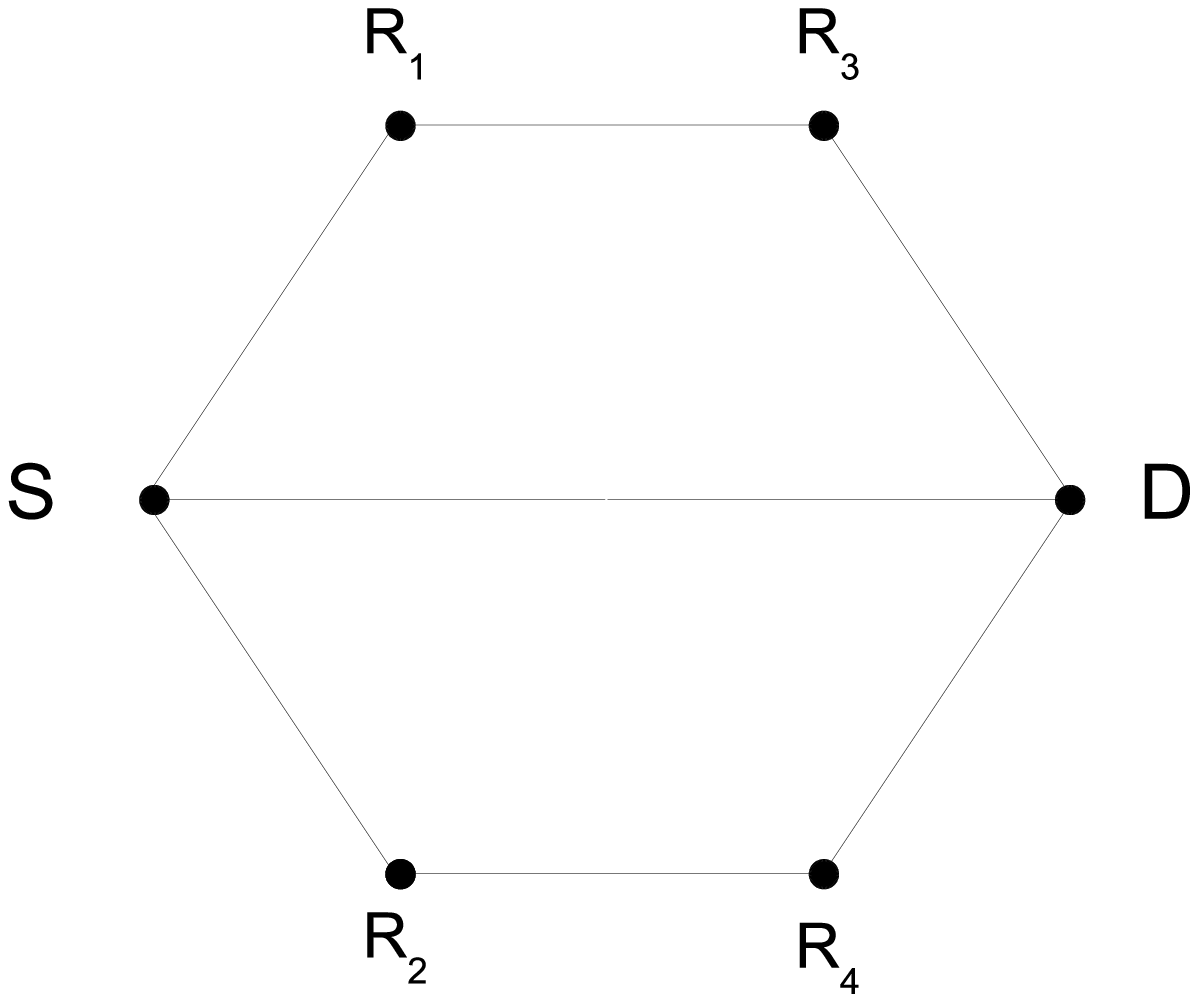}}
  \subfigure[A KPP(I, D) network with $K=2$]{\label{fig:kppid_eg}\includegraphics[width=40mm]{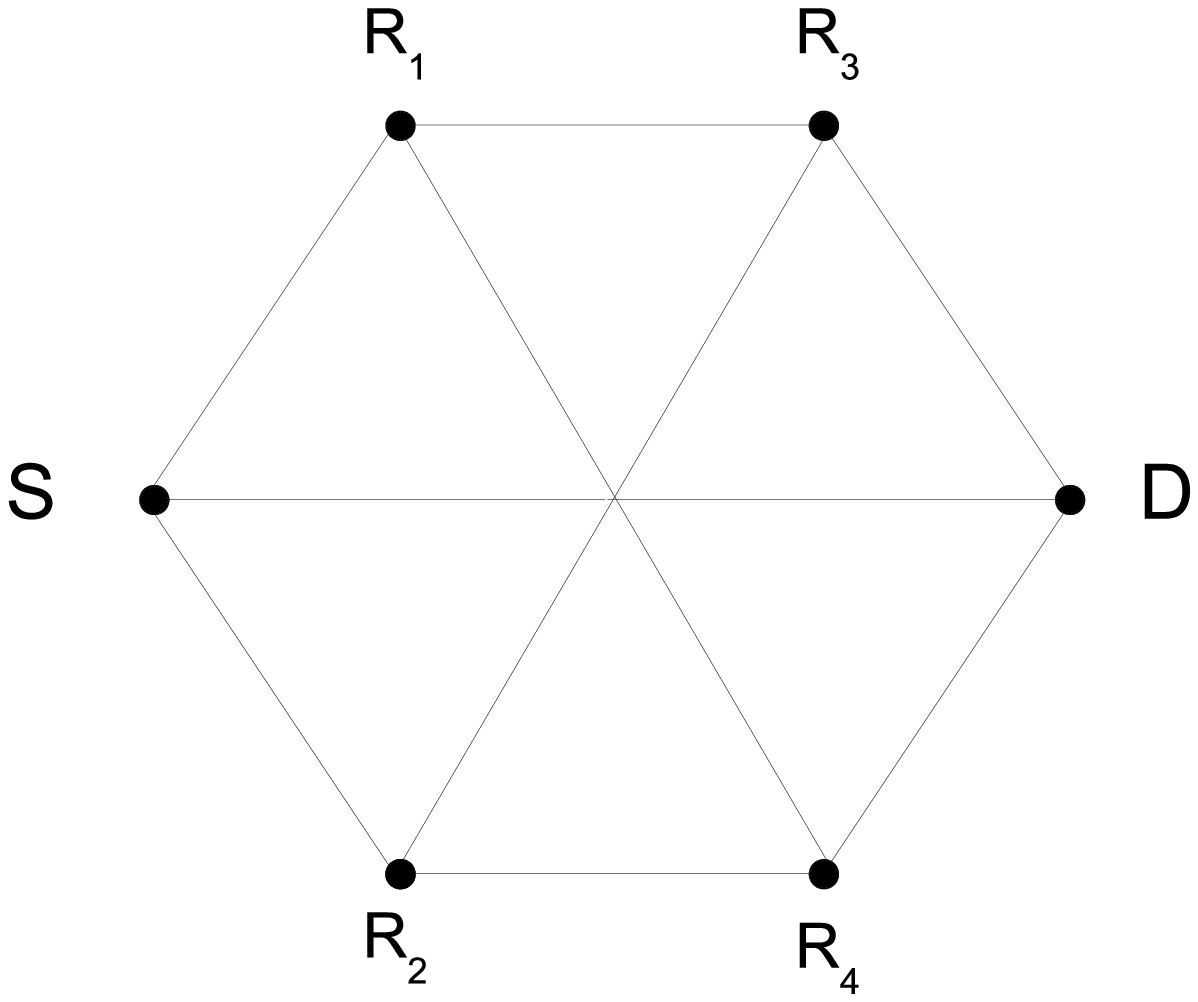}}
  \caption{Examples of KPP networks.}
  \label{fig:kpp_examples}
\end{figure}

Fig.~\ref{fig:kpp_examples} below provides examples of all four
variants of KPP networks.

For a KPP(D), KPP(I) or KPP(I, D) network, we consider the union
of the $K$ node-disjoint paths as the backbone KPP network. While
there may be many choices for the K node-disjoint paths, we can
choose any one such choice and call that the backbone KPP network.
These K relaying paths in these networks are referred to as the K
\emph{backbone} paths. A \emph{start node} and \emph{end node} of
a backbone path are the first and the last relays respectively in
the path. There are precisely $K$ start nodes in a KPP network,
which are connected at one end to the source. This remains the
case even for KPP(I) and KPP(I, D) networks. Similarly, sink node
is connected to exactly $K$ end nodes in KPP, KPP(I) and KPP(I, D)
networks.

In a general KPP network, let $P_i,i=1,2,...,K$ be the $K$
backbone paths. Let $P_i$ have $n_i$ edges. The $j$-th edge on the
$i$-th path $P_i$ will be denoted by $e_{ij}$ and the associated
fading coefficient by $g_{ij}$.

\subsubsection{Layered Networks}
A second way of generalizing  a two-hop relay network is to view
the two-hop network as a network comprising of a single layer of
relays. The immediate generalization is to allow for more layers
of relays between source and sink, with the proviso that any link
is either between nodes in adjacent layers or connects two nodes
in the same layer. We label this class of multi-hop relaying
networks as {\em layered networks}:

\bdefn \label{defn:layered} Consider a ss-ss single-antenna
bidirectional network. A network is said to be a layered network
if there exists a partition of the vertex set $V$ into subsets
$V_0,V_1,...,V_L,V_{L+1}$ such that \bit \item $V_0,V_{L+1}$
denote the singleton sets corresponding to the source and sink
respectively and for all $0 < i < L+1$, $|V_i| \geq 2$  . \item If
there is an edge between a node in vertex set $V_i$ and a node in
$V_j$, then $|i-j| \leq 1$. We assume $|V_i| > 1, i=1,2,..,L$ \eit

We will refer to $V_1,...,V_L$ as the relaying layers of the
network. A layered network is said to be fully-connected (fc) if
for any $i$, $v_1 \in V_i$ and $v_2 \in V_{i+1}$, $(v_1,v_2)$ is
an edge in the network. For fc layered networks, we include an
additional condition that the relaying layers have at least two
relays in each layer, i.e., $V_i > 1, i=1,2,...,L$. \edefn

\begin{figure}[h]
  \centering
  \subfigure[A layered network with with 4 relaying layers]{\label{fig:layered_eg.eps}\includegraphics[height=25mm]{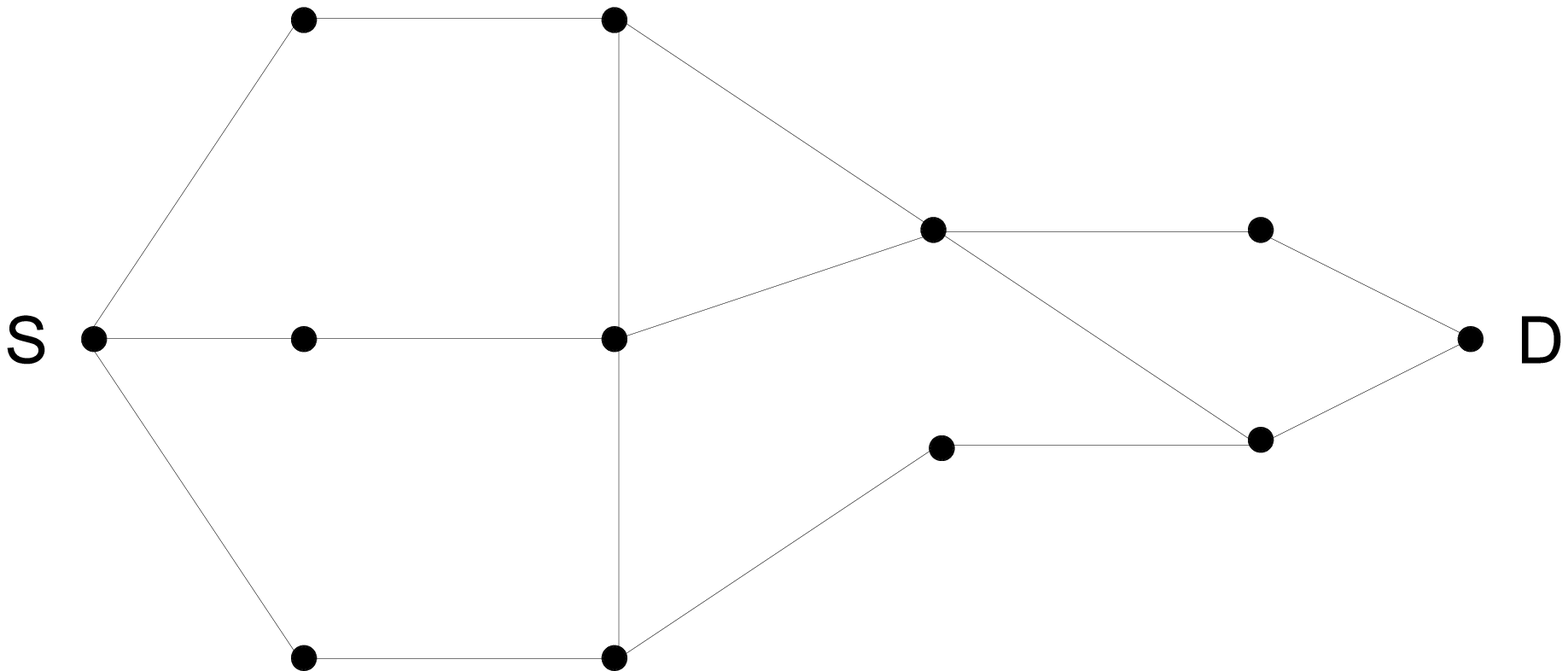}}
  \subfigure[A fully-connected layered network]{\label{fig:layered_fc_eg}\includegraphics[height=25mm]{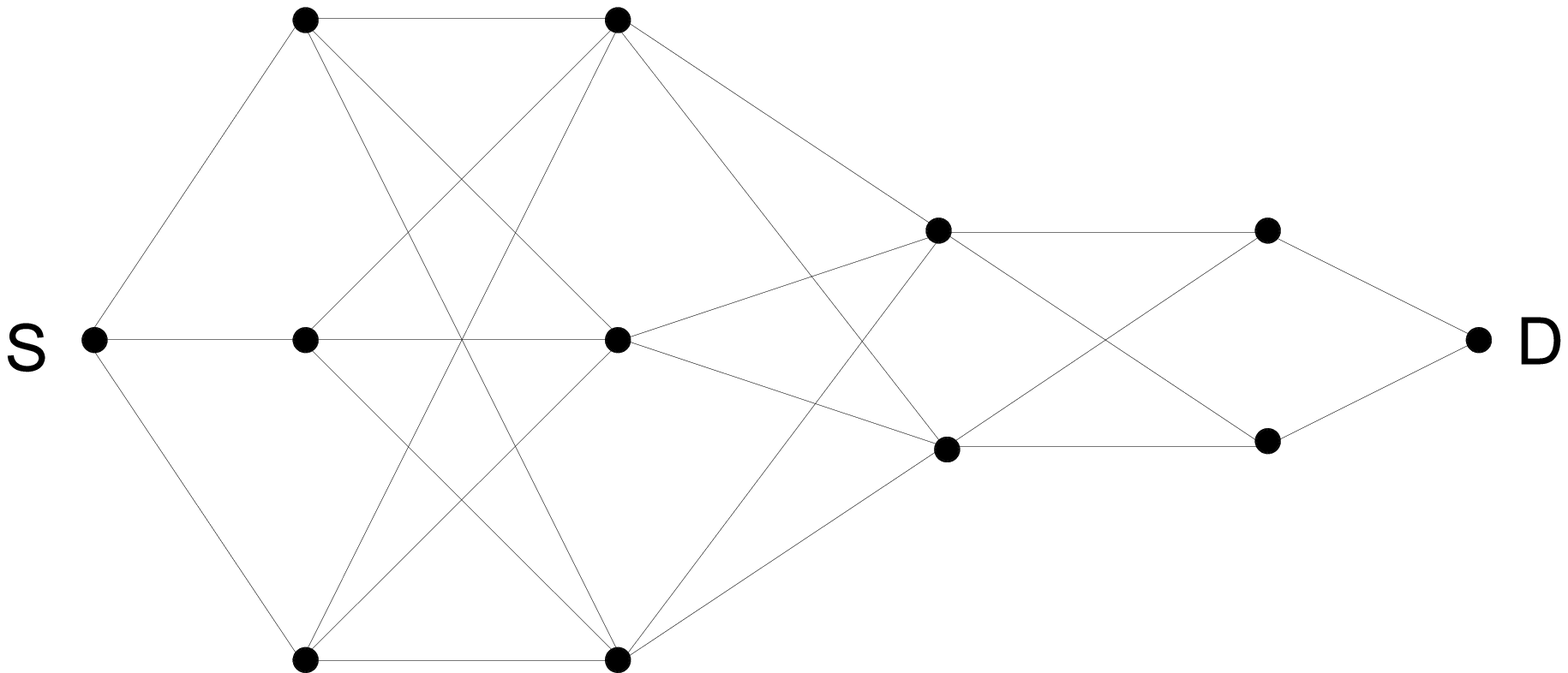}}
  \subfigure[A fully-connected layered network with intra-layer links]{\label{fig:layered_fc_general}\includegraphics[height=25mm]{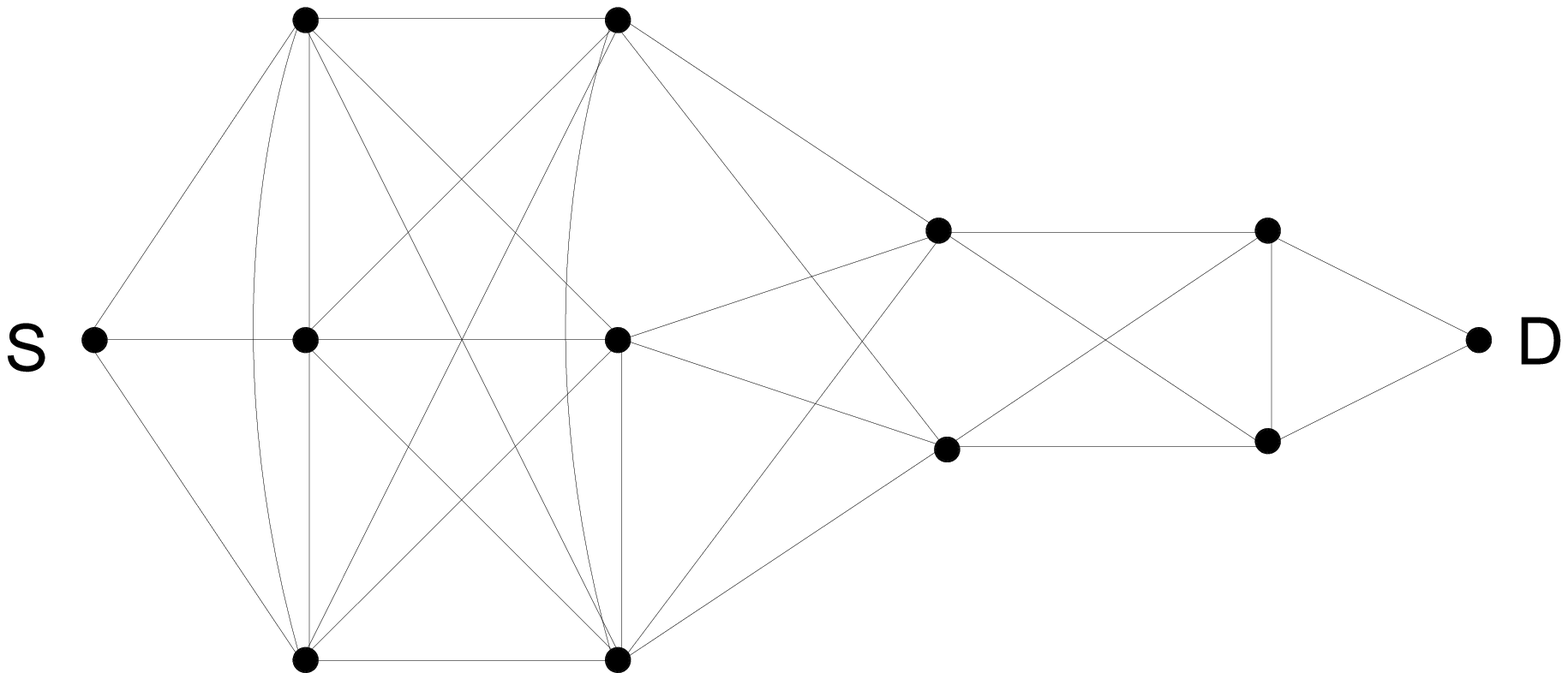}}
  \caption{Examples of layered networks}
  \label{fig:layered_regular_examples}
\end{figure}

It must be noted that a fc layered network may or may not have
links within a layer. Therefore, whenever we say fc layered
network, it applies to both networks that have intra-layer links
and those that do not have such links. Examples of both these
types of networks are shown in Fig.~\ref{fig:layered_fc_eg} and
Fig.~\ref{fig:layered_fc_general}.

\bnote \label{rem:layered_general} The definition of a layered
network is general enough to accommodate all ss-ss networks
without a direct link. This is because, any ss-ss network can be
re-drawn as a layered network with a single layer comprising of
all relays in the network and interconnections between relays. For
this general case, we give a certain achievable DMT. \enote

Every layered network will have a layer containing only the
source, and a second layer containing only the sink. In
Fig.\ref{fig:layered_regular_examples}, examples of layered
networks are given. Layered networks were also considered in
\cite{BorZheGal}, \cite{YanBelNew} and \cite{VazHea}. In
particular, \cite{BorZheGal} considered layered networks having an
equal number of relay nodes in all layers. We will refer to such
layered networks as regular networks and we will formally define
them below.

\bnote In this remark, we characterize the intersection of KPP(I)
networks and layered networks. First we observe that one is not
contained in the other. Consider the subgraph of a given KPP(I)
network graph, consisting of all the nodes of the original network
except for the source and the sink. This subgraph will have the
property that the number of node-disjoint and edge-disjoint paths
is equal to the number of relay nodes immediately adjacent to the
source. This is a key property of KPP(I) networks, which in
general, does not hold for layered networks. On the other hand,
there can be cross links between the parallel paths in a KPP(I)
network in such a way that the network cannot be viewed as being
layered.  However, these two classes of networks are not mutually
exclusive and in fact, we term networks that lie in the
intersection of the two classes as regular networks. \enote

\begin{figure}[h!]
\centering
\includegraphics[height=45mm]{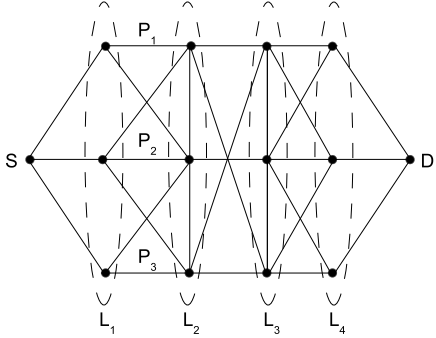}
\caption{A regular network with $4$ layers and $3$ paths
\label{fig:regular_eg}}
\end{figure}

\bdefn \label{defn: regular} A $(K, L)$ Regular network is defined
as a KPP(I) network that is also, simultaneously, a layered
network with $L$ layers of relays (see Fig.~\ref{fig:regular_eg}).
\edefn

\bnote The two-hop relay network [Fig.\ref{fig:classical_relay}]
is a KPP(I,D) network with $K = M$, $M$ being the number of
relays. In the absence of a direct link, the two-hop relay network
is a KPP(I) network with $K=M$. In fact, the two-hop relay network
without direct link is also a layered network with a single layer
of relays, thereby making it a $(M,1)$ regular network. On the
other hand, if we have a two-hop relay network with direct link
but make the additional assumption of relay isolation, then it is
a KPP(D) network with $K = M$. \enote

\subsection{Setting and Channel Model \label{sec:channel_model}}

We use uppercase letters to denote matrices and lowercase letters
to denote vectors/scalars. Vectors and scalars are differentiated
between each other by the context. Irrespective of whether it is a
scalar, vector or a matrix, boldface letters are used to denote
random entities.

Between any two adjacent nodes $v_x$ and $v_y$ of a wireless
network, we assume the following channel model.

\beq
    \bold{y} \ = \bold{ H {x} + {w}} \ ,
    \label{eq:channel_model}
\eeq where $\bold{y}$ corresponds to the received signal at node
$v_y$, $\bold{w}$ is the noise vector, $\bold{H}$ is the channel
matrix and $\bold{x}$ is the vector transmitted by the node $v_x$.

\subsubsection{Assumptions} We follow the literature in making the assumptions
listed below. Our description is in terms of the equivalent
complex-baseband, discrete-time channel.

\ben \item All channels are assumed to be quasi-static and to
experience Rayleigh fading and hence all fade coefficients are
i.i.d., circularly-symmetric complex Gaussian $\mathbb{C}\mathcal
{N} (0,1)$ random variables. \item The additive noise at each
receiver is also modeled as possessing an i.i.d.,
circularly-symmetric complex Gaussian $\mathbb{C}\mathcal {N}
(0,1)$ distribution. \item Each receiver (but none of the
transmitters) is assumed to have perfect channel state information
of all the upstream channels in the network. \footnote{However,
for the protocols proposed in this paper, the CSIR is utilized
only at the sink, since all the relay nodes are required to simply
amplify and forward the received signal.} \een

\subsection{Background}

We refer the reader to Section II-A.$1$ of the companion paper~\cite{Part1} for
a background on the diversity-multiplexing gain tradeoff (DMT).

\subsubsection{Cut-set Bound on DMT}

For each of the networks described in this paper, we can get an
upper bound on the DMT, based on the cut-set upper bound on mutual
information \cite{CovTho}. This was formalized in \cite{YukErk} as
follows:

\blem \label{lem:CutsetUpperBound} Let $r$ be the rate of
multiplexing gain at which communication between the source and
the sink is taking place. Given a cut $\omega$, there is a channel
matrix $\bold{H}_{\omega}$ connecting the input terminals of the
cut to the output terminals. Let us call the DMT corresponding to
this $\bold{H}_{\omega}$ matrix as the DMT of the cut,
$d_{\omega}(r)$. Then the DMT between the source and the sink is
upper bounded by \[ d({r}) \leq \ \min_{\omega \in \Lambda } \
\{d_{\omega}(r)\} \ ,\] where $\Lambda$ is the set of all cuts
between the source and destination.\elem

\bnote Note that the cut-set bound does not take into account
half-duplex operation of the network  and therefore applies
equally to both full and half-duplex networks.  This clearly
presents a greater challenge for half-duplex protocols. \enote

\bdefn \label{defn:DMT_Matrix} Given a random matrix $\bold{H}$ of
size $m \times n$, we define the \emph{DMT of the matrix}
$\bold{H}$ as the DMT of the associated channel $\bold{y = Hx +
w}$ where $\bold{y}$ is a $m$ length received column vector,
$\bold{x}$ is a $n$ length transmitted column vector and
$\bold{w}$ is a $\mathbb{C} \mathcal{N}(0,I)$ column vector. We
denote the DMT by $d_H(.)$ \edefn

\subsubsection{Amplify and Forward Protocols \label{sec:af}}\footnote{This section is the same as Section II-A.3 in the first part of the paper \cite{Part1} and is included here for ease of
reference.}

An AF protocol $\wp$ is a protocol $\wp$ in which each node in
the network operates in an amplify-and-forward fashion.  Such
protocols induce a linear channel model between source and sink of
the form: \beq
    \bold{y} \ = \bold{ H(\wp) {x} + {w} } \ ,
    \label{eq:channel_model}
\eeq where $\bold{y} \in \mathbb{C}^m$ denotes the signal received
at the sink, $\bold{w}$ is the noise vector, $\bold{H(\wp)}$ is
the $(m \times n)$ induced channel matrix and $\bold{x} \in
\mathbb{C}^n$ is the vector transmitted by the source. We impose
the following energy constraint on the transmitted vector
$\bold{x}$ \beqan \text{Tr}(\Sigma_x) \ := \
\text{Tr}(\mathbb{E}\{\bold{x} \bold{x}^{\dagger}\}) & \leq & n
\rho \eeqan where $\text{Tr}$ denote the trace operator, and we
will regard $\rho$ as representing the SNR on the network. We will
assume a symmetric power constraint on the relays and the source.
However it will turn out that given our high SNR perspective, the
exact power constraint is not very important. We consider both
half and full-duplex operation at the relay nodes.

Our attention here will be restricted to amplify-and-forward (AF)
protocols since as we shall see, this class of protocols can often
achieve the DMT of a network.  More specifically, our protocol
will require the links in the network to operate according to a
schedule which determines the time slots during which a node
listens as well as the time slots during which it transmits. When
we say that a node listens, we will mean that the node stores the
corresponding received signal in its buffer. When a node does
transmit, the transmitted signal is simply a scaled version of the
most recent received signal contained in its buffer, with the
scaling constant chosen to meet a transmit power constraint.
\footnote{More sophisticated linear processing techniques would
include matrix transformations of the incoming signal, but turns
out to be not needed here.} In particular, nodes in the network
are not required to decode and then re-encode.   It turns
out~\cite{AzaGamSch}, that the value of the scaling constant does
not affect the DMT of the network operating under the specific AF
protocol.  Without loss of accuracy therefore, we will assume that
this constant is equal to $1$.

It follows that, for any given network, we only need specify the
schedule to completely specify the protocol. This will create a
virtual MIMO channel of the form ${\bf y}\ = \ {\bf H}{\bf x} \ +
\ {\bf w}$ where ${\bf H}$ is the effective transfer matrix and
${\bf w}$ is the noise vector, which is in general colored.

\subsection{Certain Results from the Companion Paper\label{sec:basic_results}}

In the companion paper \cite{Part1}, we developed basic results
that will be instrumental in deriving the DMT of certain classes
of networks in this paper. A few important results among them are
given here for reference.

We proved that the correlated noise encountered at the sink of
many multi-hop networks is white in the scale of interest. This
result will be assumed throughout this paper and is formalized in
the theorem below.

\bthm \cite{Part1} \label{thm:noise_white} Consider a channel of
the form $\bold{y} = \bold{Hx} + \bold{z}$. Let
$\bold{h_1,h_2,...,h_L}$ be $L$, possibly dependent, Rayleigh
random variables. Let $\bold{G}_i,i=1,2,..,M$ be $N \times N$
matrices in which each entry is a polynomial function of the
random variables $\bold{h_1,h_2,...,h_L}$. Let $\bold{z} =
\bold{z}_0 + \sum_{i=1}^{M} \bold{G}_i\bold{z}_i$ be the noise
vector. Let $\{\bold{z}_i\}$ be i.i.d. circularly symmetric,
$n$-dimensional complex gaussian $\mathbb{C}\mathcal{N}
(\underline{0},I)$ random vectors. The random matrix $\bold{H}$ in
general depends on the random variables $\bold{h_i}$. Then
$\bold{z}$ is white in the scale of interest, i.e., \beqan
\lefteqn{Pr(\log\det(I+ \rho
\bold{H}\bold{H}^{\dagger}\Sigma^{-1}) \leq r\log\rho) } \\
& \doteq & Pr(\log\det(I+ \rho \bold{H}\bold{H}^{\dagger}) \leq
r\log\rho) \eeqan \ethm

\vspace{0.05in} We also proved a result pertaining to the DMT of
block lower triangular(blt) matrices as given below.

\bthm \cite{Part1} \label{thm:main_theorem} Consider a random blt
matrix $\bold{H}$ having component matrices $\bold{H}_{ij}$ of
size $N_i \times N_j$. Let $M := \sum_{i=1}^{N} N_i$ be the size
of the square matrix $\bold{H}$.

Let $\bold{H}^{(0)}$ be the diagonal part of the matrix $\bold{H}$
and $\bold{H}^{(\ell)}$ denote the last sub-diagonal matrix of
$\bold{H}$. Then, \ben \item $d_H(r) \geq d_{H^{(0)}}(r)$. \item
$d_H(r) \geq d_{H^{(\ell)}}(r)$.

\item In addition, if the entries of $H^{(\ell)}$ are independent
of the entries in $H^{(0)}$, then $d_H(r) \geq d_{H^{(0)}}(r) + d_{H^{(\ell)}}(r)$ \een \ethm

In this paper, we will frequently use two results on the DMT of
parallel channel that are proved in the companion paper
\cite{Part1}.

\blem \cite{Part1} \label{lem:parallel_channel} Consider a
parallel channel with $M$ links, with the $i$th link having
representation ${\bf y_i} = \bold{H_i}{\bf x_i} + \bold{w_i}$, and
let $d_i(\cdot)$ denote the corresponding DMT. Then the DMT of the
overall parallel channel is given by \beq d(r) =
\inf_{(r_1,r_2,\cdots,r_M): \ \sum_{i=1}^{M} r_i = r} \
\sum_{i=1}^{M} {d_i(r_i)} . \label{eq:parallel_dmt} \eeq \elem

\vspace{0.05in} \blem \cite{Part1} \label{lem:parallel_dependent}
Consider a parallel channel with $M$ links and repeated channel
matrices. More precisely, let there be $N$ distinct channel
matrices $H^{(1)},H^{(2)},...,H^{(N)}$, with $H^{(i)}$ repeating
in $n_i$ sub-channels, such that $\sum_{i=1}^{N}n_i = M$. Then the
DMT of such a parallel channel is given by, \beq d(r) =
\inf_{(r_1,r_2,\cdots,r_M): \ \sum_{i=1}^{N} \ n_i r_i = r} \
\sum_{i=1}^{N} {d_i(r_i)} . \label{eq:parallel_repeated_coeffs}
\eeq \elem \vspace{0.1in} We also established that diversity of
any flow in a multi-terminal network equals the min-cut between
the source and the sink of that flow. The result is given in the
below theorem.

\bthm \cite{Part1} \label{thm:mincut} Consider a multi-terminal
fading network with nodes having multiple antennas with edges
connecting antennas on two different nodes having i.i.d.
Rayleigh-fading coefficients. The maximum diversity achievable for
any flow is equal to the min-cut between the source of the flow
and the corresponding sink. Each flow can achieve its maximum
diversity simultaneously. \ethm

We established an achievable DMT region for full-duplex networks,
which is summarized in the following theorem.

\bthm \cite{Part1} \label{thm:FD_No_Direct_Path} Consider a ss-ss
full-duplex network with single antenna nodes. Let the min-cut of
the network be $M$. Let the network satisfy \emph{either} of the
two conditions below: \ben \item The network has no directed
cycles, or \item There exist a set of $M$ edge-disjoint paths
between source and sink such that \emph{none} of the $M$ paths
have shortcuts. \een Then, a linear DMT $d(r) = M(1-r)^{+}$
between a maximum multiplexing gain of $1$ and maximum diversity
$M$ is achievable. \ethm

\subsection{Results \label{sec:results}}

In this paper, we characterize the DMT of KPP networks and its
variants called KPP(D) and KPP(I) networks. We also provide an
achievable DMT for layered networks.  In many cases, the
achievable DMT equals the cut-set bound and is thereby optimal.
All the strategies are of half-duplex nature. We give explicit
protocols and code constructions for all cases.  Some of these
results were presented in conference versions of this paper
\cite{WPMC,ITA,ISIT1,ISIT2} (see also \cite{Arxiv,TechReport}).

The principal results of the paper are the following(see Table I).
\ben \item For KPP, KPP(I) and KPP(D) networks, we propose an
explicit protocol whose achievable DMT coincides with the cut-set
bound. \item For fc layered networks, we construct protocols that
achieve a DMT that is linear between the maximum diversity and
maximum multiplexing gain points. This DMT is optimal if the
number of layers is strictly less than $4$.\item For general
layered networks, we give a sufficient condition for the
achievability of a linear DMT between the maximum diversity and
the maximum multiplexing gain in
Lemma~\ref{lem:General_layered_network}. \item For KPP and layered
networks with multiple antenna nodes, we examine certain protocols
and establish achievable DMT for these protocols.  \item In
Section~\ref{sec:code_design}, we give explicit codes with short
block-lengths based on cyclic division algebras that achieve the
best possible DMT for all the schemes proposed above. We also
prove that full diversity codes for all networks in this paper can
be obtained by using codes that give full diversity on a Rayleigh
fading MIMO channel. \een

\begin{table*}
\label{tab:summary} \caption{Principal Results Summary}
\begin{center}
\begin{tabular}{||c|c|c|c|c|c|c|c||}
\hline \hline &&&&&&&\\
Network  & No of    & No of      & Direct & Upper bound on  & Achievable & Is upper bound& Reference  \\
         & sources/ & antennas    &  Link  & Diversity/DMT   & Diversity/DMT& achieved? &\\
         &  sinks   & in nodes        &        & $d_\text{bound}(r)$  & $d_\text{achieved}(r)$& &\\
\hline \hline
&&&&&&&\\
KPP(K $\geq$ 3) & Single & Single  & $\times$ & $K(1-r)^+$ & $K(1-r)^+$ & $\checkmark$& Theorem~\ref{thm:KPP}\\
&&&&&&&\\
\hline
&&&&&&&\\
KPP(D)(K $\geq$ 4) & Single & Single & $\checkmark$ & $(K+1)(1-r)^+$ & $(K+1)(1-r)^+$ & $\checkmark$& Corollary~\ref{cor:KPPD}\\
&&&&&&&\\
\hline
&&&&&&&\\
KPP(I)(K $\geq$ 3) & Single & Single & $\times$ & $K(1-r)^+$ & $K(1-r)^+$ & $\checkmark$ & Theorem~\ref{thm:KPP_I}\\
&&&&&&&\\
\hline
&&&&&&&\\
Fully & Single & Single  & $\times$ & Concave & $M(1-r)^+$ &  A linear DMT& Theorem~\ref{thm:fully_connected_layered}\\
Connected&&&&in general&& between $d_{\max}$ and &\\
Layered&&&&&& $r_{max}$ is achieved. &\\
&&&&&& $\checkmark$ for $L<4$ & Corollary~\ref{cor:optimal_layered} \\
\hline
&&&&&&&\\
Any & Single & Single  & $\times$ & Cut-set & $M(1-r)^+$ &  A linear DMT &Lemma~\ref{lem:General_layered_network}\\
network &&&&Bound&& between $d_{\max}$ and &\\
satisfying&&&&&& $r_{max}$ is achieved &\\
Lemma~\ref{lem:General_layered_network}&&&&&&&\\
\hline
&&&&&&&\\
$(K,L)$ Regular & Single & Single & $\times$ & $K(1-r)^+$ & $K(1-r)^+$ & $\checkmark$ & Theorem~\ref{thm:knregular_dmt}\\
&&&&&&&\\
\hline \hline
\end{tabular}
\end{center}
\end{table*}

\subsection{Outline \label{sec:outline}}

In Section~\ref{sec:half_duplex_isolated}, we focus on half-duplex
KPP networks and present protocols achieving optimal DMT for $K
\geq 3$. We extend this result to KPP(D) networks at the end of
this section. In Section~\ref{sec:half_duplex}, KPP(I) networks
with half-duplex relays are considered, and schemes achieving
optimal DMT are presented for KPP(I) networks allowing certain
types of interference. In Section~\ref{sec:layered}, we consider
layered networks and show that a linear DMT between maximum
multiplexing gain and maximum diversity is achievable. In
Section~\ref{sec:multiple_antenna}, we consider multi-antenna
layered and KPP networks and give an achievable DMT. Finally, in
Section~\ref{sec:code_design}, we give explicit CDA based codes of
low complexity for all the DMT optimal protocols.

\section{Half-Duplex Networks with Isolated Paths - KPP Networks \label{sec:half_duplex_isolated}}

In this section, we consider KPP and KPP(D) networks with
single-antenna nodes operating under the half-duplex constraint.
Discussion on KPP(I) nertworks is deferred to a later section.

\subsubsection{The Cut-Set Bound for KPP Networks}
\label{sec:kpp_cutset}

The cut-set (Lemma~\ref{lem:CutsetUpperBound}) upper bound on DMT
for the class of KPP networks is given by: \beqn
    d(r) \ \leq \ K(1-r)^{+} ,
\eeqn   and our aim is to design protocols that attain this
cut-set bound.

In this section we will present protocols for KPP networks, that
permit the cut-set bound to be attained for all $K \geq 3$.  For
the case $K=2$, we present a lower bound to the DMT that can be
achieved using an AF protocol.

\subsection{Full-Duplex KPP Networks}

To start with, we will consider full-duplex KPP networks and
review the results of the first part of this paper \cite{Part1} as
applied to KPP networks. Theorem~\ref{thm:FD_No_Direct_Path} gives
a lower-bound on the DMT of certain classes of networks with
full-duplex relays. The result applies to KPP and KPP(I) networks,
even with multiple antenna nodes, and is given in the corollary
below.

\bcor \label{cor:FD_KPP} For the full-duplex KPP networks without
direct link (i.e. KPP and KPP(I) networks), with potentially
multiple antenna nodes, a DMT of $d_{\max}(1-r)^{+}$ is
achievable. \ecor

When nodes have a single antenna, $K(1-r)^{+}$ turns out to be the
optimal DMT. However, the characterization of optimal DMT becomes
tricky when the relays are half-duplex. Most articles in the
literature deal with half-duplex networks by first analyzing the
networks from a full-duplex perspective and then translating the
results by conceding rate loss by a factor of $2$. In this and the
next section, we will show that half-duplex multi-hop networks
(specifically KPP networks) operating under a suitable schedule
can achieve the same DMT performance as full-duplex networks.

\subsection{Protocols Achieving the Cut-Set Bound for $K \geq 3$\label{sec:miso_protocols}}

As made clear in the introduction, we restrict our attention in
this paper to the class of AF protocols under which node
operations are restricted to scaling and forwarding. To completely
specify the manner in which the network is operated, it remains
only to identify a schedule of operation.

In all of the schedules considered here, node operations are
periodic with a period of $N$ time slots.   Thus all edge
activations are periodic as well, and we will refer to $N$ as the
cycle length of the protocol. We shall describe all our protocols
in a simple manner, as an edge coloring scheme. Let $C = \{c_1,
c_2, \ldots, c_N\}$ be the set of $N$ colors used in the scheme.
Each color in $C$ signifies a distinct time slot within the
$N$-slot protocol. There is a natural order between any two
colors, inherited from the time-slots they represent within a
cycle.
 All the edges in the network are assigned a subset of colors from
the set $C$. The subset of colors assigned to the edge $e_{ij}$
will be denoted by $A_{ij}$. Each color in $A_{ij}$ represents the
time instants during which the edge $e_{ij}$ is active.
\footnote{We assume that the network is in operation for
sufficient amount of time, so that if an edge is active, the node
at beginning of the edge always has a symbol to transmit.}

\bdefn \label{defn:orth_protocol} A half-duplex protocol is said
to be an orthogonal protocol if at any node, at a given time
instant, only one of the incoming or outgoing edges is active. An
orthogonal protocol for a KPP network is said to have the
equi-activation property if each edge along a given backbone path
in the KPP network is activated an equal number of times. \edefn

\bnote The definition of orthogonal protocol is similar in spirit
to the definition in the networking literature \cite{KodNan},
where a network is said to have orthogonal channels if
interference is avoided at all nodes and each node is permitted to
communicate with at most one other node at any given time. \enote

All of the orthogonal protocols employed in this paper will
satisfy the equi-activation property and hence, whenever in the
sequel we speak of an orthogonal protocol, we will mean an
orthogonal protocol that in addition, satisfies the
equi-activation property.

\bprop \label{prop:protocol_color} Every orthogonal protocol can
be described as an edge coloring of the network satisfying the
following constraints. Conversely, every edge coloring satisfying
the following constraints describes an orthogonal protocol.

\beqa  A_{i1} \cap A_{j1} & = & \phi, i  \neq  j. \\
A_{i{n_i}} \cap A_{jn_j} & = & \phi, i \neq j. \\
A_{ij} \cap A_{i(j+1)} & = & \phi, j =1,2,...,{n_i}-1. \\
|A_{ij}| & = & m_i, j = 1,2,...,n_i. \eeqa \eprop

\begin{figure}[h!]
\centering
\includegraphics[width=60mm]{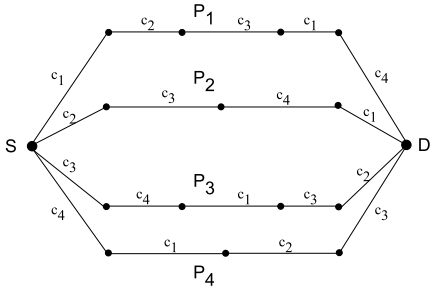}
\caption{KPP network orthogonal protocol
\label{fig:kpp_illustration}}
\end{figure}

The first constraint corresponds to the fact that for an
orthogonal protocol, only one outgoing edge is active at the
source. Similarly the second constraint corresponds to the fact
that for an orthogonal protocol, only one incoming edge is active
at the sink. The third constraint captures the half-duplex nature
of the protocol. The last constraint corresponds to the
equi-activation property of the orthogonal protocols considered
here. A coloring which respects all the above constraints for an
example KPP network is given in Fig.~\ref{fig:kpp_illustration}.
In this example, the protocol has a cycle length of $4$
time-slots, represented by the colors $c_1, c_2, c_3,$ and $c_4$
in order. Within a given cycle, each edge is activated during
exactly one time-slot, specified by a color, $c_i,$ $1 \leq i \leq
4$.

\bdefn \label{defn:rate_orth_protocol} The rate, R of an
orthogonal protocol is defined as the ratio of the number of
symbols transmitted by the source to the total number of time
slots. In terms of the notation above, we have \beqn R =
\frac{\sum_{i=1}^{K} m_i }{N}.  \eeqn \edefn

\begin{figure}[h!]
\centering
\includegraphics[width=60mm]{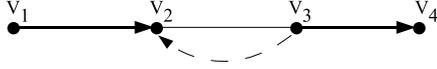}
\caption{Back-flow on a path \label{fig:backflow}}
\end{figure}

\vspace{0.1in} \bdefn Consider a KPP network. Let $v_1, v_2, v_3,
v_4$ be four consecutive vertices lying on one of the $K$ parallel
paths leading from the source to the sink. Let $v_1$ and $v_3$
transmit, thereby causing the edges $(v_1, v_2)$ and $(v_3, v_4)$
to be active. Due to the broadcast and interference constraints,
transmission from $v_3$ interferes with the reception at $v_2$.
This is termed as back-flow, and is illustrated in
Fig.\ref{fig:backflow} \edefn

Back-flow can be avoided by ensuring that there are at least two
inactive edges between any two active edges along a backbone path.
We formalize this below:

\bnote \label{note:no_backflow} An orthogonal protocol avoids
back-flow if the edge coloring satisfies: \beqan
    A_{ij} \cap A_{i(j+2)} = \phi, j = 1,2,...,{n_i}-2.
\eeqan \enote

\vspace*{0.1in}

From the remark above, it is evident that any three adjacent edges
$e_{ij}, e_{i(j+1)},$ and $e_{i(j+2)}$ will map to disjoint sets
of colors when the coloring scheme corresponds to an orthogonal
protocol that avoids back-flow.  Moreover, repetition of the same
set of colors along every third edge is permitted under the
constraints of an orthogonal protocol avoiding back-flow. This
suggests an easy way of arriving at an edge coloring that
satisfies the constraints.   To color the edges along a backbone
path in the KPP network, we will first select three disjoint
subsets of colors, order these and then repeat them cyclically
along the edges of the path moving from source to sink.    For
reasons that will become apparent later, we may deviate from this
rule with respect to the coloring of the last edge along any
backbone path. More formally, for any given path $P_i, i \in [K]$,
let $G_i = ( G_{i0},G_{i1},G_{i2} )$ be the set of three colors
repeated cyclically starting from the first node and let $F^i$
specify the subset assigned to the last edge $e_{i{n_i}}$ along
path $P_i$. Then the corresponding edge coloring is given by
 \beqa {A_{ij}} & = &
    \left \{ \begin{array}{ccc}
        G_{i (j \pmod 3)} ,& &  \ \mbox{$j \neq n_i$} \\
        F^i ,& &  \ \mbox{$j = n_i$} \\
    \end{array}
    \right. \ .  \label{eqref:color_class}
\eeqa Henceforth, we will use $G^i,F^i$ as descriptors of the
orthogonal protocol.   We show next how by adding a few additional
constraints, we obtain a protocol that achieves the cut-set bound.

\blem \label{lem:dmt_opt} Consider a KPP network. If an orthogonal
protocol satisfies the following constraints:

\ben \item The rate of the protocol is equal to one. \item In
every cycle, the sink receives an equal number of symbols from
each of the $K$ backbone paths. \item The protocol avoids
back-flow. \een

Then the protocol achieves the cut-set bound, i.e., \beqn d(r) =
K(1-r)^+  . \eeqn \elem

\bpf Consider a situation when a KPP network starts operating
under an orthogonal protocol satisfying the constraints above.
Since the sink receives an equal number of symbols from each of
the $K$ backbone paths, symbols from the source should be
transmitted via each of the $K$ backbone paths equally often.
Also, the protocol being of rate $1$, a symbol is transmitted from
the source in every time-slot. For this reason, we treat the
transmission from the source in blocks constituting of $K$
symbols, each denoted by a vector
$\bold{x}=[\bold{x}_1,\cdots,\bold{x}_{K}]^t$, in which a symbol
$\bold{x}_i$ takes on backbone path $P_i$. Clearly, the delay
encountered by each symbol depends on the length of the path it
traverses. However, it also depends on the schedule of edge
activations dictated by the protocol. For example, in a path
$P_1$, consider two adjacent edges $e_{1j}$ and $e_{1(j+1)}$ with
a common node $v$. If $e_{1j}$ and $e_{1(j+1)}$ are assigned with
colors $c_j$ and $c_{j+2}$(Here, we assume that time-slots are in
the same order as that of index of the color), then $v$ incurs a
delay of $1$ time-slot due to the schedule. Thus each component
$\bold{x}_i$ experiences the sum of delays along all the nodes in
the path $P_i$.

This difference in delay experienced by each component
$\bold{x}_i$ in $\bold{x}$ may be large enough so that the
information corresponding to $\bold{x}_i$ arrives at the sink node
during one cycle of the protocol, whereas the information
corresponding to $\bold{x}_{i+1}$ may arrive at the sink node
during some other cycle. Therefore, there is a need to synchronize
these information symbols so that they arrive at the same cycle of
the protocol. This can be done by adjusting the delays of each
path at the source (by a multiple of $K$) so that all the symbols
$\bold{x}_1,...,\bold{x}_K$ in a block of data are received in the
same cycle at the destination. However, the activation of the
edges connecting to the sink need not be in the same order as that
in which edges connected to the source are activated. Therefore,
within a cycle, the symbols might be received in a different
order. Hence, we shuffle the input appropriately within each cycle
(it corresponds to multiplication by a permutation matrix, $P$),
so that we get an input output relation between the input
$\bold{x}$ and the output $\bold{y}$ as,

\beqa
\bold{y} & = & \left[ \begin{array}{cccc} \bold{g}_1 &  &  & \\
  & \bold{g}_2 &  &  \\
   &   & \ddots &   \\
 &  &  & \bold{g}_K \\
\end{array}
\right]\bold{x} + \bold{w}, \ \label{eq:kpp_channel} \eeqa

where $\bold{g}_i$ denotes the product fading coefficient of the
path $P_i$, and $\bold{w}$ is the noise vector at the sink, which
is white in the scale of interest.

Note that the initial delay, D, encountered in the transmission is
not accounted for in the channel model given in
\eqref{eq:kpp_channel}. However, when the network operates
continuously for long duration $T$, this delay does not affect in
terms of loss in rate since the rate loss factor of
$\frac{T-D}{T}$ can be made arbitrarily small when $T$ becomes
large. We will always make this assumption whenever we are dealing
with KPP, KPP(I), KPP(D) networks.

The DMT of the above matrix, $\bold{H}$ can be easily computed to
be equal to \beqa d(r) & = & K(1-r)^{+}. \eeqa \epf

\bnote In later sections, whenever we refer to orthogonal
protocols for KPP networks satisfying conditions in
Lemma~\ref{lem:dmt_opt}, we assume that the protocol will induce a
channel as given in (8), neglecting the initial delay. This is
justified in the proof of the above lemma. \enote

We will now show that the requirements of Lemma~\ref{lem:dmt_opt}
can be met in the case of KPP networks with $K \geq 4$:  \bthm
\label{thm:K_geq_4} For any KPP network with $K \geq 4$, there
exists an explicit protocol achieving the cut-set bound on DMT.
\ethm

\bpf  We now establish an orthogonal protocol satisfying the
conditions in Lemma~\ref{lem:dmt_opt} for the case when $K \geq
4$.  We choose the cycle length of the protocol to be equal to the
number of parallel paths $K$. By our earlier observations, it
suffices to identify the coloring subsets $\left\{
(G_{i0},G_{i1},G_{i2} ), \ \ \{F_i\} \mid 1 \leq i \leq K
\right\}$ used by the protocol. These subsets are identified
below: \beqa G_i & = & [ \{c_i\}, \{c_{i+1}\},\{c_{i+2}\} ], \\
F_i & = &  \{ c_{i+3} \} . \eeqa   Verification that this coloring
scheme meets all the constraints of Lemma~\ref{lem:dmt_opt} is
straightforward.  It follows that this orthogonal protocol
achieves the cut-set bound . \epf

The previous theorem handles the case when there are $4$ or more
backbone paths in the KPP network. When $K=3$ it is not possible
to identify efficient protocols that avoid back-flow. The lemma
below reassures us that this does not prevent us from achieving
the cut-set bound on the DMT.

\blem \label{lem:back_flow_lower_bound} Consider a KPP network
operating under an orthogonal protocol, which on neglecting the
effect of back-flow, induces a block-diagonal channel matrix.
IThen, the DMT of the protocol taking into account the effect of
back-flow, is lower bounded by the DMT when neglecting the effect
of back-flow . \elem

\bpf  The presence of back-flow creates entries in the strictly
lower-triangular portion of the induced channel matrix. Since the
DMT of a lower triangular matrix is lower bounded by the DMT of
the corresponding diagonal matrix (by Theorem~\ref{thm:main_theorem}), we have that the system with back-flow will yield a
DMT no worse than the one without back-flow. \epf

We will now exploit Lemma~\ref{lem:back_flow_lower_bound} to
construct a protocol achieving the cut-set bound for the KPP
network with $K=3$.

\bthm \label{thm:k3_miso} For any KPP network with $K = 3$, there
exists an explicit protocol achieving the cut-set bound on DMT.
\ethm

\bpf For every parallel path, $P_i$, define
   \beqa {a_i} & = &
    \left\{ \begin{array}{ccc}
        1 ,& &  \ \mbox{$n_i = 1$  mod $3$} \\
        0 ,& &  \ \mbox{$n_i \neq 1$ mod $3$} \\
    \end{array}
    \right.  \ .
    \eeqa

Without loss of generality we assume that the paths are ordered
such that for the first $l$ paths, $a_i=1$ followed by the paths
for which $a_i=0$. We give a protocol for various possible values
of $l$. \bit

\item \emph{Case 1: ($l = 0, 1, \ or \ 3$)}

In this case, we will give a protocol that avoids back-flow, uses
all paths equally, and achieves rate 1. By
Lemma~\ref{lem:dmt_opt}, this protocol will achieve the cut-set
bound.

We will specify the orthogonal protocol by specifying the
activation sets $G_i = ( G_{i0},G_{i1},G_{i2} )$ and $F_i$ for all
$i$. We begin by setting $F_i=G_{i(n_i-1 \pmod  3)}$. The set of
colors used is $C = \left\{ c_0, c_1, c_2\right\}$.
\vspace*{0.1in}

For $l=0$,

$G_i =
    \left\{ \begin{array}{ccc}
        \mbox{$( \{c_i\}, \{c_{i+2}\},\{c_{i+1}\} )$} ,& &  \ \mbox{$n_i = 0$  mod $3$} \\
        \mbox{$( \{c_i\}, \{c_{i+1}\},\{c_{i+2}\} )$} ,& &  \ \mbox{$n_i = 2$  mod $3$} \\
    \end{array}
    \right. \ .$

\vspace*{0.2in}

For $l=1$,

$G_1 = ( \{c_0\}, \{c_1\},\{c_2\} )], $

\vspace*{0.05in}

$G_2 =
    \left\{ \begin{array}{ccc}
        \mbox{$( \{c_1\}, \{c_0\},\{c_2\} )$} ,& &  \ \mbox{$n_2 = 0$  mod $3$} \\
        \mbox{$( \{c_1\}, \{c_2\},\{c_0\} )$} ,& &  \ \mbox{$n_2 = 2$  mod $3$} \\
    \end{array}
    \right. \ ,$

\vspace*{0.05in}

$G_3 =
    \left\{ \begin{array}{ccc}
        \mbox{$( \{c_2\}, \{c_0\},\{c_1\} )$} ,& &  \ \mbox{$n_3 = 0$  mod $3$} \\
        \mbox{$( \{c_2\}, \{c_1\},\{c_0\} )$} ,& &  \ \mbox{$n_3 = 2$  mod $3$} \\
    \end{array}
    \right. \ .$

\vspace*{0.2in}

For $l=3$,

$G_i = ( \{c_i\}, \{c_{i+1}\},\{c_{i+2}\} ) . $

\vspace*{0.1in}

\item \emph{Case 2: ($l = 2$)}

For $l = 2$, it turns out that there is no orthogonal protocol
that satisfies all the conditions of Lemma~\ref{lem:dmt_opt}. To
handle this situation, we shall now come up with a protocol which
satisfies conditions $1$ and $2$ of Lemma~\ref{lem:dmt_opt}, but
does not satisfy condition $3$, i.e., the protocol will not avoid
back-flow. Then, we utilize Lemma~\ref{lem:back_flow_lower_bound}
to establish that the DMT for this protocol is equal to the
cut-set bound.

Consider the protocol having the following descriptor:

$G_i = ( \{c_i\}, \{c_{i+1}\},\{c_{i+2}\} )$ and $F_i=G_{i(n_i-1
\pmod  3)}$.

\vspace*{0.1in}

After this assignment, we make the following modifications to
$A_{ij}$,

\vspace*{0.1in}

$A_{3(n_3)} = \{c_2\},$

$A_{3(n_3-1)} = \{c_0\}$, if $n_3 = 2 \pmod 3$.

\vspace*{0.1in}

It can be checked that the protocol satisfies the first two
conditions of Lemma~\ref{lem:dmt_opt}. While condition $3$ is not
satisfied, because back-flow is present, by
Lemma~\ref{lem:back_flow_lower_bound}, the presence of back-flow
does not worsen the DMT and therefore, the protocol achieves the
cut-set bound on DMT. \eit \epf

\subsection{The Case of Two Parallel Paths}
We now proceed to the handle the last remaining case, namely the
case when $K=2$. As in the previous case, it turns out that it is
impossible to have cut-set bound achieving orthogonal protocols
that avoid back-flow. Furthermore, it is not even possible to
construct a rate-$1$ orthogonal protocol with or without back-flow
in general. We will first determine an upper bound to the rate of
any orthogonal protocol for a KPP network with $K=2$. This will
then be followed by the presentation of an orthogonal protocol
which achieves this upper bound.

\bthm \label{thm:k2_max_rate} For any KPP network with $K=2$, the
maximum
 achievable rate for any orthogonal protocol is given by   \beqa R_{max} & \leq &
    \left\{ \begin{array}{ccc}
        1 ,& &  \ \mbox{$n_1 + n_2 = 0$ \ mod \ $2$} \\
        \frac{2n_2 - 1}{2n_2} ,& &  \ \mbox{$n_1 + n_2 = 1$ \ mod \ $2$}
    \end{array}
    \right. , \label{eqref:k2_rmax}
    \eeqa
    where $n_1 \leq n_2$.
\ethm

\bpf Any given orthogonal protocol can be represented as a
coloring of the edges satisfying the conditions in
Prop.~\ref{prop:protocol_color}. For this case of $K=2$, it will
be found convenient to relabel the $n_1+n_2$ edges in the network
so as to form a cycle $l_1,l_2,...,l_{n_1+n_2}$ of length
$n_1+n_2$. The specific relabeling is given by \beqan l_j & = &
    \left\{ \begin{array}{cc}
        e_{1j} \ ,& \ j \leq n_1 \\
        e_{2(n_2+n_1+1-j)} \ , &  \ n_1 < j \leq n_1+n_2
    \end{array}
    \right. \ .
\eeqan We associate edge $l_1,l_2,...,l_{n_1+n_2}$ with color
$D_1,D_2,...,D_{n_1+n_2}$ so that\beqan D_j & = &
    \left\{ \begin{array}{cc}
        A_{1j} \ ,&  \ j \leq n_1 \\
        A_{2(n_2+n_1+1-j)} \ ,& \ n_1 < j \leq n_1+n_2
    \end{array}
    \right. \ .
\eeqan with a single constraint, \beqa D_j \cap D_{(j+1) \ \pmod \
(n_1+n_2)} = \phi, \ \ \forall j=1,2,...,n_1+n_2,
\label{eqref:con0} \eeqa that simultaneously meets the first three
conditions laid out in Prop.~\ref{prop:protocol_color}.

Now suppose we have a coloring scheme with N colors. Then each
color can be assigned to at most $\lfloor\frac{n_1+n_2}{2}\rfloor$
edges. This is because if more edges were assigned with the same
color, then the half-duplex constraint would be violated. So by
counting edge-color pairs in two different ways, we obtain the
bound,  \beqa
  \sum_{i = 1}^{2} \ \sum_{j = 1}^{n_i} \ |A_{ij}| & \leq & \left \lfloor\frac{n_1+n_2}{2} \right \rfloor N, \nonumber \\
  \text{ie.}, n_1m_1 + n_2m_2 & \leq & \left \lfloor\frac{n_1+n_2}{2} \right \rfloor N, \label{eqref:con1}
  \eeqa where $m_i = |A_{ij}| \forall j \in [n_i]$.
The half-duplex constraint implies that,
 \beqa
  2{m_1} \leq {N}, \label{eqref:con2}\\
  2{m_2} \leq {N} \label{eqref:con3}.
  \eeqa

To find the maximum rate, we pose the maximization problem:
maximize rate R = $(\frac{m_1}{N}+\frac{m_2}{N})$ subject to
(\ref{eqref:con1}), (\ref{eqref:con2}), and (\ref{eqref:con3}).
This can be easily solved to obtain, \beqan
  \frac{m_1}{N} & = & 0.5, \\
  \frac{m_2}{N} & = & \frac{1}{n_2}\left \lfloor\frac{n_1+n_2}{2} \right\rfloor \ - \
  \frac{n_1}{2n_2}.
\eeqan

As a result, the maximum rate of the protocol is upper-bounded as
   \beqa R_{max} & \leq &
    \left\{ \begin{array}{cc}
        1 , &  \ n_1 + n_2 = 0 \pmod 2 \\
        \frac{2n_2 - 1}{2n_2} ,& \ n_1 + n_2 = 1 \pmod 2
    \end{array}
    \right. ,  \label{eqref:kequals2}
    \eeqa
    where $n_1 \leq n_2$.

\epf

\bconstr \label{constr:k2_max_rate} This construction establishes
an orthogonal protocol for $K = 2$ which achieves the maximum
rate. By Prop.~\ref{prop:protocol_color}, it is sufficient to
specify the coloring subsets $A_{ij}$ $ \forall i,j$ or
equivalently, to specify the subsets $D_j$.

\emph{Case 1:} $(n_1+n_2) = 0 \pmod{2}$

For this case we choose $2$ as the cycle length of the protocol in
our construction. Accordingly let the set of colors be $C = \{c_0,
c_1\}$. Set \beqan D_j & = &
    \left\{ \begin{array}{ccc}
        \{c_0\} \ , & \ j = 1,3,...,n_1+n_2-1 \\
        \{c_1\} \ , & \ j = 2,4,...,n_1+n_2
    \end{array}
    \right. ,
\eeqan which essentially corresponds to coloring the cycle formed
by the network alternately with the two colors $c_0$ and $c_1$.

\emph{Case 2:} $(n_1+n_2) = 1 \pmod{2}$

The coloring prescribed in Case $1$ does not work here, since the
cycle is of odd length. Therefore, we resort to a different
coloring in this case.

We have the set of colors $C = \{c_1, c_2,...,c_N\}$, where
$N=2n_2$. We will add colors to $D_j$ using the following
algorithm.

\ben \item Step 1: $D_j \leftarrow \phi \ \forall j \in
\{1,2,...,n_1+n_2\}$.

\item Step 2: Now we will add colors to each of the set $D_j$
using the following algorithm. In the algorithm, whenever we refer
to $D_j$, with $j \notin \{0, 1, 2, \ldots, n_1+n_2\}$, we mean
$D_j = D_{j \pmod{n_1+n_2}}$ and with $j = 0$, we mean $D_j =
D_{n_1+n_2}$.

$\{$

$\hspace{0.1in}t \leftarrow 1;$

$\hspace{0.1in} \text{for} \ k = 1 \ \text{to} \ n_2 \ \text{in
steps of} \ 1:$

$\hspace{0.1in}\{$

$\hspace{0.2in} \text{for} \ i = 1 \ \text{to} \ n_1+n_2-1 \
\text{in steps of} \ 1:$

$\hspace{0.2in}\{$

$\hspace{0.3in} \text{if} \ i \ \text{is odd}, D_{i-k+1}
\leftarrow D_{i-k+1} \cup \{c_t\};$

$\hspace{0.3in} \text{if} \ i \ \text{is even}, D_{i-k+1}
\leftarrow D_{i-k+1} \cup \{c_{t+1}\};$

$\hspace{0.2in}\}.$

$\hspace{0.2in}t \leftarrow t+2;$

$\hspace{0.1in}\}.$

$\}.$

\een \econstr

\bprop \label{prop:2PP_Constr_Rate} The orthogonal protocol shown
in Construction~\ref{constr:k2_max_rate} achieves the maximum rate
given in Theorem ~\ref{thm:k2_max_rate}. \eprop

\begin{figure*}
  \subfigure[Time slots 1 and 2]{\label{fig:illustration_1}\includegraphics[width=40mm]{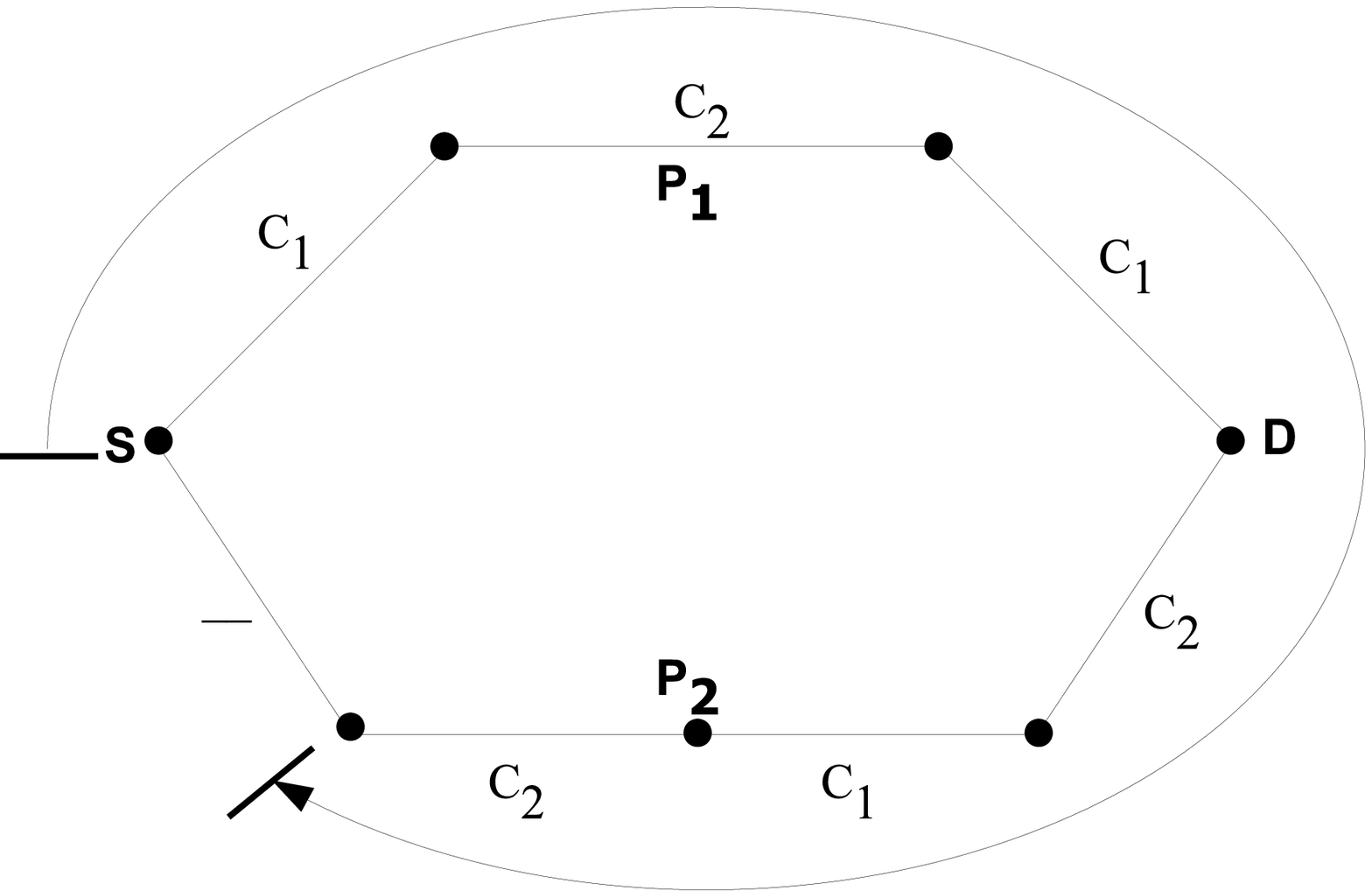}}
  \subfigure[Time slots 3 and 4]{\label{fig:illustration_2}\includegraphics[width=40mm]{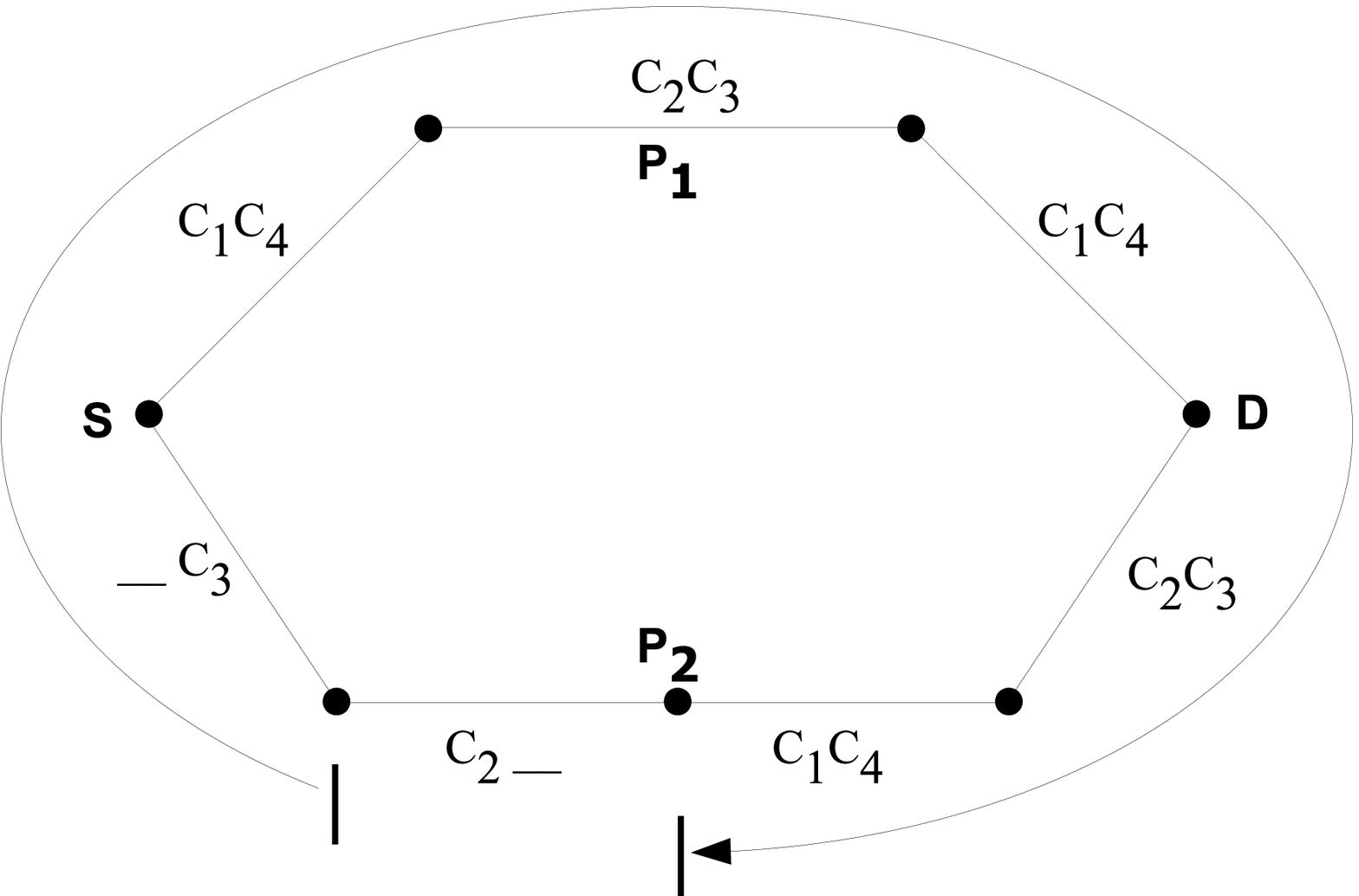}}
  \subfigure[Time slots 5 and 6]{\label{fig:illustration_3}\includegraphics[width=40mm]{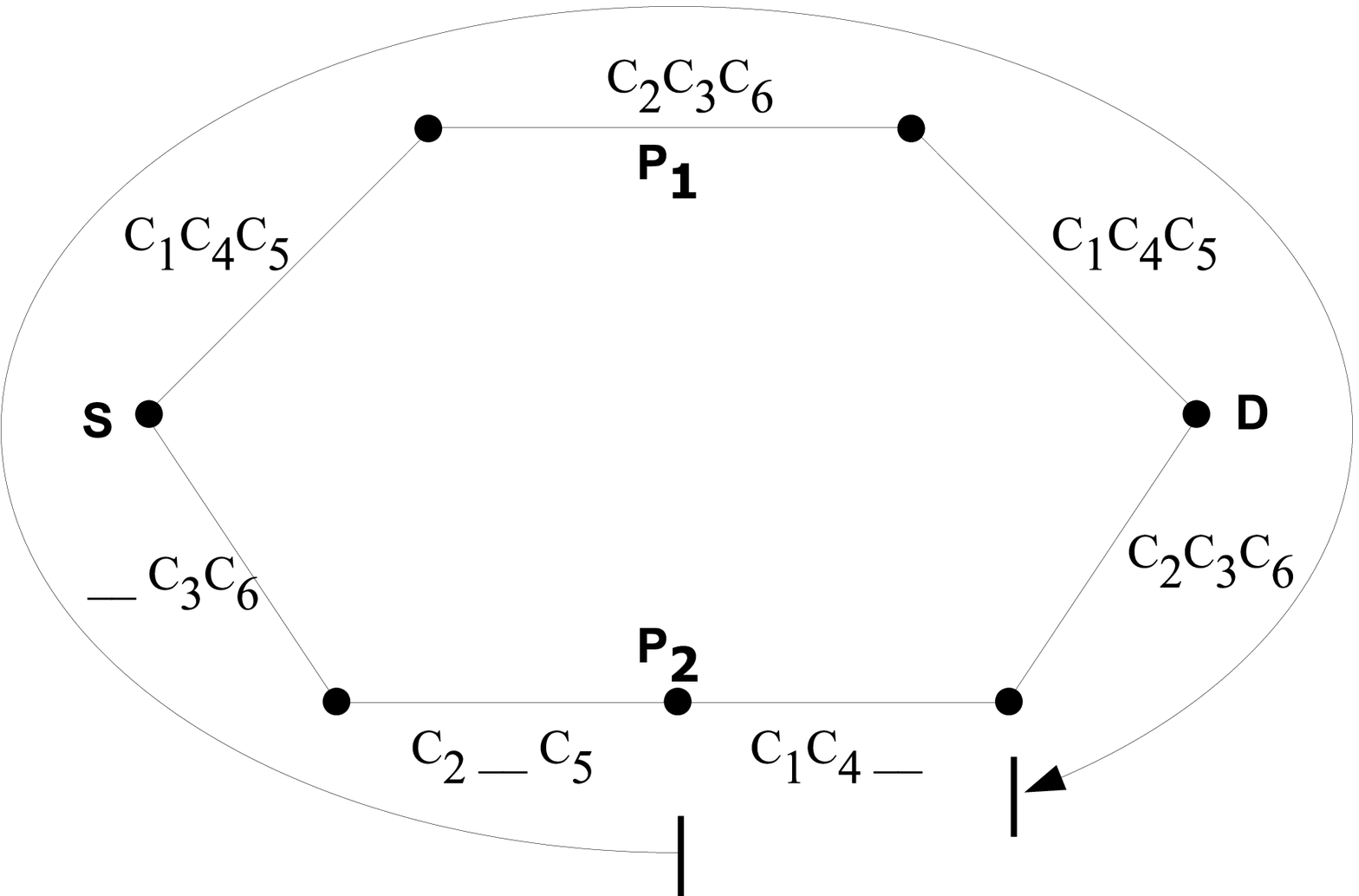}}
  \subfigure[Time slots 7 and 8]{\label{fig:illustration_4}\includegraphics[width=40mm]{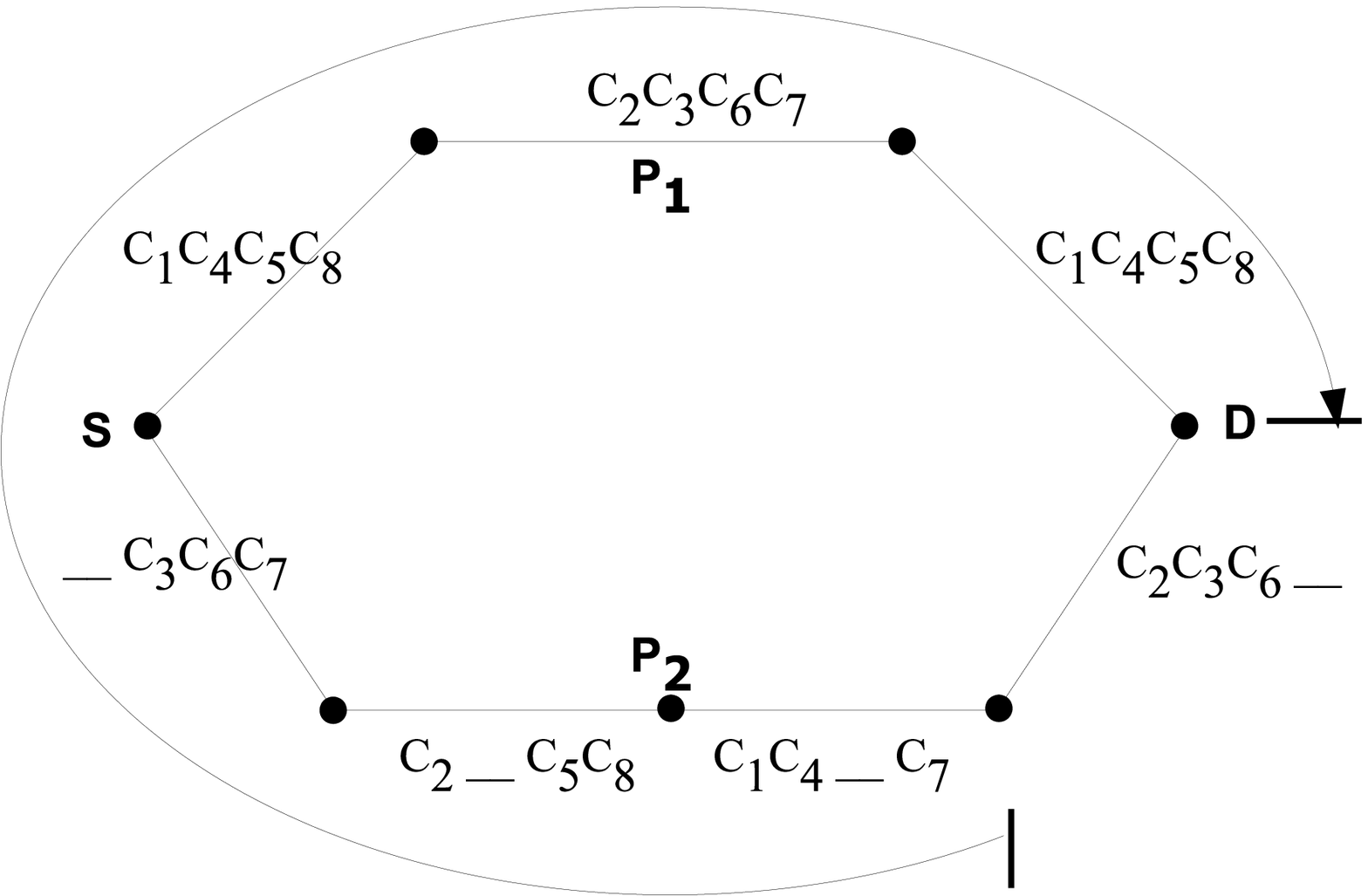}}
  \caption{Protocol Illustration: ($n_1$, $n_2$) = (3,4)}
  \label{fig:illustration_set_1}
\end{figure*}

\bpf For \emph{Case 1}, it is clear that the algorithm achieves
rate $1$. For \emph{Case 2}, the algorithm can be summarized as
follows: During the $i$-th iteration, we fix our starting point as
the $i$-th link in the longer path. We have two colors at the
$i$-th iteration, $C_{2i+1}$ and $C_{2i+2}$, which we will
associate alternately with the edges in the circular loop starting
from the $(i-1)$-th edge in the longer path $P_2$. This coloring
respects the half-duplex, broadcast and interference constraints,
because all the constraints reduce to a single one, i.e., adjacent
edges in the network viewed as a cycle shall have distinct colors.
After the $n_2$ iterations are over, we have mapped $2n_2$ colors
to the network. The shorter path gets $n_2$ colors on each edge,
whereas the longer path gets $n_2-1$ colors on each edge. Hence
the rate of the resultant protocol will be equal to $\frac{2n_2 -
1}{2n_2}$. This is illustrated with an example, $(n_1, n_2) =
(3,4)$ in Fig.~\ref{fig:illustration_set_1}.

\epf

\bthm \label{thm:2PP_DMT} For a KPP network with $K=2$, if the two
path lengths are equal modulo $2$, then the DMT achieved by the
orthogonal protocol of Construction~\ref{constr:k2_max_rate} meets
the cut-set bound, i.e., $d(r) = 2(1-r)^{+}$. \ethm

\bpf The proof follows from Lemma~\ref{lem:dmt_opt},
Theorem~\ref{thm:k2_max_rate} and
Prop.~\ref{prop:2PP_Constr_Rate}.\epf

\bthm \label{thm:KPP} For a KPP network, there exists an
orthogonal protocol achieving the cut-set bound if $K \geq 3$ or
$K=2$ and $n_1 = n_2 \pmod 2$. \ethm

\bpf Follows from Theorem~\ref{thm:K_geq_4},
Theorem~\ref{thm:k3_miso} and Theorem~\ref{thm:2PP_DMT}. \epf

\subsection{KPP Networks with Direct Link\label{sec:kppwd}}

\bthm \label{thm:KPP_D}For KPP(D) networks, the cut-set bound on
DMT is achievable whenever there is an orthogonal protocol for the
backbone KPP network satisfying the conditions of
Lemma~\ref{lem:dmt_opt}. \ethm

\bcor \label{cor:KPPD} For KPP(D) networks, the cut-set bound on
DMT is achievable whenever $K \geq 4$. \ecor

\bpf (of Theorem~\ref{thm:KPP_D})

The same protocol in the backbone KPP network yields an induced
channel matrix, which is a diagonal matrix with the $K$ path gains
appearing cyclically along the diagonal. In the presence of a
direct link, as is the case here, clearly it is the path gain
$\bold{g}_d$ of the direct link that will appear along the
diagonal of the induced channel matrix. The path gains of the
backbone paths will appear in general below the main diagonal.
However it is not hard to see that by suitably delaying symbols
along each of the $K$ paths, the path gains $\bold{g}_i,
i=1,2,\ldots,K$ can be made to appear cyclically along a single
sub-diagonal, say the $D$-th sub-diagonal.

\begin{figure}[h!]
\centering
\includegraphics[width=60mm]{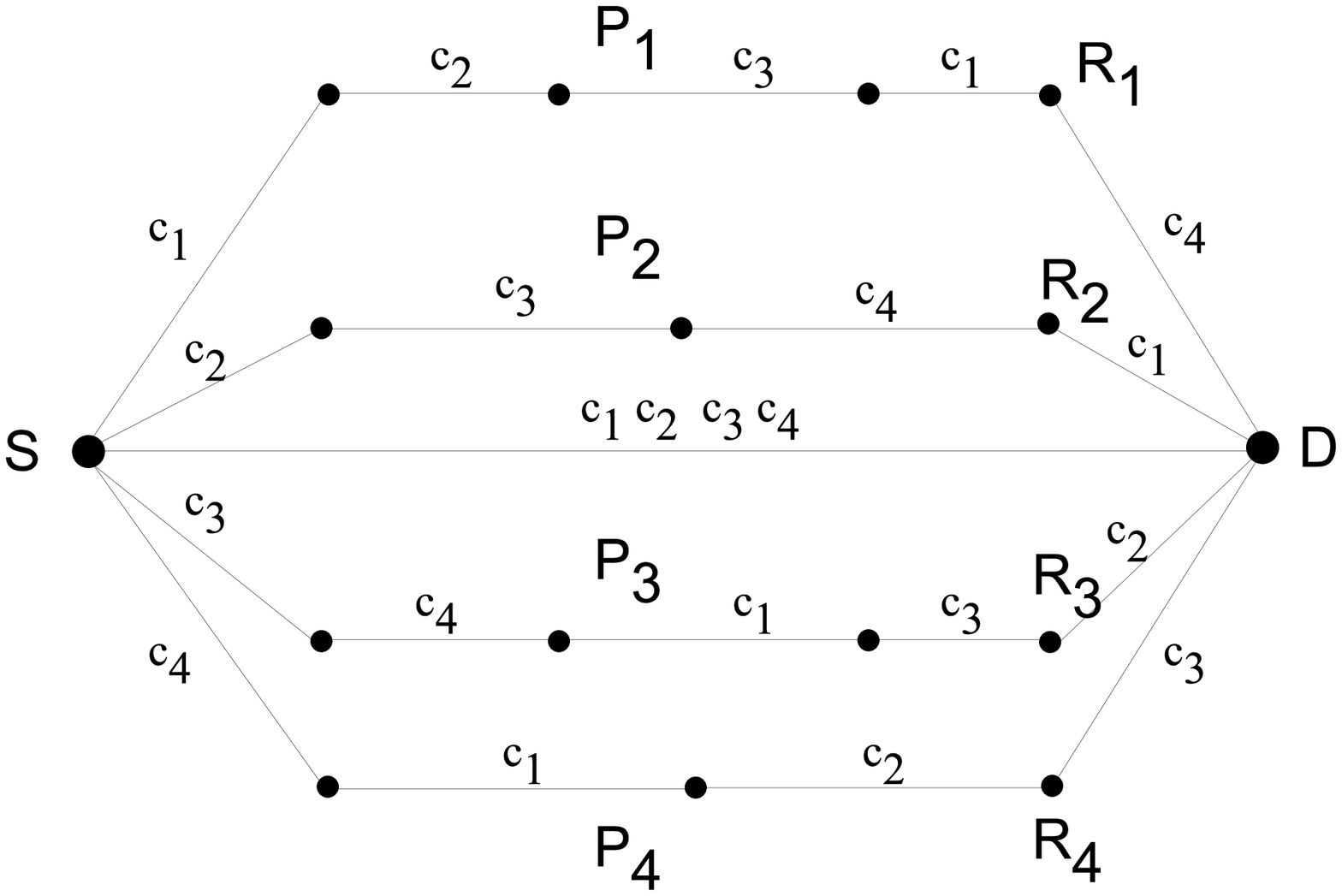}
\caption{KPP(D) network protocol \label{fig:kppd_illustration}}
\end{figure}

For instance, consider a KPP(D) network with $4$ relaying paths
(see Fig.~\ref{fig:kppd_illustration}). The figure describes an
orthogonal protocol with the aid of a coloring using a set of
colors $\{ c_1, c_2, c_3, c_4\}$. The cycle length of the protocol
is $4$, and $c_i$ represents the $i$th time-slot within a cycle.
Unlike the case in KPP networks, there is no initial delay between
the start of transmission at the source and the reception at the
sink, due to the presence a direct link. However, the delays
encountered by symbols in various backbone paths are potentially
different. In this example, delay in paths $P_1$ and $P_3$ are $7$
time-slots, whereas that in paths $P_2$ and $P_4$ are $3$
time-slots. So if the fading coefficient along the path $P_i$ is
${\bf g}_i$ and that along the direct link is ${\bf g}_d$, then
the input-output relation between the first $10$ transmitted and
received symbols would be of the form:
\beqan {\bf y} & = & {\bf H}{\bf x} + {\bf w}, \text{ where} \\
H & = & \left[ \begin{array}{cccccccccc} {\bf g}_d &     &      &     &     &     &     &  & & \\
                                                 & {\bf g}_d &      &     &     &     &     & &  & \\
                                                 &     &  {\bf g}_d &     &     &     &     & & & \\
                                                 &     &      & {\bf g}_d &     &     &     & & & \\
                                                 & {\bf g}_2 &      &     & {\bf g}_d &     &     & & & \\
                                                 &     &      &     &     & {\bf g}_d &     & & & \\
                                                 &     &      & {\bf g}_4 &     &     & {\bf g}_d & & & \\
                                             {\bf g}_1 &     &      &     &     &     &     & {\bf g}_d  & & \\
                                                 &     &      &     &     & {\bf g}_2 &     &     & {\bf g}_d & \\
                                                 &     &  {\bf g}_3 &     &     &     &     &     & & {\bf g}_d  \\
\end{array}
\right]. \label{eq:kppd_channel} \eeqan

Here ${\bf x}$, ${\bf y}$, and ${\bf w}$ represent the input,
output and noise vectors respectively, each of length $10$.

By adding a delay of $4$ time-slots at $R_2$ and $R_4$, the
effective delay encountered by symbols traveling via paths $P_2$
and $P_4$ can be made to equal $7$ time-slots. The cycle length of
the protocol being $4$, a delay of $4$ can be introduced by idling
the respective nodes for one cycle of the protocol. This will
result in a new channel matrix ${\bf H}^\prime$ between the same
input and output vectors ${\bf x}$ and ${\bf y}$, where \beqn
{\bf H}^\prime = \left[ \begin{array}{cccccccccc} {\bf g}_d &     &      &     &     &     &     &  & & \\
                                                 & {\bf g}_d &      &     &     &     &     & &  & \\
                                                 &     &  {\bf g}_d &     &     &     &     & & & \\
                                                 &     &      & {\bf g}_d &     &     &     & & & \\
                                                 &  &      &     & {\bf g}_d &     &     & & & \\
                                                 &     &      &     &     & {\bf g}_d &     & & & \\
                                                 &     &      &  &     &     & {\bf g}_d & & & \\
                                             {\bf g}_1 &     &      &     &     &     &     & {\bf g}_d  & & \\
                                                 & {\bf g}_2 &      &     &     &  &     &     & {\bf g}_d & \\
                                                 &     &  {\bf g}_3 &     &     &     &     &     & & {\bf g}_d  \\
\end{array}
\right]. \eeqn

So the effective channel matrix due to an orthogonal protocol in a
KPP(D) network will be of the form, \beqn
{\bf H} = \left[ \begin{array}{cccccccc} {\bf g}_d &     &      &     &     &  & &    \\
                                    & {\bf g}_d &      &     &     &  &&\\
                                    &     & \ddots  &     &     &     &     &\\
                                 {\bf g}_1 &  &      &     &  &     &     &  \\
                                     & {\bf g}_2    &      &     &     &  &     & \\
                                              &     & \ddots     &     &    &  \ddots   &     &  \\
                                                   &     & & {\bf g}_K  & & & {\bf g}_d &  \\
                                                 &   &  &     & {\bf g}_1    &     &     &  {\bf g}_d  \\
\end{array}
\right]. \eeqn

Next, consider the situation when the network is operated for a
duration of $M = mK + D$ time-slots, for some positive integer
$m$.   We now invoke Theorem~\ref{thm:main_theorem} to the induced
channel matrix ${\bf H}$ to arrive at a lower bound on the DMT as,
\beqan d_H(r) & \geq & d_{H^{(0)}}(r) + d_{H^{(\ell)}}(r) \\
 & \geq & (1-\frac{r}{M})^{+} + K (1-\frac{r}{M-D})^{+}. \\
\Rightarrow d(r) & = & d_H(Mr) \\
 & \geq & (1-{r})^{+} + K (1-\frac{M}{M-D}r)^{+}, \eeqan
 which, as $M$ tends to infinity, becomes \beqa d(r) \geq (K+1)
 (1-r)^{+}. \label{eq:KPPD_DMT} \eeqa

For KPP(D) networks, the cut-set upper bound on DMT
(Lemma~\ref{lem:CutsetUpperBound}) yields $d(r) \leq (K+1)
(1-r)^{+}$. Combining this with the DMT lower bound
\eqref{eq:KPPD_DMT}, we get

$d(r) = (K+1) (1-r) ^{+}$. \epf

\bnote Since a two-hop relay network possessing $N$ relays with
direct link and with relays isolated is a KPP(D) network with
$K:=N$, the DMT optimal strategy for these family of networks is
given by Corollary~\ref{cor:KPPD}. This turns out to be the same
strategy as SAF protocol given in \cite{YanBelSaf} for these
networks. \enote

\section{Half-Duplex KPP(I) Networks \label{sec:half_duplex}}

In this section, we move on to consider multi-hop networks with
interference. We consider KPP(I) networks with single antenna
nodes operating under the half-duplex constraint. We will show
that even here, the cut-set bound can be achieved using AF
protocols. The cut-set bound (Lemma~\ref{lem:CutsetUpperBound})
gives the same DMT upper-bound $d(r) \leq K(1-r)^{+}$, as in the
case of KPP networks. In this section, we will demonstrate that
this DMT is in fact achievable.

In the case of a KPP(I) networks, we note that any protocol for
the backbone KPP network automatically induces a protocol on the
KPP(I) network. Although a protocol is orthogonal with respect to
the backbone KPP network, it will most likely result in a protocol
on the KPP(I) network that is not orthogonal because in the
presence of interfering links, interference avoidance is no longer
guaranteed. The aim here is to come up with a protocol for the
backbone KPP network that induces a derived protocol on the KPP(I)
network such that it will result in a DMT-achieving channel matrix
for the KPP(I) network. As explained in the case of KPP networks,
without loss of optimality, we can neglect the initial delay here
also while considering the effective channel matrix induced by the
protocol between the source and the sink. We begin by introducing
the notion of causal interference.

\subsection{Causal Interference\label{sec:causal_interference}}

\bdefn \label{def:Causal_Interference} Let ${\bf H}$ denote the channel matrix induced by an AF protocol $\wp$
employed in a KPP(I) network $\mathcal{N}$. Let ${\bf H}^{(0)}$ be the diagonal part of ${\bf H}$. Let ${\bf
H}_{bb}$ denote the channel matrix induced by $\wp$ when it is employed on the backbone of $\mathcal{N}$. Then
$\wp$ is said to be a causal protocol (i.e., a protocol with causal interference) if the interference admitted
by the protocol is causal in nature, i.e.,  \ben \item ${\bf H}$ is lower triangular and
\item ${\bf H}_{bb}$ = ${\bf H}^{(0)}$. \een \edefn

\bnote Note that condition 2) in the definition above is satisfied
only if the protocol $\wp$ is derived from a protocol that is
orthogonal with respect to the backbone KPP network.  For this
reason, all the protocols $\wp$ for KPP(I) networks encountered in
this section will be derived from orthogonal protocols for the
backbone KPP network.  Thus throughout the remainder of this
section, whenever the word protocol appears in the context of
KPP(I) networks, it should be interpreted to mean a protocol
derived from an orthogonal protocol for the backbone KPP network.
\enote

\blem \label{lem:causal_interference} Consider a KPP(I) network operating under a causal protocol $\wp$. Let the
induced channel matrices on the KPP(I) and backbone KPP networks be given by ${\bf H}$ and ${\bf H}_{bb}$
respectively. Then the DMT of $d_H(r) \ \geq \ d_{H_{bb}}(r)$. Furthermore, if the protocol achieves the cut-set
bound on the backbone KPP network, i.e., $d_{H_{bb}}(r) \ = \ K(1-r)^{+}$, then it also achieves the cut-set
bound on the KPP(I) network, i.e., $d_{H}(r)=K(1-r)^{+}$. \elem

\bpf  Since the interference caused by the protocol is causal,
${\bf H}$ is lower triangular. We also have that ${\bf H}^{(0)} =
 {\bf H}_{bb}$ is the diagonal part of ${\bf H}$. The lemma now
follows because by Theorem~\ref{thm:main_theorem}, the DMT of a
lower triangular matrix is lower bounded by the DMT of the
corresponding diagonal matrix, i.e., $d_H(r) \ \geq \
d_{H_{bb}}(r)$. The rest follows since the same cut-set bound also
applies to the KPP(I) network. \epf

We now give a sufficient condition for a protocol to be causal.

\bprop \label{prop:Causal_Interference_prelim} Consider a KPP(I) network under a protocol $\wp$.  Then $\wp$ is
causal if the unique shortest delay experienced by every transmitted symbol is through a backbone path.\eprop

\bpf The proof is straightforward. \epf

\bprop \label{prop:Causal_Interference} Consider a KPP(I) network under a protocol $\wp$ derived from a protocol
that achieves the cut-set bound on the KPP network. If the unique shortest delay experienced by every
transmitted symbol is through a backbone path, then $\wp$ achieves the optimal DMT of $d(r) = K(1-r)^{+}$.
\eprop

\bpf The proof is clear by combining Prop.~\ref{prop:Causal_Interference_prelim} and
Lemma~\ref{lem:causal_interference}. \epf

\subsection{Optimal DMT for Regular Networks \label{sec:regular_dmt}}

For the class of regular networks, the sufficient condition given
in Prop.~\ref{prop:Causal_Interference} is easily satisfied,
leading to the following result.

\bthm \label{thm:knregular_dmt} The optimal DMT $d(r) = K(1-r)^{+}$ of (K, L) regular networks is achievable.
\ethm

\bpf  Consider a (K, L) regular network.  This network can be
regarded as a KPP(I) network in which each backbone path has an
equal number of edges. We operate the network using a protocol
that is derived from an orthogonal protocol specified for the
backbone KPP network. The orthogonal protocol for the backbone KPP
is specified by giving the coloring sets. The cycle length of the
protocol is $K$, and hence we use the set of colors $C = \{c_0,
c_1,\ldots, c_{K-1}\}$, and assume $ \forall \ell >K, c_{\ell} =
c_{\ell \pmod K}$. The coloring set for edge $e_{ij}$ is given by,
\beq A_{ij} = \{ c_{i+(j-1)} \}, \ \ 1 \leq i \leq K, \ 1 \leq j
\leq L+1 .\eeq

\begin{figure}[h!]
\centering
\includegraphics[width=60mm]{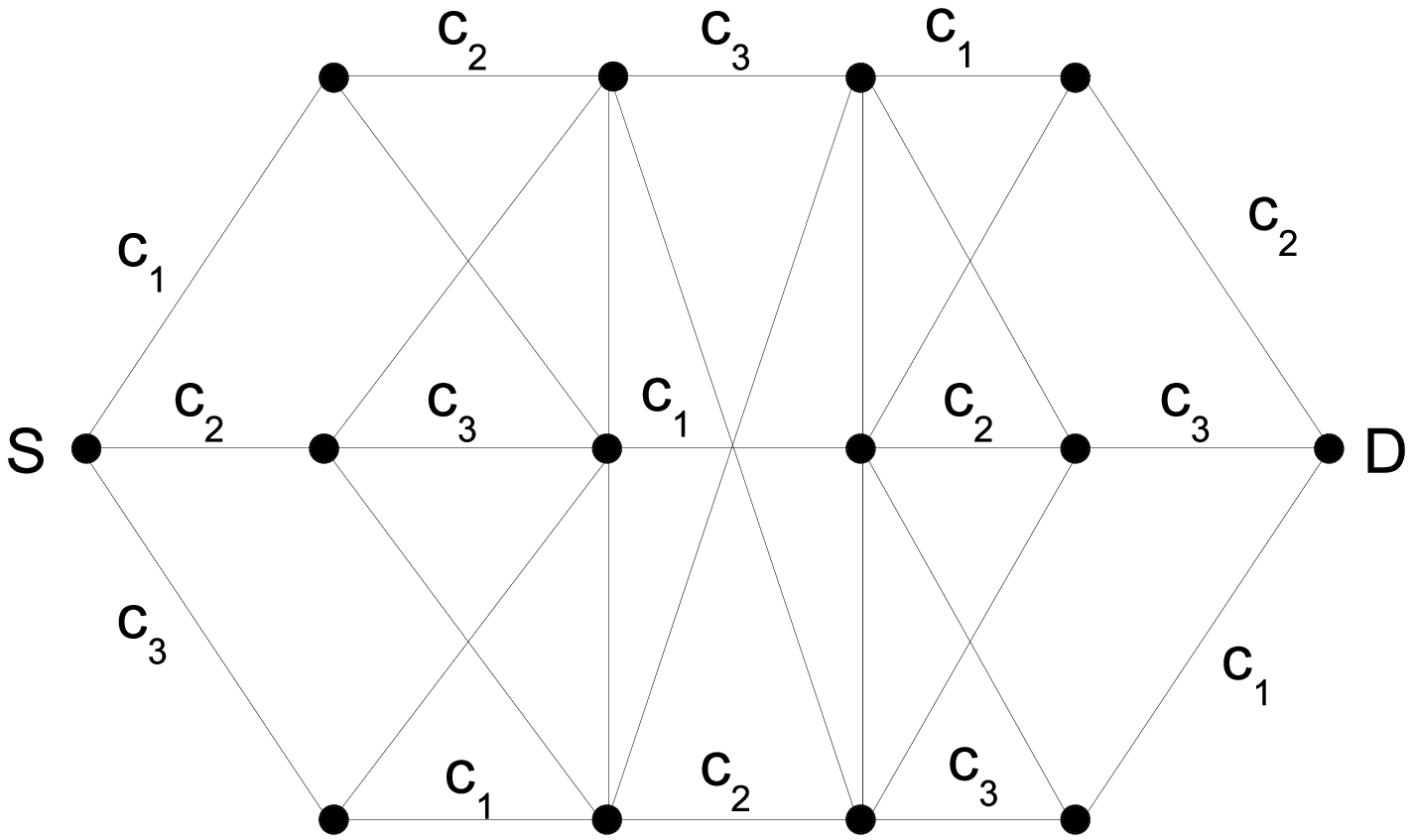}
\caption{Regular network with the protocol
\label{fig:regular_colored}}
\end{figure}

An example of this coloring scheme for the network in
Fig.~\ref{fig:regular_eg} is given in
Fig.~\ref{fig:regular_colored}.

It can be verified that the shortest delay encountered by every
symbol is through a backbone path. It can also be checked that the
protocol satisfies conditions of Lemma~\ref{lem:dmt_opt} and thus
achieves the cut-set bound on the backbone KPP network. Thus the
conditions of Prop.~\ref{prop:Causal_Interference} are satisfied
and hence proposed protocol achieves the cut-set bound on the (K,
L) regular network. \epf

\bcor \label{cor:two_hop_DMT} For the two-hop relay network
without direct link, the optimal DMT is achieved irrespective of
the presence or absence of links between relays. \ecor

\bpf The two-hop relay network without the direct link is a (K,1) regular network, where $K$ denotes the number
of relays in the network. This holds irrespective of the presence of links between relays, since they only
contribute to intra-layer links. Thus Theorem~\ref{thm:knregular_dmt} implies this corollary. \epf

\bnote The result in Corollary~\ref{cor:two_hop_DMT} was also
proven in a parallel work \cite{GhaBayKha}. The protocols used in
this paper and in \cite{GhaBayKha} are essentially the same as the
slotted-amplify-and-forward (SAF) protocol \cite{YanBelSaf},
except that the protocol is applied to a network that does not
have a direct link. However, the proof techniques used here and in
\cite{GhaBayKha} are very different. \enote

\subsection{Causal Interference for KPP(I) Networks with $K=3$\label{sec:schedule_delay}}

In this subsection, we construct a protocol satisfying the conditions of Prop.~\ref{prop:Causal_Interference}
and thereby, the protocol will be optimal. The protocol will turn out to require the introduction of suitable
delays at intermediate nodes in the network. We begin by considering KPP(I) networks with $K = 3$. Subsequently,
we will use the theory developed here for $K = 3$ to handle the case $K>3$.

\blem \label{lem:KPP3_causal} For any KPP(I) networks with $K =
3$, there exists a causal protocol that achieves the cut-set bound
when used on the back-bone KPP network. \elem

\bpf The proof is deferred to Appendix~\ref{app:causal}. \epf

\bthm \label{thm:KPP_2_3_Causal} For any KPP(I) network with with
$K=3$, there exists an explicit protocol that achieves the cut-set
bound on the DMT.\ethm

\bpf In Lemma~\ref{lem:KPP3_causal}, we showed how to construct an
orthogonal protocol that admits causal interference. This lemma
along with Lemma~\ref{lem:causal_interference} proves the theorem.
\epf

\subsection{Optimal DMT of KPP(I) networks\label{sec:kppi_networks}}

In the last section, we have shown how to design protocols for KPP(I) networks which admit causal interference.
This leads to the following results on the optimal DMT of KPP(I) networks with $K \geq 3$.

\bthm \label{thm:KPP_I} Consider any KPP(I) network with $K \geq
3$. The cut-set bound on the DMT $d(r) = K(1-r)^+$ is achievable.
\ethm

\bpf For $K=3$, it follows from Theorem~\ref{thm:KPP_2_3_Causal}.

Now, we will consider the case when $K > 3$. First we use the
procedure illustrated in Lemma~\ref{lem:Switch_Removal} in order
to remove all non-contiguous $k$-switches for $2 \leq k \leq K$.
This will lead us to a modified KPP(I) network which does not
contain any non-contiguous $k$-switches. Consider a $3$PP
sub-network of this KPP(I) network. From
Lemma~\ref{lem:KPP3_causal}, it follows that we can design a causal protocol for such a network which results in an
induced channel matrix with three product coefficients
corresponding to the three back-bone paths along the diagonal.
There are now $K \choose 3$ possible $3$PP subnetworks of the
modified KPP(I) network. If each of these subnetworks is activated
in succession, it would yield a lower triangular matrix with all
the $K$ product coefficient $g_i$ repeated $K-1 \choose 2$ times
along the diagonal. By Theorem~\ref{thm:main_theorem}, the DMT of this matrix is no worse than that of the
diagonal matrix alone. The diagonal matrix has a DMT equal to
$K(1-r)^+$ after rate normalization. Therefore a DMT of $d(r) \geq
K(1-r)^{+}$ can be obtained. However, since $d(r) \leq K(1-r)^{+}$
by the cut-set bound, we have $d(r) = K(1-r)^{+}$. \epf

\section{Half-Duplex Layered Networks \label{sec:layered}}

In this section, we consider half-duplex layered networks with
single-antenna nodes. As in the case of KPP(I) networks, an
achievable DMT for layered networks with full-duplex relays comes
as an immediate consequence of
Theorem~\ref{thm:FD_No_Direct_Path}. It is given in the below
corollary.

\bcor \label{cor:FD_Layered} For the full-duplex layered networks,
a linear DMT between the maximum diversity and maximum
multiplexing gain can be achieved. \ecor

Half-duplex layered networks are typically treated by first
considering full-duplex operation and then activating alternate
layers in order to satisfy the half-duplex constraint
\cite{YanBelNew}, \cite{VazHea}. However, this leads to a rate
loss of a factor of two. We will demonstrate in this section that
there exists schedules for which this rate loss is not incurred.

We will prove that the same result holds good for fully-connected
(fc) layered networks even when the relays are of half-duplex
nature. First, we will establish a sufficient condition for a
layered network, such that a linear DMT of $d_{\max}(1-r)^{+}$ is
achievable. Also, we will prove that $d_{\max}(1-r)^{+}$ is always
achievable for the fc layered networks.

\subsection{Linear DMT in Layered Networks}

Similar to the case of KPP networks, the basic idea is to activate
$d_{\max}$ paths from the source to the sink using AF protocol. In
the case of KPP networks, clearly we have $d_{\max} = K$
node-disjoint paths, and we could activate them without
interference between one another. In the case of layered networks,
we begin with identifying node-disjoint paths from all possible
paths from the source to the sink. A path from source to sink in a
layered network is said to be \emph{forward-directed} if all the
edges in the path are directed from one layer to the next layer
towards the sink (i.e., no edge in the path goes from one layer to
the previous layer and there is no edge which starts and ends in
the same layer). Identification of node-disjoint forward-directed
paths will allow us to schedule the edges in the network in a
similar fashion as how the parallel paths in KPP network were
scheduled. In the following lemma, we propose a technique to
identify node-disjoint paths in a layered networks.

\bdefn \label{def:Bipartite_Graph} Given a set of forward-directed
paths $P$ in a layered network, the bipartite graph corresponding
to $P$ is defined as follows: \bit \item Construct a bipartite
graph with vertices corresponding to paths in $P$ on both sides.
\item Connect a vertex associated with path $P_i$ on the left to a
vertex associated with $P_j$ on the right if the two paths are
node disjoint. \eit \edefn

\blem \label{lem:Bipartite_Graph} Consider a set of paths $P = \{
P_i,i=1,2,\ldots,N \} $ in a given layered network. Let the
product of the fading coefficient on the $i$-th path $P_i$ be
${\bf g}_i$. Construct the bipartite graph corresponding to $P$.
If there exists a complete matching in this bipartite graph, then
these paths can be activated in such a way that the DMT of this
protocol is greater than or equal to the DMT of a parallel channel
with fading coefficients ${\bf g}_i,i=1,2,...,N$ with the rate
reduced by a factor of $N$, i.e., $d(r) \geq d_{H_d}(Nr)$, where
$H_d \ = \ \text{diag}({\bf g}_1,{\bf g}_2,\ldots,{\bf g}_N)$.
\elem

\bpf Suppose there is a complete matching $\pi$ on the graph
constructed as above. The complete matching specifies for every
path on the left $P_i$, a partner on the right $P_{\pi_i}$. The
length of each path and therefore the delay is equal to $D:=L+1$
since the network is layered with $L$ relaying layers. Consider
operating the network under the following protocol.

\emph{Step 1:} Set $i=1$. Consider the path $P_i$ along with its
partner path $P_{\pi_i}$ as a two parallel path KPP(I) network
with backbone paths of equal length. This sub-network is in fact a
$(2, L)$ regular network with $L$ being the number of layers in
the original network. The sub-network is activated using the
protocol specified in the proof of Theorem~\ref{thm:knregular_dmt}
for $T >> 2$ cycles, and since each protocol cycle is of duration
$2$ time slots, the total length of activation is $2T$ time slots.
Since the protocol is causal, it will induce a lower triangular
channel matrix between the input and the output with channel gains
${\bf g}_i$ and ${\bf g}_{\pi_i}$ alternating along the diagonal.

\emph{Step 2:} Repeat \emph{Step 1} for $i= 2, 3, \ldots, N$.

The induced channel matrix ${\bf H}$ will be a lower triangular
$2NT \times 2NT$ matrix with each of ${\bf g}_1, {\bf g}_2,\ldots,
{\bf g}_N$ appearing for $2T$ time-slots.

Now the DMT of the protocol $d(r)$ can be related to the DMT of
the diagonal part ${\bf H}^{(0)}$ as $d(r) = d_H(2NTr) \geq
d_{H^{(0)}}(2NTr)$ by applying Theorem~\ref{thm:main_theorem}. If
${\bf H}_d$ is a $N\times N$ diagonal matrix comprising of ${\bf
g}_1,{\bf g}_2,\ldots,{\bf g}_N$ along the diagonal, then
$d_{H^{(0)}}(r) = d_{H_d}(\frac{r}{2T})$ by applying Lemma~\ref{lem:parallel_dependent}. This yields $d(r) \geq
d_{H_d}(Nr)$. \epf

We utilize the above lower bound to obtain a sufficient condition
that guarantees that a linear DMT between the maximum diversity
and multiplexing gain is achieved on a general layered network in
the following lemma.

\blem \label{lem:General_layered_network} For a general layered
network, a DMT of $d(r) \geq M(1-r)^{+}$ is achievable whenever
the network has $M$ forward-directed edge-disjoint paths from the
source to the sink and the bipartite graph corresponding to the
set of edge-disjoint paths $e_i$, $i=1,2,...,M$ has a complete
matching. \elem

\bnote Since an arbitrary network without a direct link can be
represented as a general layered network by
Remark~\ref{rem:layered_general}, this lemma applies to arbitrary
networks. If in addition to the requirements in the lemma above,
the network has the number of forward-directed edge-disjoint paths
equal to the min-cut, then a linear DMT of $d_{\max}(1-r)^{+}$ is
achievable. \enote

\bpf (of Lemma~\ref{lem:General_layered_network}) By using
Lemma~\ref{lem:Bipartite_Graph} we will be able to get a DMT of
$d(r) \geq d_{H_d}(Mr)$. But since the paths are edge-disjoint,
the fading coefficients are independent, and we get $d_{H_d}(r) =
(M - r)^{+}$ by applying Lemma~\ref{lem:parallel_dependent}.
Therefore, we get, $d(r) \geq M(1-r)^{+}$.  \epf

Since the min-cut $M$ is equal to the maximum diversity of the
network by Theorem~\ref{thm:mincut}, a DMT of
$M(1-r)^{+}$ signifies a DMT that is linear between the maximum
diversity and the maximum multiplexing gain $1$.

\subsection{Fully-Connected Layered Networks}

While we have established a sufficient condition for achieving a
linear DMT between maximum diversity and multiplexing gain for an
arbitrary layered network, even for many fc layered networks, this sufficient condition is not satisfied. For
example, a fc layered network with $(3,2,3)$ nodes does not
satisfy the sufficient condition in
Lemma~\ref{lem:General_layered_network}. However, as we shall see
in the sequel, this can be remedied and a linear DMT is obtainable
for arbitrary fc layered networks with number of nodes in any
relaying layer being at least $2$. A supporting lemma is needed
before we proceed to prove this main result.

Consider an fc layered network with $L$ relaying layers.  Let
$\bold{h}^{(i)}_j$ be the $i$-th fading coefficient in the $j$-th
hop. Let there be $R_i$ relay nodes in the $i$-th layer. Then the
number of forward-directed paths $N$ is equal to \beqa N & = &
\prod_{i=1}^L R_i . \eeqa

\blem \label{lem:complete_matching} Consider an fc layered network
with $L$ relaying layers. Let there be $R_i$ relay nodes in the
$i$-th layer. Let $P = \{P_1,..,P_N\}$ be the set of all forward
directed paths in the network. Then the bipartite graph
corresponding to $P$ has a complete matching. \elem

\bpf We will prove this by producing an explicit complete matching
on the bipartite graph. Let the layered network have $L$ layers.
Let us fix an (arbitrary) ordering on the relays in each hop. Let
the relays in the $j$-th hop be indexed $0,1,...,R_j-1$.

There is a one-to-one correspondence between the set of all
forward-directed paths and the $L$ tuples $(b_{1},...,b_{L})$,
where $b_{j}$ denotes the index of the relay in the $j$-th layer
visited by that path. For any given path $P$ associated to the $L$
tuple $(b_{1},...,b_{L})$, consider a path $P^{\prime}$ associated
to the tuple $(c_{1},...,c_{L})$ where $c_i = b_i + 1 \pmod{R_i}$.
Clearly these two paths are node-disjoint, because $R_i \geq 2, \
\forall i$ by definition of fc layered network. The collection of
all such pairings constitutes a complete matching on the set of
all forward-directed paths. \epf

Consider an fc layered network with $N$ forward-directed paths and
$L$ relaying layers. Let ${\bf g}_i$ denote the product fading
coefficient corresponding to the forward-directed path $P_i$ for
$i=1,2,\ldots,N$. Now, $\{{\bf g}_i\}$ are mutually correlated
because each ${\bf g}_i$ is the product of Rayleigh fading
coefficients of links in that path, and a link may belong to
multiple paths. Next we compute the DMT of a parallel channel with
$N$ sub-channels in which $i$th sub-channel corresponds to the
$i$th forward-directed path in the fc layered network. Clearly,
the coefficient of the $i$th sub-channel is ${\bf g}_i$, which is the
product of $L+1$ Rayleigh coefficients, one each chosen from $L+1$
different sets of Rayleigh coefficients. Here each of the $L+1$
sets corresponds to the set of fading coefficients of all links
connecting two adjacent layers of relays.

\begin{figure}[h!]
\centering
\includegraphics[height=40mm]{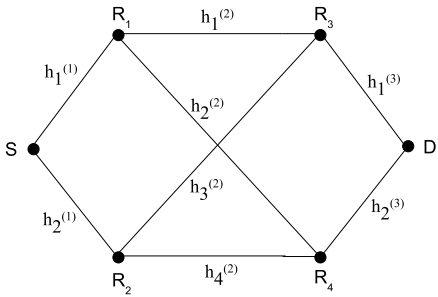}
\caption{Three hop layered network\label{fig:three_hop_fc}}
\end{figure}

For instance, consider an fc layered network with two layers of
relays. \footnote{This network was also used in \cite{AveDigTse1}
for illustration.} (see Fig.~\ref{fig:three_hop_fc}) There are
four forward-directed paths between the source and the sink. The
product coefficients corresponding to these four paths are
\beqa {\bf g}_1 & = & {\bf h}^{(1)}_1 {\bf h}^{(2)}_1 {\bf h}^{(3)}_1 \\
{\bf g}_2 & = & {\bf h}^{(1)}_1 {\bf h}^{(2)}_2 {\bf h}^{(3)}_2 \\
{\bf g}_3 & = & {\bf h}^{(1)}_2 {\bf h}^{(2)}_3 {\bf h}^{(3)}_1 \\
{\bf g}_4 & = & {\bf h}^{(1)}_2 {\bf h}^{(2)}_4 {\bf h}^{(3)}_2.
\eeqa

Our interest is to compute the DMT of the parallel channel with
$4$ sub-channels with coefficients ${\bf g}_1,\ldots, {\bf g}_4$.
The DMT can not be computed using Lemma~\ref{lem:parallel_channel} since the individual links are not independent.
Intuitively, the diversity of this channel is at most two, since
if ${\bf h}^{(1)}_1$ and ${\bf h}^{(1)}_2$ are both in deep fade,
then the overall channel is likely in deep fade. We present the
lemma below, which computes the DMT of general parallel channels
with product coefficients having a certain structure. As a result
of this lemma, it will follow that the example channel has a
diversity of two.

\bdefn \label{def:balanced} A subset $S$ of a product set
$\mathcal{H} \subset \{ H^{(1)} \times H^{(2)} \times \cdots
\times H^{(L+1)} \}$ is said to be
$(\nu_1,\nu_2,...,\nu_{L+1})$-balanced if every element in
$H^{(i)}$ appears precisely $\nu_i$ times as the $i$th coordinate
of an element in $S$. Clearly we have $|S| = \nu_i |H^{(i)}|$.
\edefn

\blem \label{lem:product_channel_DMT} Let $H^{(i)} = \{ {\bf
h}^{(i)}_{1},{\bf h}^{(i)}_{2},\ldots,{\bf h}^{(i)}_{M_i} \},
i=1,2,...,L+1$ be sets of i.i.d. Rayleigh fading coefficients. Now
let $\mathcal{H} \subset \{ H^{(1)} \times H^{(2)} \times \cdots
H^{(L+1)} \}$ be a $(N_1,N_2,...,N_{L+1})$-balanced set and let
$N=|\mathcal{H}|$. Let $ N_{\max} = \max_{i=1,2,\ldots,N} N_i $
and $M_{\min} = \min_{i=1,2,\ldots,L+1} M_i $. Let $\psi:
\mathcal{H} \rightarrow G$ be the product map such that $\psi( (
a_1, a_2,...,a_{L+1} ) ) = \prod_{j=1}^{L+1} {a}_{i}$. Now let
$\psi(\mathcal{H}) = \{ {\bf g}_1, {\bf g}_2, \ldots, {\bf g}_N
\}$, where ${\bf g}_i = \prod_{k=1}^{L+1} {\bf h}^{(k)}_{e(i,k)}
$, with $e(i,k)$ being a map from $[N] \rightarrow [M_k]$ for a
fixed $k \in [L+1]$.

Let ${\bf H}$ be a $N \times N$ diagonal matrix with the diagonal
elements given by ${\bf g}_i$. The DMT of the parallel channel
${\bf H}$ is a linear DMT between a diversity of
$\frac{N}{N_{\max}}$ and a multiplexing gain of $N$, \beq d(r) =
\frac{(N-r)^{+}}{N_{\max}}. \eeq \elem

\bpf The proof is deferred to
Appendix~\ref{app:product_channel_DMT}. \epf

Having thus established the DMT of a parallel channel with product
fading coefficients, we now proceed to utilize this lemma to
compute a lower bound on the DMT of fc layered networks.

\bthm \label{thm:fully_connected_layered} For a fully-connected
layered network, a linear DMT between the maximum diversity and
maximum multiplexing gain of $1$ is achievable. \ethm

\bpf Consider an fc layered network with $L$ layers. Let there be
$R_i$ antennas in the $i$-th layer for $i=0,1,...,L+1$. We
consider the source as layer $0$, and sink as the $L+1$th layer so
that $R_0=R_{L+1}=1$. Let $M_i := R_{i-1}R_i, i=1,2,...,L+1$ be
the number of fading coefficients between the $(i-1)$-th and $i$th
layer of relays.

Let $H^{(i)}:=\{{\bf h}^{(i)}_{j}, j=1,2,..,M_i \}$ be fading
coefficients of links connecting nodes in the $(i-1)$th layer to
those in the $i$th layer, $i=1,2,...,L+1$. Let $N$ be the total
number of forward-directed paths from source to sink, and $P_i,i
\in [N]$ be the various forward-directed paths. Let $P$ denote the
set of all these forward-directed paths. Then $|P| = N =
\prod_{i=1}^{L} R_i$. Let ${\bf g}_i$ be the product fading
coefficient on path $P_i$. Let $M_{\min} = \min_{i=1}^{L+1} M_i$
and $N_{\max} = \max_{i=1}^{L+1} N_i$. Note that $M_{\min}$
corresponds to the value of the min-cut.

By Lemma~\ref{lem:complete_matching}, the bipartite graph
corresponding to $P$ has a complete matching. The set of paths $P$
satisfies the criterion of Lemma~\ref{lem:Bipartite_Graph} and
therefore, we can obtain a DMT of \beqa d(r)& \geq & d_{H_d}(Nr)
\label{eq:layered_mid} \eeqa where ${\bf H}_d$ is an $N \times N$
diagonal matrix whose $i$th diagonal entry is ${\bf g}_i$. Now, we
need to compute $d_{H_d}(r)$. We associate with every path $P_i$
an $L+1$ tuple of fading coefficients $\Theta_i =
(\theta_{i1},\theta_{i2},\ldots,\theta_{i(L+1)})$, where
$\theta_{ik} \in H^{(k)}$ is the fading coefficient of the $k$th
link in path $P_i$. Now ${\bf g}_i$ is related to $\Theta_i$ as
${\bf g}_i = \prod_{k=1}^{L+1} \theta_{ik} $. Note that the
collection of fading coefficient tuples $\Theta_i$ corresponding
to all the paths $P_i \in P$ is a
$(N_1,N_2,\ldots,N_{L+1})$-balanced subset of the cartesian
product set $ \{ H^{(1)} \times H^{(2)} \times \cdots \times
H^{(L+1)} \}$, where

\beqa N_{\ell} & = &
\frac{\prod_{j=0}^{L+1}R_{j}}{R_{\ell-1}R_{\ell}} \eeqa represents
the number of tuples $\Theta_i$ in which $h^i_{j}$ appears as a
component for any $j \in [M_{\ell}]$.

We can now see that the parallel channel matrix ${\bf H}_d$
satisfies the conditions of Lemma~\ref{lem:parallel_channel}, which can be applied to obtain the DMT of
${\bf H}_d$ as \beqa d_{H_d}(r) & = & \frac{(N-r)^{+}}{N_{\max}}.
\eeqa Substituting this back in \eqref{eq:layered_mid}, we obtain
a lower bound to the DMT of the protocol as \beqa d(r) & \geq &
d_{H_d}(Nr) \\
& = & N \frac{(1-r)^{+}}{N_{\max}} \\
\Rightarrow d(r) & \geq & M_{\min}(1-r)^{+}. \eeqa

By Theorem~\ref{thm:mincut}, we have that the maximum
diversity is equal to the min-cut $d_{\max} = M_{\min}$ in any
network. Therefore a DMT of \beqa  d(r) & \geq & d_{\max}(1-r)^{+}
\eeqa is achievable. \epf

Therefore for the two-layer fc layered network example considered
in Fig.~\ref{fig:three_hop_fc}, a DMT of $2(1-r)^{+}$ is
achievable using the strategy proposed in the theorem. However
this DMT is also the upper bound using the cut-set bound.
Therefore the DMT for this example network is identically equal to
$2(1-r)^{+}$. This is not particular to this example. For fc
layered networks with $L < 4$, the min-cut is either at the source
side or at the sink side, and hence we have the following
corollary:

\bcor \label{cor:optimal_layered} For an fc layered
network with $L < 4$, the cut-set bound on DMT is achievable.\ecor

\bpf Let $n_i$ be the number of relays in layer $i$. Consider a
layered network with $L=1$, i.e., there is only one layer. The DMT
upper bound is $n_1(1-r)^{+}$ from the cut-set bound, which is
achieved using the proposed strategy. For $L=2$, the cut-set bound
on DMT is $\min \{n_1,n_2 \} \ (1-r)^{+}$, which is achieved. For
$L=3$, it can be seen that $d_{\max} = \min \{n_1,n_2\}$ and that
the DMT upper bound is $\min \{n_1,n_2\} \ (1-r)^{+}$, which is
indeed achieved. \epf

\section{Networks with Multiple Antenna Nodes\label{sec:multiple_antenna}}

In the previous sections, we considered networks with all nodes
having single antennas. In this section, we consider the general
case of networks with multiple antenna nodes. Specifically, we
consider KPP, KPP(I) and fc layered networks under both
full-duplex and half-duplex constraints. As it is difficult to
characterize the DMT completely, we present lower bounds, i.e., an
achievable DMT region.

\subsection{Full-Duplex, Fully-Connected Layered Networks\label{sec:layered_multiple}}

\begin{figure}[h!]
\centering
\includegraphics[height=65mm]{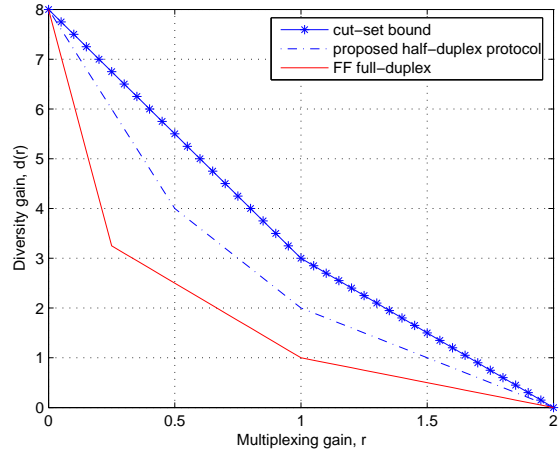}
\caption{Comparison of various protocols for $(2, 4, 2)$
network\label{fig:multiple_ant_dmt_curve}}
\end{figure}

We consider fc layered networks with multiple antennas at the
source and the sink. In the case of networks with multiple
antennas at relays, we will replace every multiple-antena relay with an equivalent
number of single-antenna relays in the same layer, and proceed to
analyze the resultant network. Clearly, any protocol on the
resultant network has a derived protocol in the original network and
clearly the DMT of the resultant network serves as a lower bound
to the DMT of the original network.

\bdefn A ss-ss layered network with multiple antennas at the
source and the sink is referred to as an $(n_0, n_1, \ldots, n_L, n_{L+1})$
network if the network has $L$ layers, with the source
having $n_0$ antennas, the sink having $n_{L+1}$ antennas, and the
$i$-th layer of relays having $n_i$ single-antenna nodes. \edefn

In \cite{YanBelNew}, parallel AF and flip-and-forward (FF)
protocols have been proposed for the $(n_0, n_1, \ldots, n_{L+1})$
network with full-duplex operation and directed antennas. The
parallel AF protocol aims to achieve the full diversity for the
network, whereas FF achieves the extreme points of maximum
multiplexing gain and the maximum diversity gain. In
\cite{YanBelNew}, it has been shown that the FF protocol achieves
a better DMT than does an AF protocol. However, the DMT curves of
both these protocols lie some distance away from the cut-set DMT
bound. In both parallel AF and FF protocols, the key idea is to
partition the relay nodes in each layer into subsets of nodes
called super nodes. We propose a protocol with achievable DMT
(i.e, with a lower bound to DMT) that is better than that of
existing protocols for a full-duplex, $(n_0, n_1, \ldots,
n_{L+1})$ network.

A ``partitioning'' $\mathfrak{p}$ of a layered network corresponds
to dividing the nodes in every layer into super-nodes. We refer to
the number of nodes within a super-node as the size of the
super-node. Partitioning could potentially include partitioning of
the source by which we mean dividing the transmit antennas of the
source into super-nodes. It is similarly possible to partition the
sink. Connecting any two super-nodes of size $a$ and $b$
respectively, are $ab$ edges, which we will regard as a single
super-edge. We will term the resultant network comprised of
super-nodes connected through super-edges as the super-network.
The super-network inherits from the original network the property
of being fully-connected.   For a given partitioning
$\mathfrak{p}$, let $S^{(\mathfrak{p})}_i$ be the number of
super-nodes in layer $i$. The number of super-edge-disjoint paths
in the super-network is equal to the min-cut of the super-network
given by, \beqan N^{(\mathfrak{p})} & := &
\min_{\{i=0,1,2,...,L\}} S^{(\mathfrak{p})}_i
S^{(\mathfrak{p})}_{i+1} . \eeqan

We propose a protocol which uses different partitionings depending
upon the multiplexing gain $r$ (we will refer to $r$ as the rate
by abuse of notation). \footnote{The idea of varying the protocol
parameters depending on the multiplexing gain $r$ was used in
\cite{EliVinAnaKum} for the NSDF protocol.} The basic intuition is
that, at lower rates, we can exploit the diversity of the network
by using a partitioning that results in the creation of a large
number of super-edge-disjoint paths. At higher rates, we utilize a
partitioning which supports enough degrees of freedom (which is
equal to the minimum of the number of antennas in any super-node
of an edge-disjoint path).

Let $\mathfrak{P}$ denote all possible partitionings. There can be
two different partitionings in which every layer has the same
number of super-nodes with corresponding sizes. However, we need to
choose only one of them, as the performance of the protocol
depends only on the size of super-nodes. Fix a partitioning
$\mathfrak{p}$ $\in$ $\mathfrak{P}$ on the layered network. We operate the network
using the following protocol: Activate all the
$N^{(\mathfrak{p})}$ super-edge-disjoint paths successively so
that each path is activated for $T$ time instants. During the
activation of $i$th path, we will get an induced channel matrix
${\bf H}$ that is block lower-triangular with ${\bf
H}^{(\mathfrak{p})}_i$, the product matrix for the $i$-th path
appearing as the $i$th entry along the diagonal of ${\bf H}$. The
induced matrix has entries in the lower triangular part of the
matrix due to the presence of back-flow.

Since ${\bf H}$ is blt, by Theorem~\ref{thm:main_theorem}, we can
lower bound the DMT of this matrix by the DMT of ${\bf H}^{(0)}$,
the block diagonal matrix extracted from ${\bf H}$. Let
$d_{i}^{(\mathfrak{p})}(r)$ be the DMT of ${\bf
H}^{(\mathfrak{p})}_i$, which can be computed using the techniques
for computing the DMT of product Rayleigh matrices given in
\cite{YanBelNew}. Now the DMT $d_{H^{(0)}}(r)$ can be computed
using the  parallel channel DMT in Lemma~\ref{lem:parallel_channel} to yield the DMT of the protocol as,  \beqa
d(r) & \geq & \max_{\{\mathfrak{p}
\in \mathfrak{P}\}} \ \inf_{ \left\{ \begin{array}{c} (r_1,r_2,\cdots,r_{N^{(\mathfrak{p})}}): \\
                                                        \sum_{i=1}^{N^{(\mathfrak{p})}} r_i = N^{(\mathfrak{p})}r \end{array}
                                    \right\} } \ \sum_{i=1}^{N^{(\mathfrak{p})}}
{{d_i^{(\mathfrak{p})}}(r_i)} . \nonumber \\
\label{eq:MA_FD_Layered_mid} \eeqa

Since the optimization is over the set of all possible partitions,
it might be difficult to compute the DMT in general. So we
consider a restricted case when the source and sink are not
partitioned, and all the relay layers are partitioned into the
same number of super-nodes, $S$. Under this assumption, we have
that $1 \leq S \leq n_{\min}$, where  $n_{\min} = \min \{n_i\}$ is
the minimum number of antennas in any layer. Each super-node in
the $i$th layer contains $n^S_i$ relays, where
 $n^S_i := \left \lfloor \frac{n_i}{S} \right \rfloor,
i=1,2,...,L$. The remaining $n_i \pmod S$ relays in layer $i$ are
requested to be silent. This is done for simplicity of computing
the DMT. Let $d_{(n_0,n_1,...,n_{L+1})}(r)$ denote the DMT of a
product of independent Rayleigh matrices of size $n_0 \times n_1,
n_1 \times n_2,...,n_L \times n_{L+1}$, which can be computed
using the techniques given in \cite{YanBelNew}. The DMT lower
bound in \eqref{eq:MA_FD_Layered_mid} now simplifies to \beq d(r)
\geq  \max_{S \in \{1,2,...,n_{\min}\}} \ \ S \
d_{(n_0,n^{(S)}_1,...,n^{(S)}_L,n_{L+1})}(r) .  \eeq

The strategy of Corollary~\ref{cor:FD_Layered} can be used to
obtain a DMT of $d_{\max}(1-r)^{+}$ for any full-duplex layered
network even in the presence of multiple antennas at source and
sink. By combining this strategy with the aforementioned strategy
and choosing the one with the better DMT based on $r$, we get a
DMT of
\beqa & & d(r) \ \geq \ \max \{d_{\max}(1-r)^{+}, \nonumber \\
&& \max_{S \in \{1,2,...,n_{\min}\}} \ \ S \
d_{(n_0,n^{(S)}_1,...,n^{(S)}_L,n_{L+1})}(r) \} .
\label{eq:DMT_MA_FD}\eeqa

The proposed protocol is essentially the same as \cite{YanBelNew}
except for the following differences: \bit \item The presentation
here is not restricted to directed graphs since we are able to
handle back-flow that might arise in an undirected graph by using
Theorem~\ref{thm:main_theorem} to show that back-flow does
not impair the DMT.
 \item The presentation here is not restricted to partitions of fixed size since evaluation of the
 DMT in the case of arbitrary size partitions is made possible by the use of Lemma~\ref{lem:parallel_channel},
which computes the DMT of the parallel-channel. \item Also, since
we permit the size of the partition to vary with the rate, the
additional flexibility can be used to improve upon the DMT
attained by the FF protocol. While the dependence of network
operation upon the rate could increase implementation complexity,
one could adopt an intermediate strategy in which there are a
small number of modes of operation, for example, two modes
reserved respectively for low and high-rates. \item Finally, as
will be shown in the sequel, the above results can be extended to
half-duplex networks when all the relay layers are partitioned
into the same number of super-nodes. \eit

\emph{Example 1 :} Consider a $(2, 4, 2)$ multi-antenna layered
network. The achievable DMT curve using the FF protocol, the
proposed protocol and the cut-set bound are plotted in the
Figure~\ref{fig:multiple_ant_dmt_curve}. For rates $r \leq 0.5$,
the strategy for full-duplex layered network given in
Corollary~\ref{cor:FD_Layered} without any node-partitioning
performs the best. For rates $r \geq 0.5$, partitioning the middle
layer into two super-nodes with each containing two nodes performs
better. A combination of these two strategies gives a superior DMT
performance to the existing FF protocol.

\subsection{Half-Duplex Fully-Connected Layered Networks}

We consider multi-antenna layered networks with the additional
constraint of half-duplex relay nodes. We prove that the methods
provided above for full-duplex networks can be generalized for the
half-duplex network with bidirectional links.

Consider the partitioning method stated for full-duplex layered
networks, with $S_i = S, \forall i=1,2,...,L$, i.e., the relaying
layers are partitioned into equal number of super-nodes. Let the
source and sink be un-partitioned. When the relay layer $i$ is
partitioned into $S_i$ partitions, each super-node contains $n^S_i
:= \left \lfloor {\frac{n_i}{S_i}} \right \rfloor$ relays. If it
contains more, the remaining relays are requested to be silent, as
in the full-duplex case.

The following observations are in place: Once we replace the nodes
corresponding to the same partition by a super-node, this virtual
network forms a regular network. This is because each relaying
layer has the same number of partitions and therefore the same
number of super-nodes. The resultant network being regular, we use
the protocol that is given in Theorem~\ref{thm:knregular_dmt}.
Since the paths are of equal length, the interference is causal,
making the induced channel matrix lower triangular. This has
better DMT than the corresponding diagonal matrix by
Theorem~\ref{thm:main_theorem}. This yields the same lower bound
on DMT as in the full-duplex case. Thus the DMT of the protocol
with any given partitioning in the half-duplex case is no worse
than that with full-duplex protocol under the same partitioning.
So we get,

\beqa d(r) & \geq & \max \{d_{\max}(1-r)^{+}, \nonumber \\
&& \sup_{S \in \{2,3,..,n_{\min} \}} \ \ S \
d_{(n_0,n^S_1,...,n^S_L,n_{L+1})}(r) \}. \label{eq:DMT_MA_HD}\eeqa

The justification for retaining the term $d_{\max}(1-r)^{+}$ as
part of the maximization is because the original network is
fully-connected and by adopting the
matching-forward-directed-paths strategy even in the presence of
multiple antennas at the source and sink (see
Theorem~\ref{thm:fully_connected_layered}) we can achieve the same
lower bound for half-duplex networks as well.

\emph{Example 2:} For the case of $(2, 4, 2)$ network with
half-duplex constraint, the proposed protocol achieves the same
DMT as the full-duplex case of $\emph{Example 1}$. However, the FF
protocol used naively  for a half-duplex system by activating
alternative layers during alternate time slots will entail
multiplexing-gain loss by a factor of $\frac{1}{2}$.

\subsection{KPP(I) Networks}

We consider KPP(I) networks with all nodes, including the source
and the sink, having multiple antennas.

\subsubsection{Full-Duplex KPP(I) Networks\label{sec:multiple_fullduplex}}

We consider full-duplex KPP(I) networks with multiple-antenna
nodes. In the case of single-antenna KPP(I) networks, we activated
all backbone paths for equal durations of time in order to obtain
a linear DMT in Corollary~\ref{cor:FD_KPP}. We will use a similar
protocol here except that we activate different paths for
different durations of time.

Let ${\bf H}_{ij}$ be the fading matrix on edge $e_{ij}$. Let the
product fading matrix along backbone path $P_i$ be ${\bf G}_i$.
Then ${\bf G}_i = \prod_{j=1}^{n_i} {\bf H}_{ij}$. Let the DMT
corresponding to this product matrix ${\bf G}_i$ be $d_i(r)$,
which can be computed according to formulae given in
\cite{YanBelNew}.

Since activating different paths can potentially have different
DMTs, it is not optimal in general to use all paths equally. When
one is operating at a higher multiplexing gain, one might want to
use a path with higher multiplexing gain more frequently in order
to get greater average rate. While operating at a low rate, all
the paths must be used in order to get maximum diversity. We
consider a generic case where path $i$ is activated for a fraction
$f_i$ of the duration. \footnote{A similar technique can be used
for full-duplex fc layered networks with multiple antennas to
improve the achievable DMT.} These fractions can be chosen
depending on $r$ in order to maximize $d(r)$.

By so doing, we will get a parallel channel with repeated
coefficients. The DMT of such a channel was evaluated in
Lemma~\ref{lem:parallel_dependent}. After making
suitable rate adjustments, we obtain a lower bound on the DMT of
the protocol as,  \beqa d(r) & \geq & \sup_{(f_1,f_2,\cdots,f_K)}
\ \ \inf_{\left\{
\begin{array}{c} (r_1,r_2,\cdots,r_K): \\
                 \sum_{i=1}^{K} \ f_i r_i = r \end{array} \right\} } \ \sum_{i=1}^{K} {d_i(r_i)} \nonumber \\
\label{eq:multiple_antenna_KPP_FD} \eeqa

\subsubsection{Half-Duplex KPP(I) Networks\label{sec:multiple_halfduplex}}

From Section~\ref{sec:half_duplex}, we know that under the
half-duplex constraint, there exists a protocol activating the $K$
paths equally for KPP(I) networks with $K \geq 3$ causing only
causal interference. We can use the same protocol notwithstanding
the fact that the relays contain multiple antennas. By doing so,
we will get a transfer matrix which will be blt. Also, the
diagonal entries of this channel matrix would remain the same as
though the relay nodes operate under full-duplex mode. By
Theorem~\ref{thm:main_theorem}, this gives a lower bound on
the DMT, and it is equal to DMT lower bound of the full-duplex
network in \eqref{eq:multiple_antenna_KPP_FD}. However the
protocol for KPP(I) networks for $K=3$ activates all paths for
equal fractions of time, which is equivalent to setting $f_i =
\frac{1}{3}$. Therefore even when there is half-duplex constraint,
we can achieve the same DMT given by the
\eqref{eq:multiple_antenna_KPP_FD} with $f_i = \frac{1}{3}$
instead of maximization over all possible $f_i$.

If we want to achieve different fractions of activation for
different parallel paths, then we can follow a different trick for
$K \geq 4$. In this case, we can use the $K \choose 3$ 3-parallel
path networks, but activate each $3$PP network for a different
fraction of time, employing the same equi-activation protocol as
described in Section~\ref{sec:half_duplex}. Hence a path within a
$3$PP network is activated one-third fraction of the duration for
which the $3$PP is activated. Thus we use a $3PP$ network as a
fundamental unit in the strategy, and for that reason, any
fraction of time of activation, $f_i$ for a particular path $P_i$
is limited by $\frac{1}{3}$. In many cases, $f_i > \frac{1}{3}$
may turn out to be infeasible. Moreover, one can show that, for $K
\geq 4$, all time fractions $f_i, \ 1 \leq i \leq K$ are feasible
as long as $(f_1,f_2,...,f_K) \in \mathcal{F}$ where \beqan
\mathcal{F} := \{(f_1,f_2,...,f_K): \sum_{i=1}^{K} f_i = 1, \ \ 0
\leq f_i \leq \frac{1}{3} \}. \eeqan This is shown in
Appendix~\ref{app:fraction}.

For $K \geq 4$, this yields a DMT of  \beqa d(r) & \geq &
\sup_{(f_1,f_2,\cdots,f_K)\in \mathcal{F}} \ \
\inf_{ \left\{ \begin{array}{c} (r_1,r_2,\cdots,r_K): \\
         \sum_{i=1}^{K} \ f_i r_i = r \end{array} \right\} } \ \sum_{i=1}^{K} {d_i(r_i)}
\nonumber \\ && \label{eq:multiple_antenna_KPP_HD} \eeqa

This is the same as the lower bound on the DMT for the full-duplex
case, except that we are constrained to have all activation
fractions $f_i$ to be lesser than one-third.

\section{Code Design\label{sec:code_design}}

\subsection{Design of DMT-Achieving Codes}

Consider any network and AF protocol described above, and let us
say the network is operated for $N$ slots using such a protocol to
obtain an induced channel matrix  ${\bf Y} \ = \ {\bf H}{\bf
X}+{\bf W}$ where $X$ is a $(M \times 1)$ vector and ${\bf Y},{\bf
W}$ are $P \times 1$ vectors and ${\bf H}$ is a $P \times M$
matrix. However, to achieve the DMT of this induced channel one
needs to code over both space and time, i.e, transmit a matrix
${\bf X}$ drawn from a space-time (ST) code ${\cal X}$ as opposed
to just sending a vector. In order to obtain an induced channel
with ${\bf X}$ being a $M \times T$ matrix, we do the following:
Instead of transmitting a single symbol, each node transmits a row
vector comprising of $T$ symbols during each activation. Then the
induced channel matrix takes the form: ${\bf Y} \ = \ {\bf H}{\bf
X}+{\bf W}$, where now, ${\bf X}$ is a $M \times T$ matrix, ${\bf
Y},{\bf W}$ are $M \times T$ matrices with channel matrix $H$
remaining as before. We will regard the product $MT$ as
representing the block length of the ST code ${\cal X}$ since the
transmission of code matrix ${\bf X}$ takes place over $MT$
channel uses.

Now from \cite{TavVis}, we know that if the code matrix is drawn
from an approximately universal code ${\cal X}$, then the code
${\cal X}$ will achieve the optimal DMT of the channel matrix $H$
irrespective of the statistics of the channel. Explicit minimal
delay approximately universal codes for the case when $T=M$ are
given in \cite{EliRajPawKumLu}, constructed based on appropriate
cyclic division algebras (CDA) \cite{SetRajSas}. These codes can
be used here to achieve the optimal DMT of the induced channel
matrix.

\subsection{Short DMT-optimal Code Design for Block-Diagonal Channels}

In the special case that the channel matrix $H$ is a
block-diagonal matrix, we can use MIMO parallel channel codes to
construct DMT optimal codes of shorter block length, thereby
entailing lesser decoding complexity and delay.\footnote{The same
technique can be used when, after permuting the inputs or the
outputs, the matrix $H$ is block diagonal.} In particular if
$\bold{H}$ is block diagonal with entries $\bold{H}_i,
i=1,2,...,L$ on the diagonal, with $H_i$ of size $m_i \times p_i$.
Since $H$ is of size $P \times M$, we have $M = \sum m_i, P = \sum
p_i$. Let $T := \max_{i} m_i$. Let us construct the code $X$ of
size $M \times T$ as \beqa {\bf X} & = & \left[ \begin{array}{c}
{\bf X}_1\\
{\bf X}_2\\
\vdots\\
{\bf X}_L \end{array} \right],  \label{eq:column} \eeqa where
${\bf X}_i$ is a $m_i \times m $ matrix, with ${\bf X}_1,{\bf
X}_2,...,{\bf X}_M$ forming an approximately universal MIMO
parallel channel code, then the constructed code is DMT optimal.
CDA-based ST codes construction for the rayleigh parallel channel
were provided in {\cite{YanBelRek,Lu}.  This construction was
shown to be approximately universal for the class of MIMO parallel
channels in \cite{EliKum}.   These codes are DMT optimal for every
statistical description of the parallel channel and therefore are
DMT optimal in this setting as well. Thus the parallel-channel
code ${\cal X}$ constructed here will have code matrices of size
$M \times (\max \{m_i\})$ in place of the earlier size $M \times
(\sum m_i)$.

\emph{Example 1: (MIMO-NAF)} Consider the MIMO NAF protocol for
the $N$ relay channel introduced by \cite{YanBelMimoAf} and
explained in \emph{Example-4} of section~$II.E$ in \cite{Part1}.
This protocol in essence, is the sequential concatenation of $N$
protocols for a single-relay NAF channel, each involving a
different relay. If the source has $n_s$ antennas and destination
$n_d$ antennas, the channel matrix will then be  block diagonal
with each block being of size $2n_d \times 2n_s$, i.e., $m_i :=
2n_s, i=1,2,...,N$. Therefore the code design for this block
diagonal $H$ will result in a DMT-optimal code of size $2 N n_s
\times 2n_s$, which gives essentially the same code as in
\cite{YanBelMimoAf}. In particular for the NAF protocol with a
single antenna at all nodes, this yields a code of size $2N \times
2$. Now, even if the paths are activated for unequal fractions of
time as suggested in the latter part of \emph{Example-4}, the
matrix remains block-diagonal and thus this short-code
construction technique can be used here as well.

\emph{Example 2: KPP Networks} For KPP networks, the matrix
between input and the output is a diagonal matrix of size $nK
\times nK$ with the diagonal entries comprising of the $K$ product
coefficients ${\bf g}_i$ repeated periodically. If we consider the
$K \times K$ sub-matrix ${\bf H}_{\text{sub}}$ of ${\bf H}$
comprising of all the $K$ product coefficients the DMT is still
the same, i.e., the DMT is equal to $K(1-r)^{+}$. So we will
restrict our attention to this $K \times K$ sub-matrix ${\bf
H}_{\text{sub}}$. Since this matrix is diagonal, the code
construction for block-diagonal channel above gives a parallel
channel code of size $K \times 1$ that will be DMT optimal for
this channel. For KPP networks with multiple antennas, the same
technique will yield a code of size $K n_s \times n_s$ where $n_s$
is the number of antennas at the source.

\subsection{Short DMT-optimal Code Design for KPP(I) Networks}

For KPP(I) networks the matrix is not block-diagonal and therefore
the block diagonal code construction can not be used and the
longer code construction affords a code length of $M \times M$.
Also we need $M$ very large for the initial delay overhead to be
minimal. This entails a very large block length, and indeed very
high decoding complexity. Now a natural question is whether
optimal DMT performance can be achieved with shorter block
lengths. We answer this question for KPP(I) networks by
constructing DMT optimal codes that have $T=L$ and a block length
of $L^2$, where $L$ is the period of the protocol used. We also
provide a DMT optimal decoding strategy that also requires only
decoding a $L \times L$ space time code at a time. This is a
constant which does not depend on $M$ and therefore, even if we
make $M$ large, the delay and decoding complexity are unaffected.

Consider the first $L$ inputs $x_{1},x_{2},...,x_{L}$. If the
channel matrix is restricted to these $L$ time slots alone, then
channel matrix would be a lower triangular matrix with the $L$
independent coefficients $g_i$, $i=1,2,..,K$ repeated
periodically. The DMT of this matrix, after adjusting for rate, is
$d_K(r) = K(1-r)^{+}$. So if we use a $L \times L$ DMT optimal
matrix as the input (this can be done by setting $T=L$ and using a
$L \times L$ approximately universal CDA based code for the
input), we will be able to obtain a DMT of $d_K(r)$ for this
subset of the data. This means that the probability of error for
this vector comprising of $T$ input symbols will be of exponential
order $P_e \doteq \rho^{-d_K(r)}$ if an ML decoder is used to
decode the $L \times L$ matrix. Then we cancel this portion of
input and then focus on the next $L$ inputs. Again the transfer
matrix between input and output will be lower triangular with the
same properties, yielding a DMT of $K(1-r)^{+}$. However, the
probability of first block error increases the net probability of
decoding error for the second block by a factor of two. Since this
constant factor does not matter in the scale of interest, we
conclude that the DMT achieved remains as $K(1-r)^{+}$. This can
be repeated for the whole matrix in a successive manner. Thus, the
above mentioned successive-interference-cancelation (SIC) based
technique yields an optimal DMT while reducing the decoding
complexity significantly.

\subsection{Universal Full-Diversity Codes \label{sec:univ_full_div}}

Consider a input output equation of the form ${\bf Y} ={\bf HX} +
{\bf W}$ where ${\bf X},{\bf Y},{\bf H},{\bf W}$ are $M \times M$
matrices.

Usually the code design criterion given for a input matrix to have
full diversity for rayleigh fading is that the difference of any
two possible input matrices be of full rank. In this section, we
show that such a criterion is sufficient to get full diversity
under \emph{any} statistical description of the channel matrix. By
full diversity here, we mean that the code will attain a diversity
equal to $d(0)$ for the corresponding channel.

We quote the following theorem from the theory of approximately
universal codes (Theorem 3.1 in \cite{TavVis} ):

\begin{thm} \cite{TavVis} \label{thm:au_mimo}
A sequence of codes of rate $R(\rho):= r \log  \rho $ bits/symbol
is approximately universal over the MIMO channel if and only if,
for every pair of codewords, \beq\label{eq:mimo_univ_crit}
\lambda_1^2 \lambda_2^2\cdots \lambda_{n_{\min}}^2  \geq
\frac{1}{2^{R(\rho)+o(\log \rho )}} = \frac{1}{{\rho^r} \
2^{o(\log \rho)}}, \eeq where $\lambda_1,\ldots
,\lambda_{n_{\min}}$ are the smallest $n_{\min}$ singular values
of the normalized (by $\frac{1}{\sqrt{\rho}}$) codeword difference
matrix. A sequence of codes achieves the DMT of any channel matrix
if and only if it is approximately universal.
\end{thm}

For the particular case of zero multiplexing gain in
Theorem~\ref{thm:au_mimo}, we obtain that the criterion reduces to
\beq \lambda_1^2 \lambda_2^2\cdots \lambda_{n_{\min}}^2 \geq
\frac{1}{ 2^{o(\log \rho)}}. \eeq

As a result, if for all pairs of codewords, a code satisfies the
condition that difference determinant is non-zero, i.e., \beq
\lambda_1^2 \lambda_2^2\cdots \lambda_{n_{\min}}^2 \geq L > 0,
\eeq then that code is approximately universal at rate $r=0$, and
thus will achieve, the maximum possible diversity gain $d(0)$ of
any given channel matrix.

This criterion is the same as the criterion for full diversity on
a Rayleigh channel. This means that all codes with full diversity
designed for the rayleigh fading MIMO channel are indeed full
diversity for MIMO channels having an arbitrary fading
distribution. This is summarized in the following lemma:

\blem A code having non-zero difference determinant achieves full
diversity $d(0)$ over any (arbitrary) fading channel. \elem

Therefore we can use a full-diversity code designed for a rayleigh
fading MIMO channel to get full-diversity for any KPP or Layered
network, when used along with the corresponding protocol for these
networks.

\appendices

\section{Proof of Lemma~\ref{lem:KPP3_causal} \label{app:causal}}

We begin with some definitions.

\bdefn \label{def:partition} A \emph{partition} is defined as a
set of $K$ nodes obtained by selecting precisely one node from
each of the $K$ parallel paths. \edefn

We use the term partition here in the sense of a boundary
separating one part of the graph from the other, although we do
permit edges to cross the partition. The nodes on each backbone
path can be assumed to be ordered from left to right. Therefore it
is meaningful to speak of nodes that are to the left of the
partition and nodes that are to the right of the partition. We use
this natural partial order on the set of nodes to define a partial
order on partitions.

\bdefn A partition $\mathcal{P}_1$ is said to be on the left of
another partition $\mathcal{P}_2$ if on each parallel path, if the
node corresponding to partition $\mathcal{P}_1$ is to the left of,
or the same as, the node corresponding to partition
$\mathcal{P}_2$. \edefn

\bdefn Given a subset $ S \subseteq \{ 1,2,..,K \}$, a KPP(I)
network restricted to $S$ is defined as the subgraph consisting of
nodes present in the $|S|$ parallel paths specified by the set
$S$. \edefn

\bdefn (\emph{k-Switch}) Consider a subset $S \subseteq \{
1,2,..,K \}$ of size $|S|=:k>1$, and a KPP(I) network restricted
to the subset $S$. Let $\mathcal{P}_1$ and $\mathcal{P}_2$ be two
partitions, such that $\mathcal{P}_1$ is to the left of
$\mathcal{P}_2$. Construct a bipartite graph where the vertices on
the left side of the bipartite graph are identified with the left
nodes of the partition and those on the right are associated with
nodes on the right side of the partition.    Thus there are a
total of $2\mid S \mid$ nodes in the bipartite graph. We draw an
edge in the bipartite graph, between a node on the left and a node
on the right if the corresponding nodes are connected by an edge
which does not lie on any backbone path. However, if partitions
share a node on a particular backbone path, we do not draw an edge
between them. If the subgraph comprising the nodes in $S$ on the
left and right has a complete matching, then the two partitions
are said to form a $k$-switch corresponding to the set $S$. \edefn

\bdefn A $k$-switch corresponding to the set $S$ is said to be
\emph{contiguous} if there are no nodes lying in between the two
partitions on any of the $k$ backbone paths in the network
restricted to the set $S$. Otherwise, the $k$-switch is said to be
\emph{non-contiguous}. \edefn

Examples of a $2$-switch and a $3$-switch in a 3PP network are
shown in the figures Fig.~\ref{fig:CI_TwoCycle} and
Fig.~\ref{fig:CI_ThreeCycle}.

\begin{figure}[h]
  \centering
  \subfigure[Non-contiguous 2-switch]{\label{fig:CI_Non_Cont_2switch}\includegraphics[width=50mm]{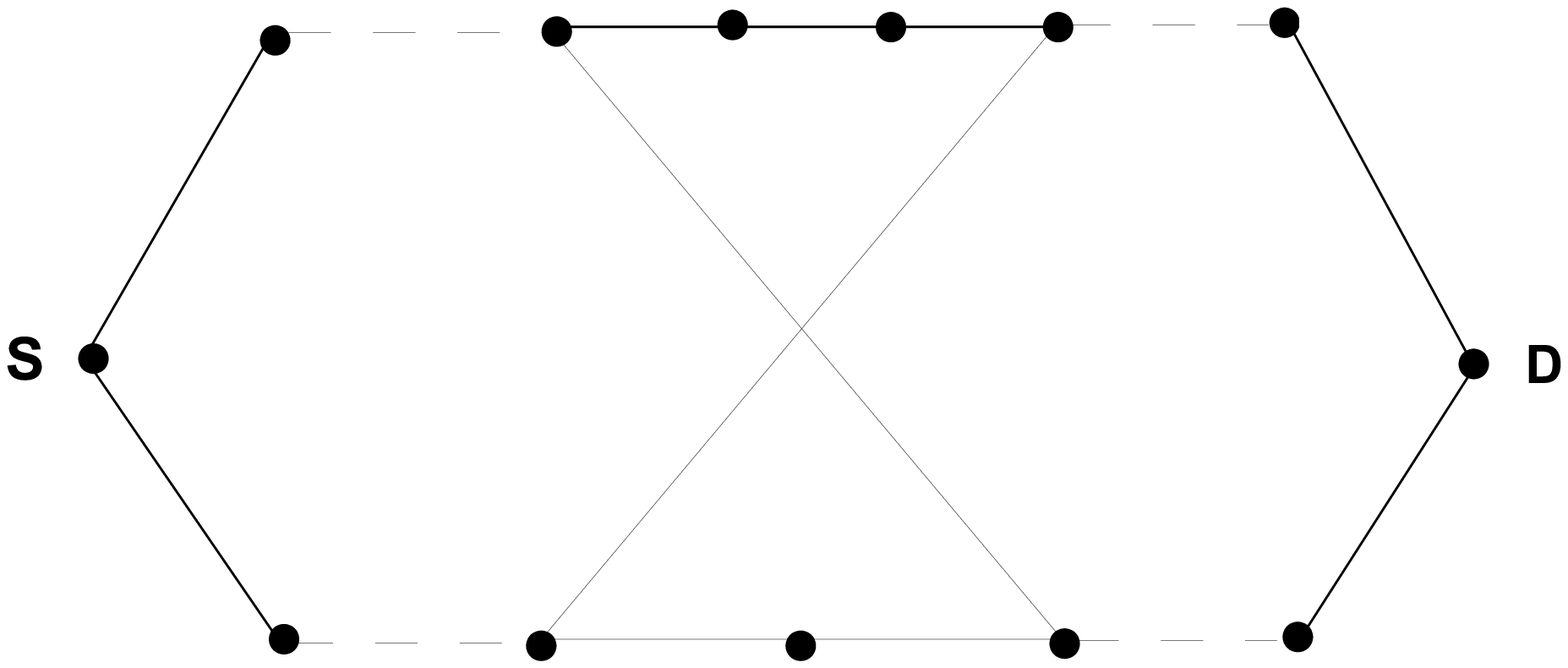}}
  \subfigure[Contiguous 2-switch]{\label{fig:CI_Cont_2switch}\includegraphics[width=50mm]{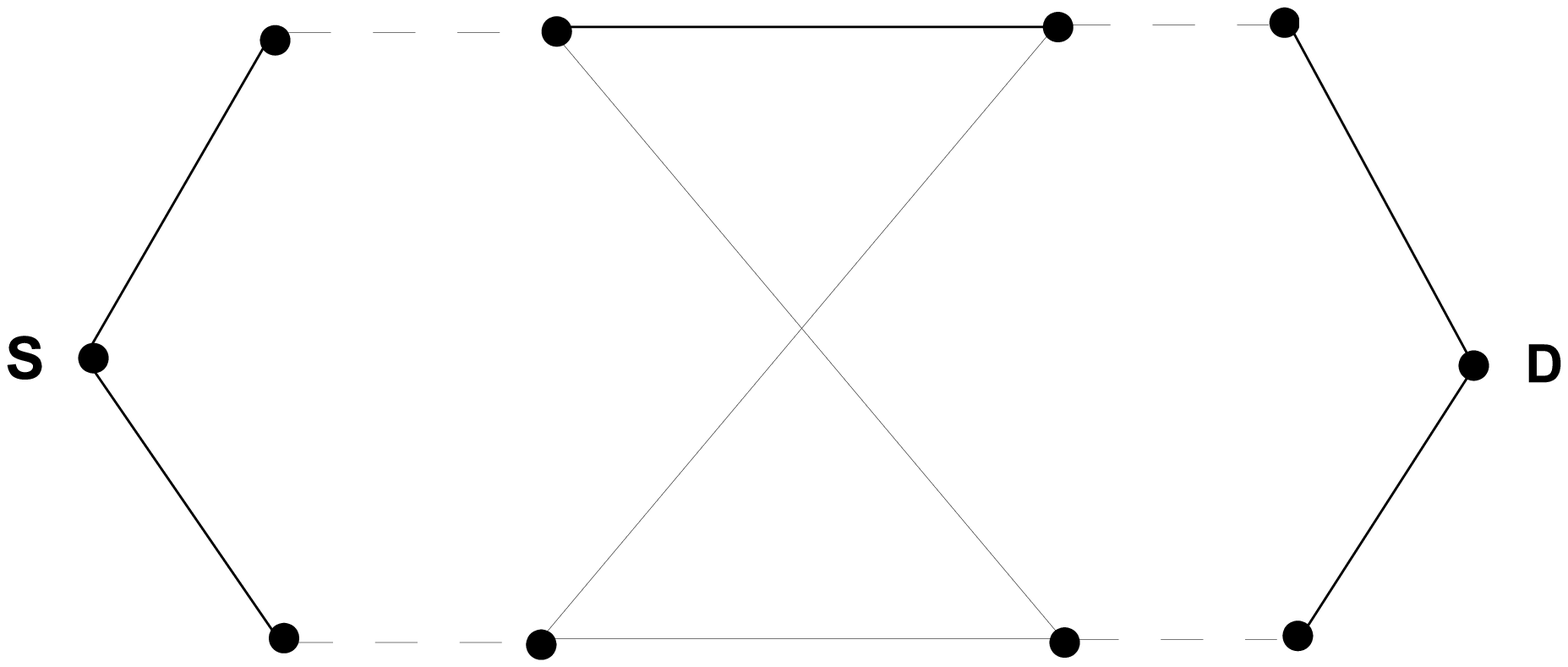}}
  \caption{Examples of 2-switches in KPP(I) networks}
  \label{fig:CI_TwoCycle}
\end{figure}

\begin{figure}[h]
  \centering
  \subfigure[Non-contiguous 3-switch]{\label{fig:CI_Non_Cont_3switch}\includegraphics[width=50mm]{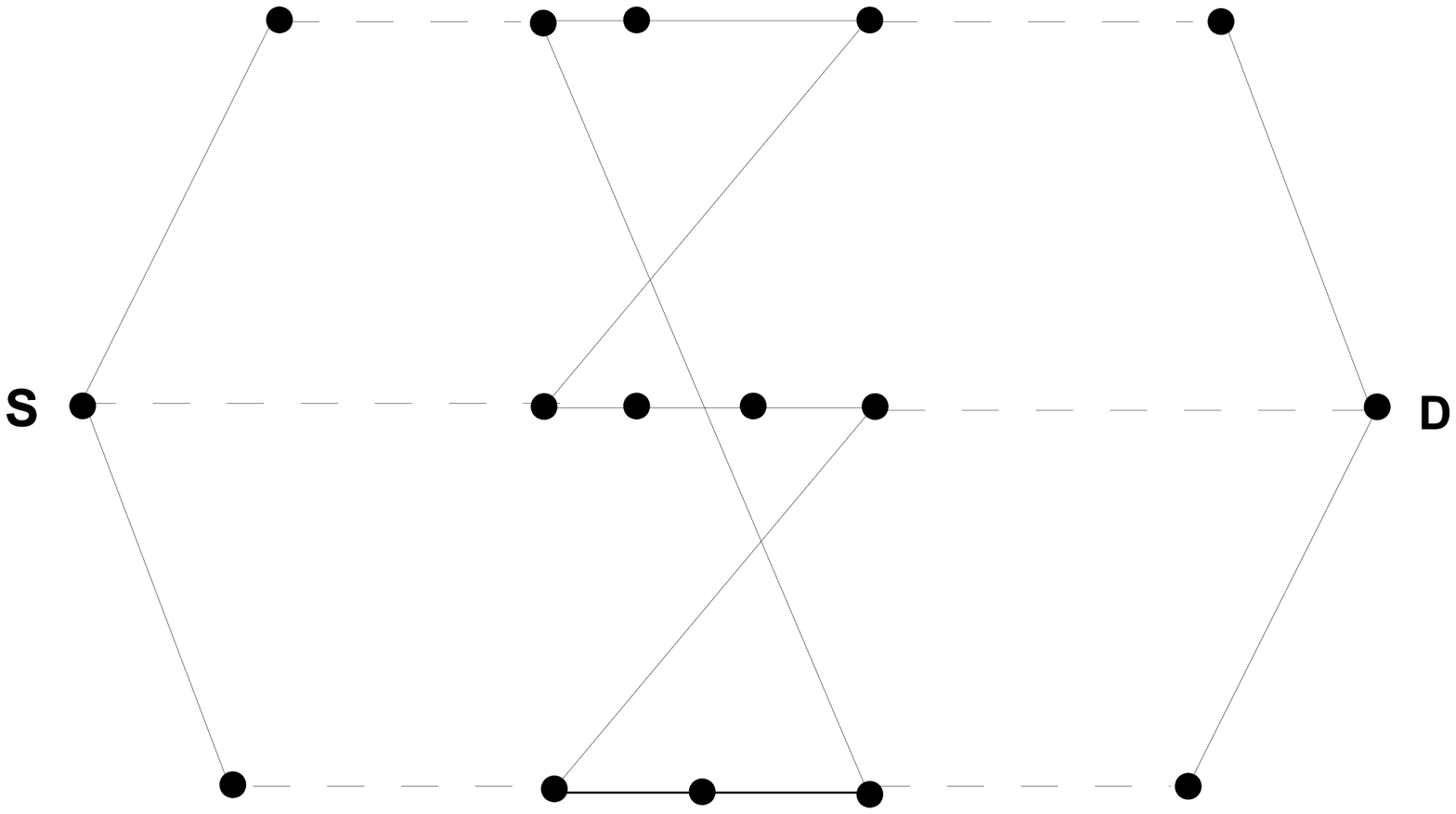}}
  \subfigure[Contiguous 3-switch]{\label{fig:CI_Cont_3switch}\includegraphics[width=50mm]{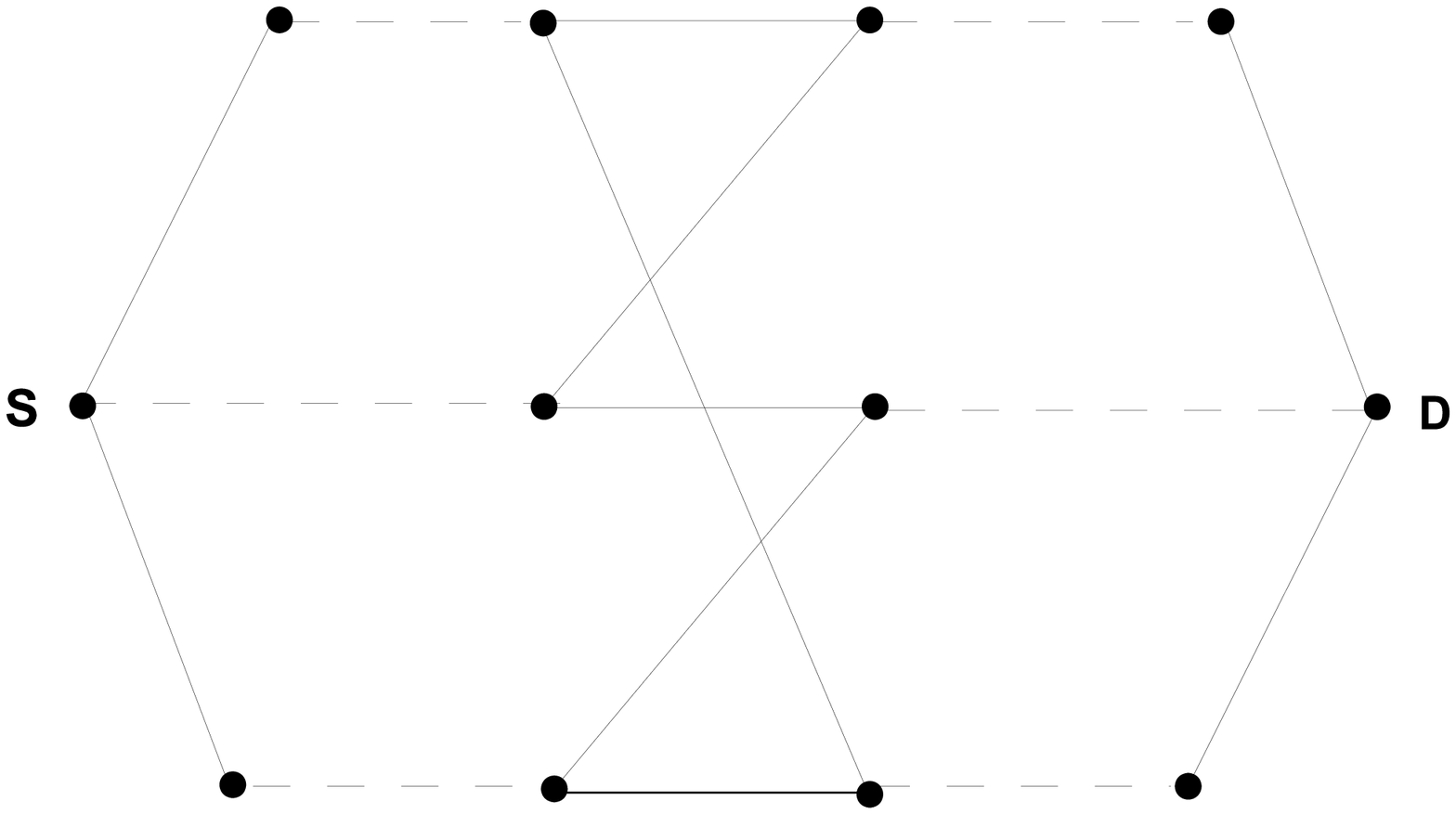}}
  \caption{Examples of 3-switches in KPP(I) networks}
  \label{fig:CI_ThreeCycle}
\end{figure}

Having the necessary definitions in place, we now proceed to
specify a protocol for a $3PP$ network in terms of delays
introduced at nodes as well as a schedule of edge activations. It
is appropriate to recall here that a KPP(I) network can be viewed
as a union of a backbone KPP network along with links that connect
between various nodes in the network, which we will term as
interference links. The particular schedule advocated here is
described below:

\subsubsection{Step 1 - Preprocessing}

We begin by deactivating nodes in backbone paths that are
encompassed by a shortcut (see Fig.) and redefine the backbone
paths to include the shortcut. By deactivating a node, we mean
that we will request the node to never transmit. Since there are
no transmissions from that node, it can be effectively deleted
from the graph. This will lead to a graph in which no backbone
path possesses a shortcut. This means that all interference links
will connect nodes in two distinct back-bone paths. Next, we
proceed to remove all the non-contiguous 2-switches, and
3-switches as explained in the lemma below.

\blem \label{lem:Switch_Removal} Any KPP(I) network can be
converted to a second KPP(I) network which does not have any
non-contiguous $k$-switches, for $2 \leq k \leq K$, by
deactivating certain nodes from the network and appropriately
redefining the backbone paths. \elem

\bpf Let us consider the given KPP(I) network. We employ the
following algorithm on the KPP(I) network to perform this
conversion:

\ben \item Identify any non-contiguous $k$-switch in the network
for any $2 \leq k \leq K$. If there are no non-contiguous
$k$-switches in the network, terminate the algorithm.

\item Given a non-contiguous $k$-switch, re-define the backbone
path in such a way that the segment of the newly-defined backbone
paths lying in between the left and right partition are precisely
the edges corresponding to the complete matching.  After redrawing
the KPP network, it can be seen that there are nodes which are a
part of none of the backbone paths. Deactivate these nodes. Since
the $k$-switch was non-contiguous, there exists at least one such
node. It can be readily verified that this reconfiguration does
not affect the number of backbone paths.

\item Repeat Step 1). \een

It remains to verify that the algorithm terminates. This is
clearly the case, since the algorithm deactivates at least one
node during each iteration and this process can not go on
indefinitely because there are only a finite number of nodes. (It
is for this reason, that this procedure excludes contiguous
switches where there will be no node to deactivate).

Note that throughout the iterative process, the number of backbone
paths has always remained fixed at $K$. \epf

The above lemma establishes that it is indeed possible to remove
all non-contiguous switches for any given KPP(I) network.

\subsubsection{Step 2 - Layering}

After preprocessing the network, we decompose the network into
sections which we call layers. As we shall see later, this
decomposition of network into layers will be very useful in
identifying a DMT optimal schedule for the network.

\bdefn Two partitions, one on left of the other, are said to
create a \emph{layer} if there are no links from any node in the
left (right) of the layer to a node inside or to the right (left)
of the layer, i.e., no links cross either partition. \edefn

\bdefn A network is said to be decomposed into a set of layers
$L_1,L_2,\ldots,L_N$ if the network can be split into layers $L_i$
such that the right partition of layer $L_i$ is the left partition
of layer $L_{i+1}$. \edefn

Now, we attempt to decompose a given KPP(I) network into a set of
layers with each layer having certain properties that will be
useful for us to obtain an efficient schedule for the network. We
note that, with the set of nodes that are connected directly
(i.e., by a single edge) to the destination as the left partition,
and the destination itself as the right partition, a natural layer
is formed. We will refer to this as the sink layer.   An analogous
definition yields the source layer.

\bnote \label{rem:links_on_partitions} The definition of a layer
does not preclude the possibility of having interference links
connecting nodes within a partition. We will prove later that such
interference links will not present non-causal interference. With
this foresight, we will neglect the links present on partitions
for now and later return to demonstrate that these do not change
the causal nature of the interference.  \enote

We next proceed to segment the entire network into a series of
layers, each following the next.  The layering process is
sequential in that we will be ready to identify the second
category of layer only after we have identified and segregated
layers belonging to the first category.

\bdefn A pair of partitions is said to form a {\em T-$3$ layer}
(short for layer of Type-3) if the partitions either form a
3-switch which is contiguous or else contains two contiguous
2-switches between different pairs of paths, see
Fig.~\ref{fig:CI_T3} for an example.  Apart from layers of this
type, we will also regard the source and sink layers as T-3
layers. \edefn

\begin{figure}[h]
  \centering
  \subfigure[Case 1]{\label{fig:CI_T3_1}\includegraphics[height=40mm]{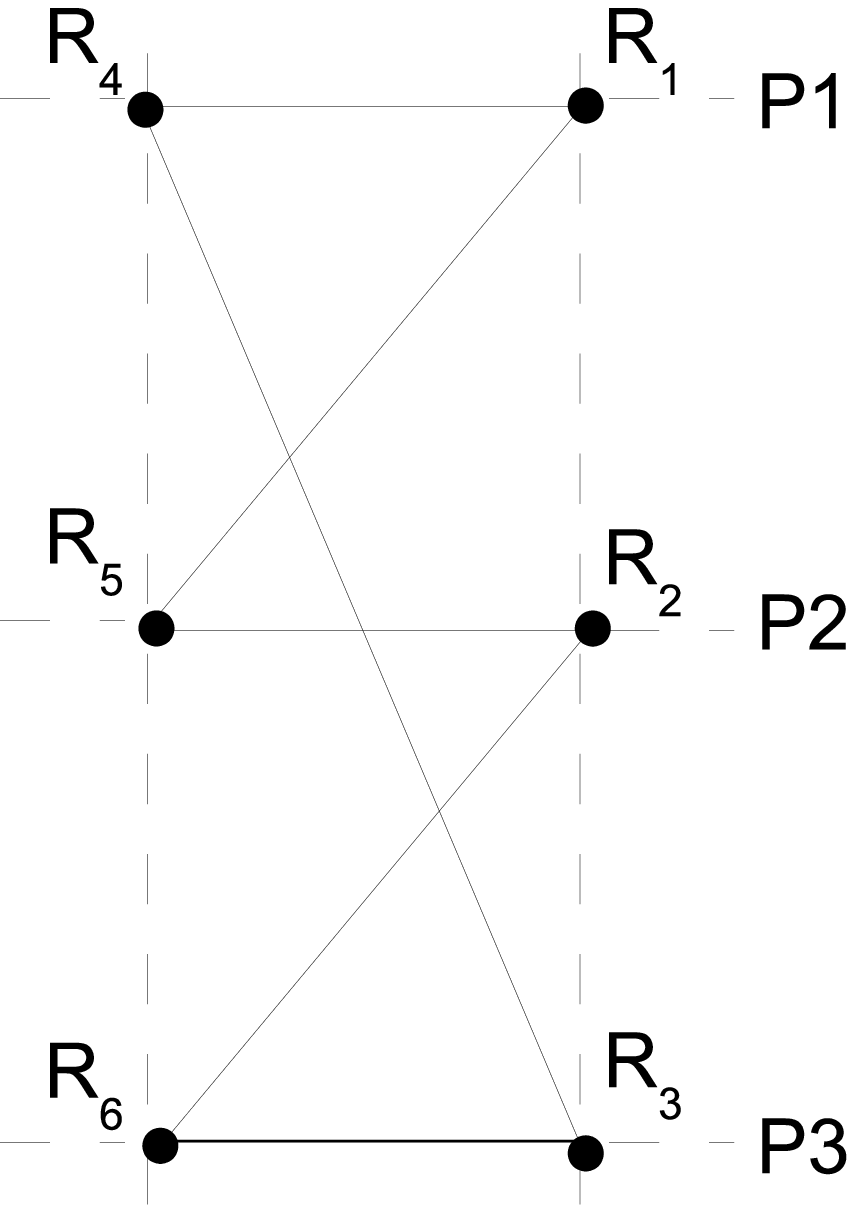}}
  \subfigure[Case 2]{\label{fig:CI_T3_2}\includegraphics[height=40mm]{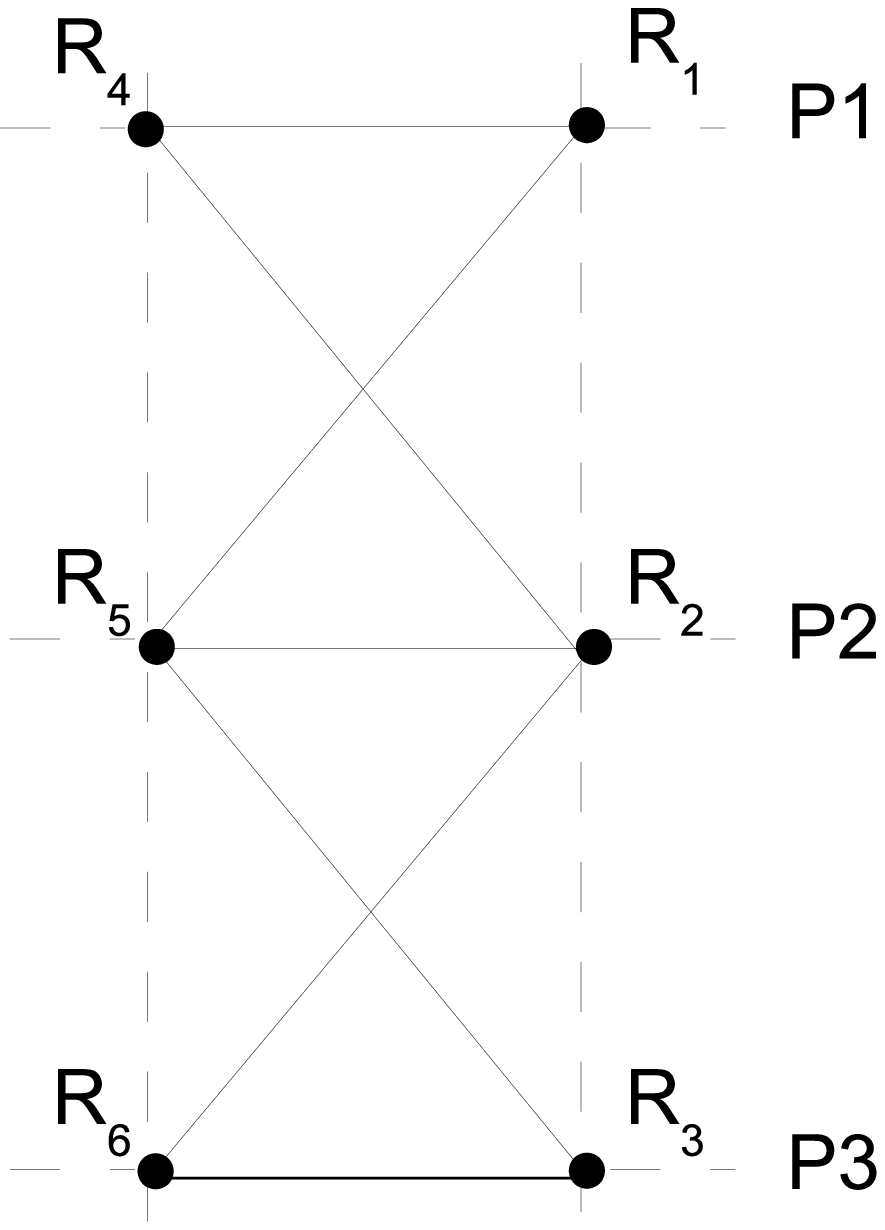}}
  \caption{T-3 layers in KPP(I) network with $K=3$}
  \label{fig:CI_T3}
\end{figure}

It must be noted that it has to be proved that a pair of
partitions comprising a contiguous $3$-switch or two contiguous
$2$-switches is indeed a layer. This can be proved by
contradiction, by showing that if the pair of partitions is not a
layer, then there must exist a non-contiguous $3$-switch or a
non-contiguous $2$-switch in the network, which have been assumed
to be removed in the previous preprocessing step.

\begin{figure}[h]
  \centering
  \subfigure[Case 1]{\label{fig:CI_T2_1}\includegraphics[height=40mm]{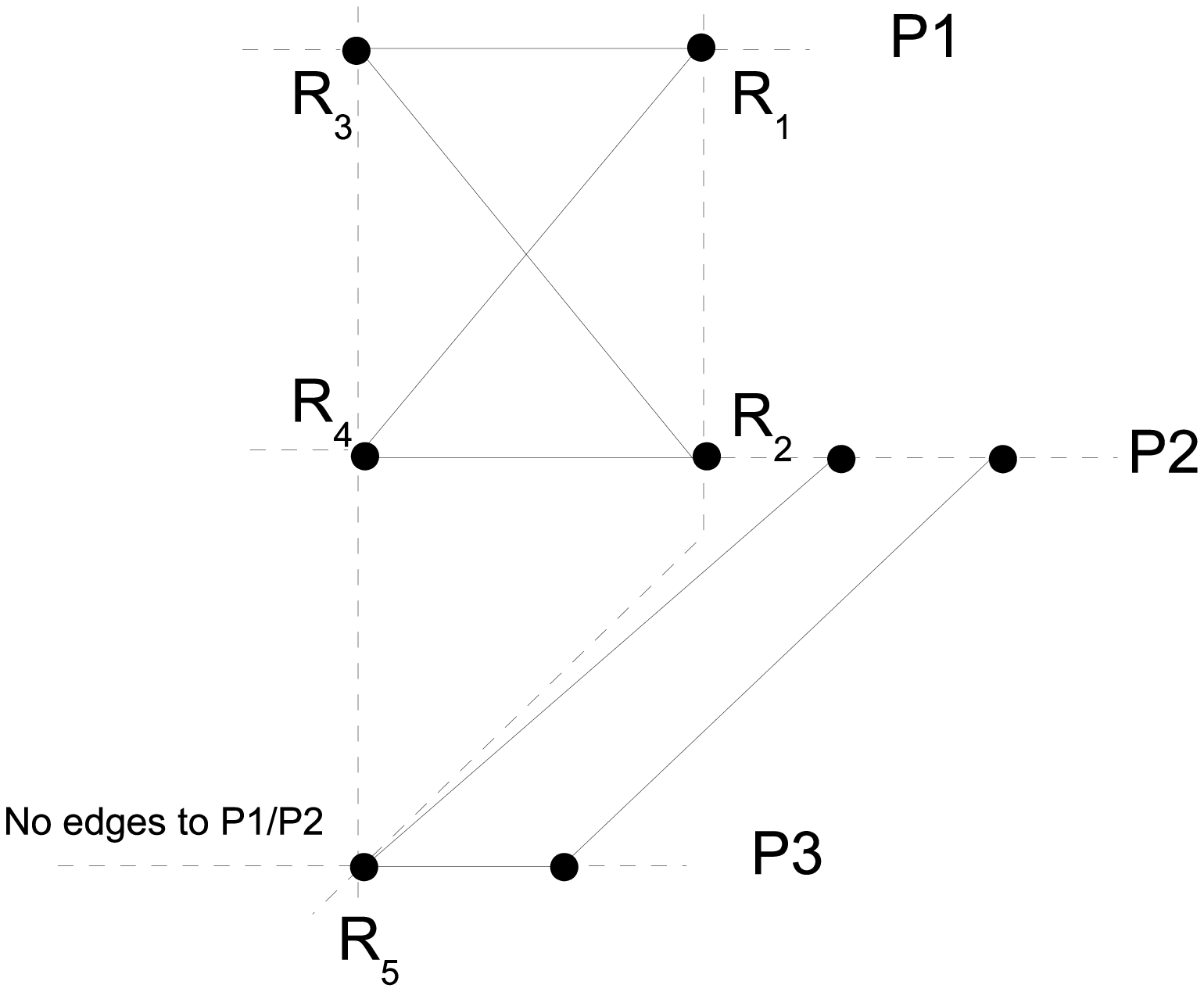}}
  \subfigure[Case 2]{\label{fig:CI_T2_2}\includegraphics[height=40mm]{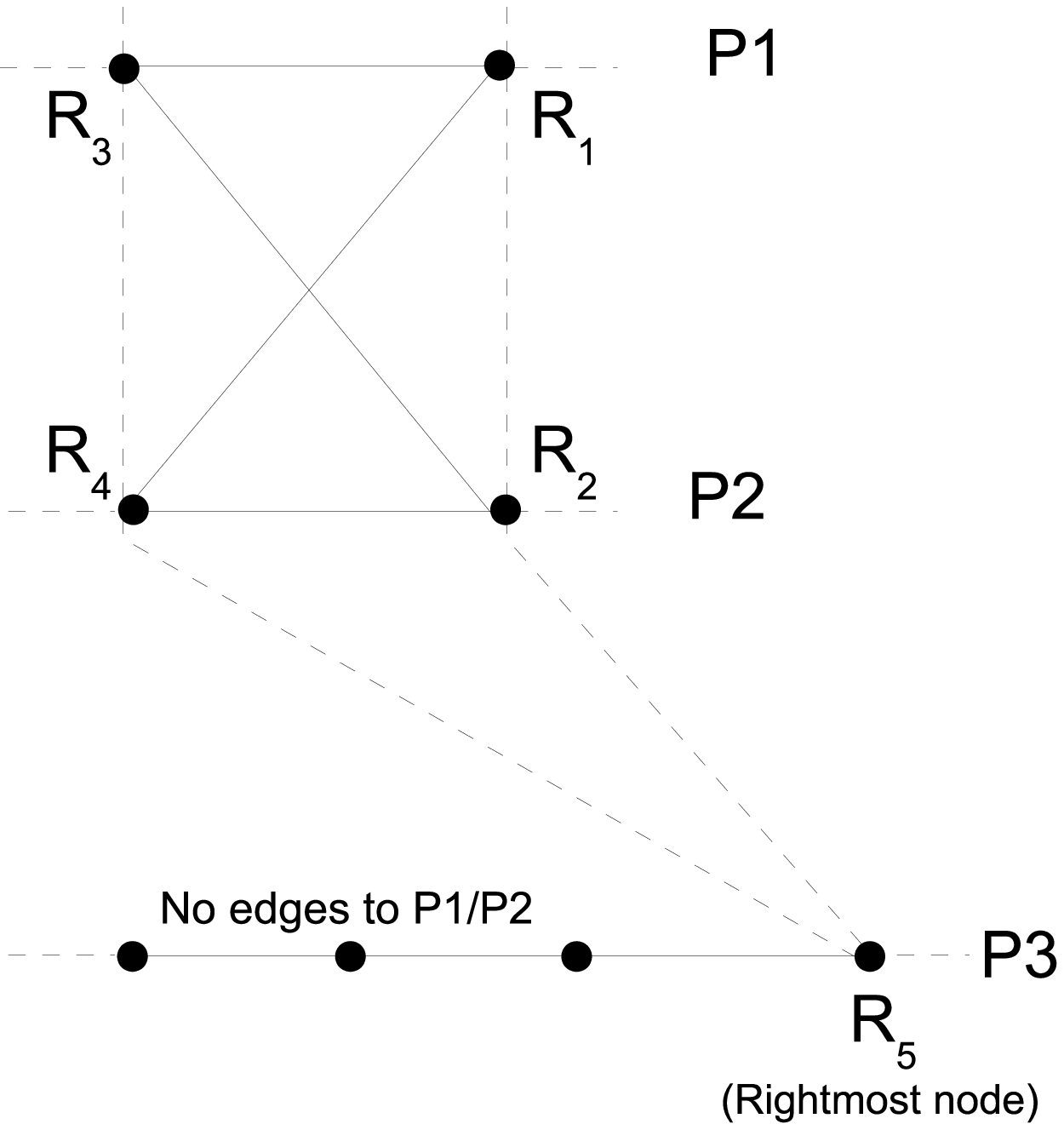}}
  \caption{Type T-$2$ layers in a KPP(I) network with $K=3$}
  \label{fig:CI_T2}
\end{figure}

If there exists no T-$3$ layer in the network, then we proceed to
identifying the next type of layer (T-$2$). Having identified a
T-$3$ layer in the network, we construct a modified graph after
removing the internal nodes from the graph. This is for the
purpose of identifying the remaining layers in the graph, and does
not imply that the internal nodes or edges are deactivated. The
obtained network graph will comprise of two disconnected
components, which we will call as fragments. We call this process
as segregating a layer, which essentially means removing all
internal components of a layer. Now we operate in the obtained
graph, and try to identify layers in the remaining components of
the graph. Once we have identified all the T-$3$ layers in the
graph, we can proceed to decompose the remaining fragments into
layers.

\bdefn Given a fragment of the network, consider a contiguous
$2$-switch between two paths (say paths $P_1$ and $P_2$). Consider
the leftmost node in $P_3$ connected to the right of the
$2$-switch on paths $P_1$ or $P_2$ (if there is no node connected
to the right of the $2$-switch, consider the rightmost node in
$P_3$ in the fragment). Choose the two nodes in the left partition
of the $2$-switch along with this node as the left partition and
choose the nodes in the right partition of the $2$-switch along
with the same node in $P_3$ as the right partition for a layer.
These two partitions can be shown to form a layer, using the fact
that the fragment does not contain any T-$3$ layers.  We call a
layer of this type as a \emph{T-$2$ layer}, see
Fig.~\ref{fig:CI_T2} for two examples of T-$2$ layers.\edefn

Continuing the sequential layering process, after all T-$3$ and
T-$2$ layers have been segregated, we are once again left with
fragments of the network.  Any such fragment does not contain a
layer of type T-$2$ or T-$3$ and is thus guaranteed not to have
any switches.  We proceed to decompose these fragments as follows:

\begin{figure}[h]
  \centering
  \subfigure[Case 1]{\label{fig:CI_T1_1}\includegraphics[height=40mm]{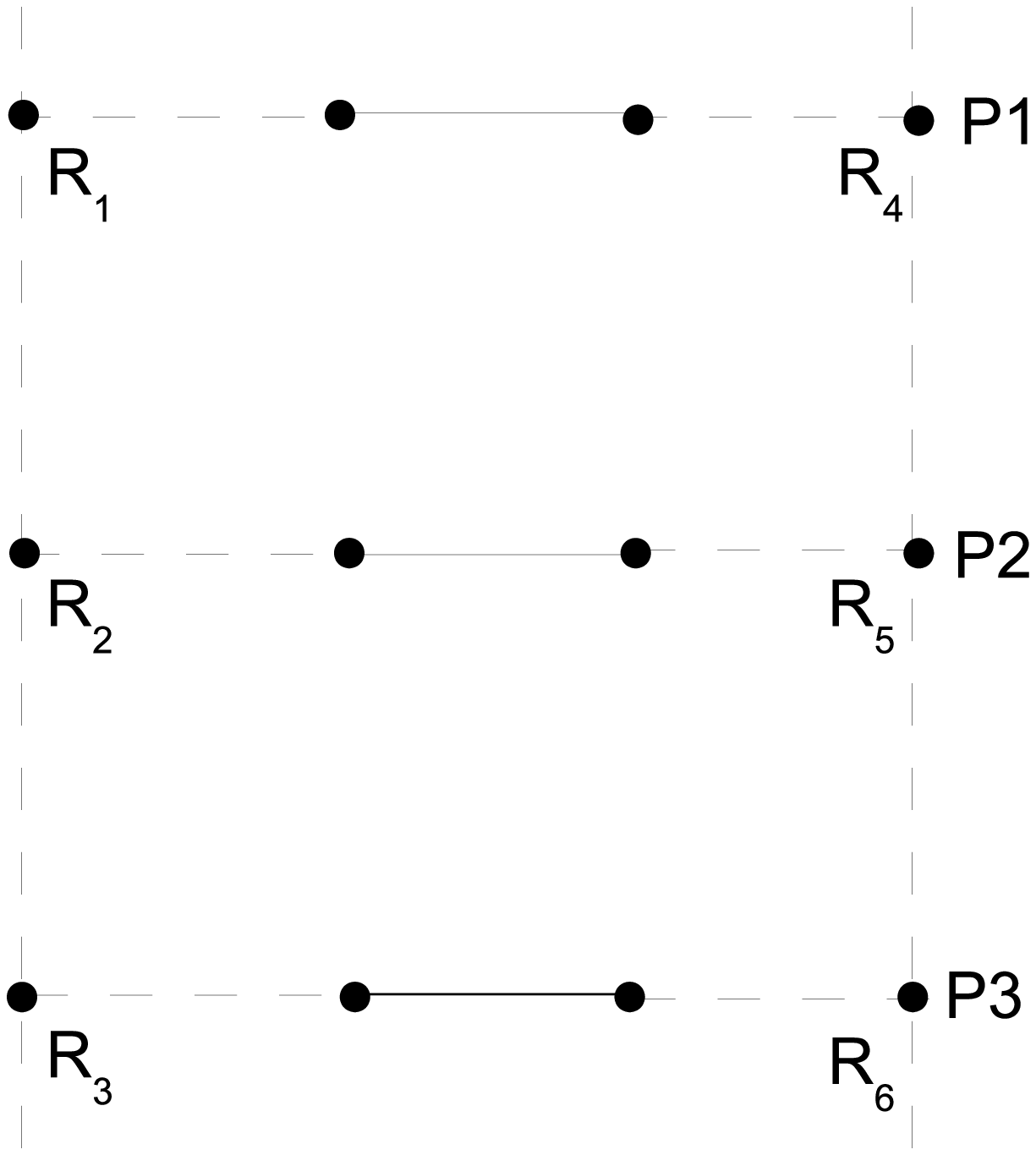}}
  \subfigure[Case 2]{\label{fig:CI_T1_2}\includegraphics[height=40mm]{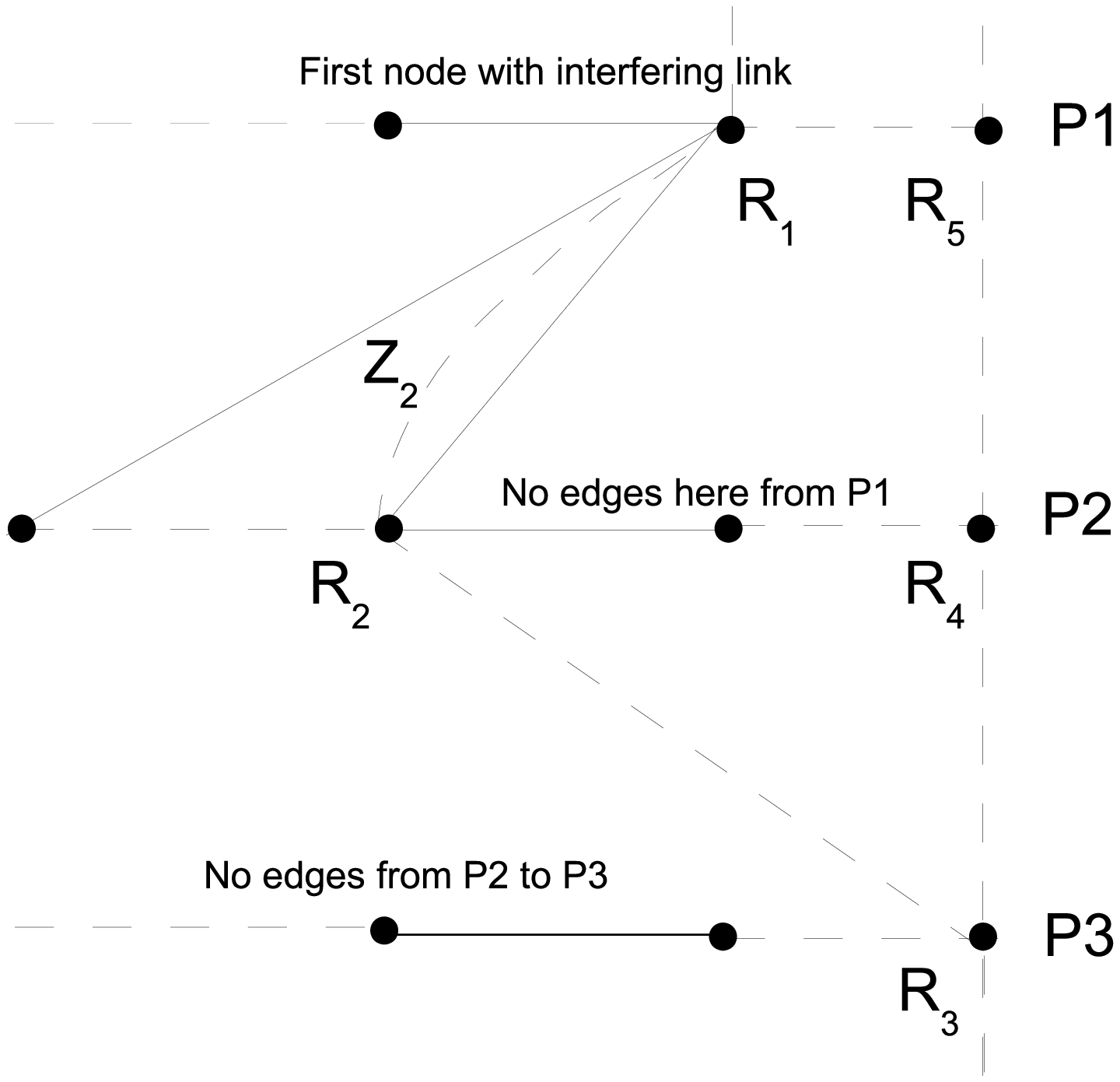}}
  \subfigure[Case 2]{\label{fig:CI_T1_3}\includegraphics[height=40mm]{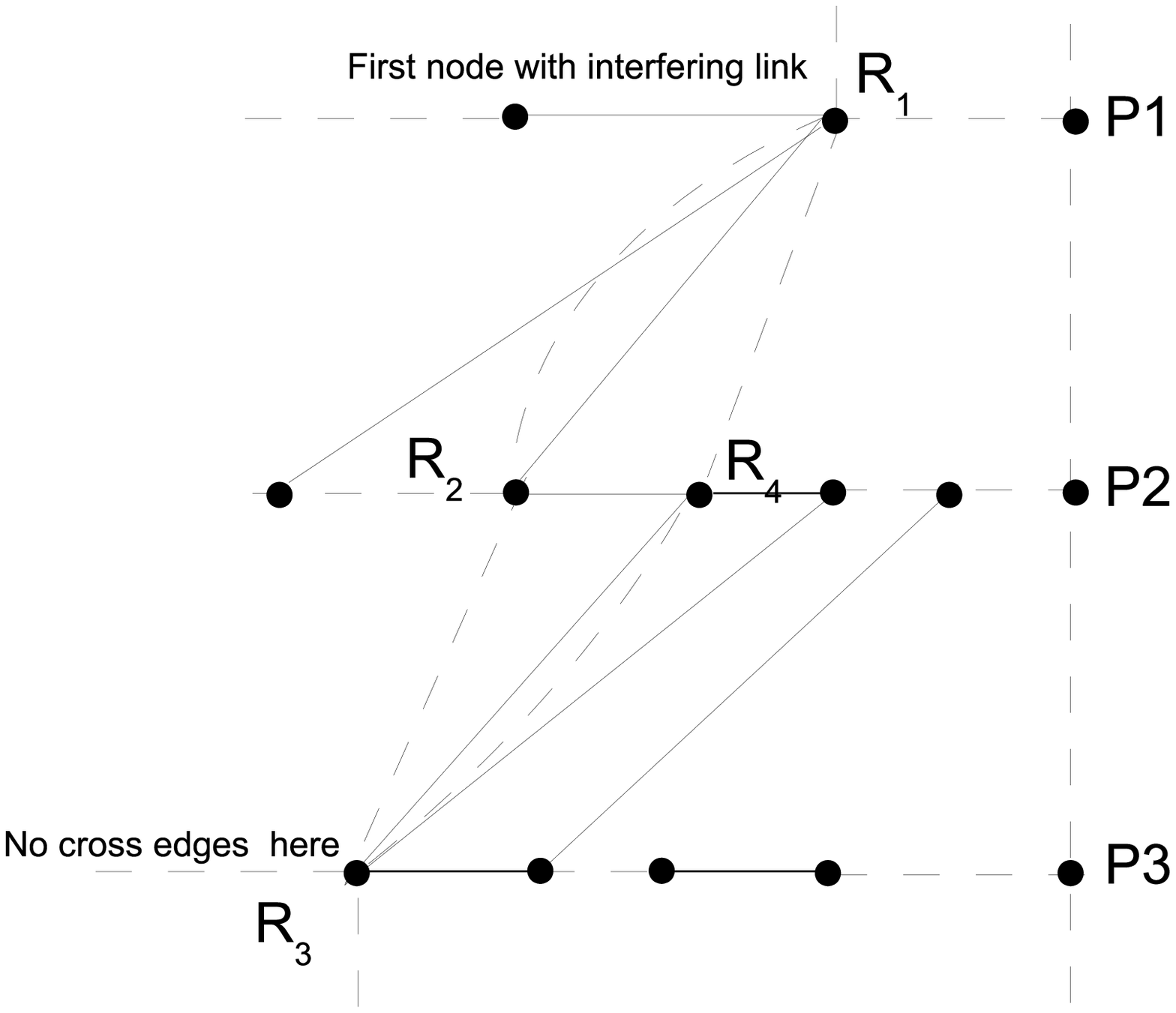}}
  \caption{T-1 layers in KPP(I) network with $K=3$}
  \label{fig:CI_T1}
\end{figure}

If there are no interference links in the given fragment, then the
whole fragment is treated as a layer. An example of such a layer
is given in Fig.~\ref{fig:CI_T1_1}. If there are interference
links, we pick a backbone path to which this link is connected and
label it as path $P_1$. Find the rightmost node in $P_1$, say
$R_1$, that is connected to an interference link. Let the other
end of this interference link be connected to a node $R_2$ in a
second backbone path, which we label as path $P_2$. Conceivably,
node $R_1$ could be connected via other interference links to
nodes in path $P_2$ that lie to the right of node $R_2$ on path
$P_2$. If this is the case, then we choose the rightmost such node
and relabel this node on path $P_2$ as node $R_2$. Then \bit \item
if there are no interference links connecting to a node on $P_2$
to the right of $R_2$, then $R_1$, $R_2$ and the rightmost node of
the remaining backbone path $P_3$ in the fragment, form a left
partition; this left partition along with the rightmost three
nodes of the fragment acting as the right partition, can be
verified to be a layer. An example of such a layer is given in
Fig.~\ref{fig:CI_T1_2} \item if there is an interference link
connected to a node on path $P_2$ that is to the right of $R_2$,
this interference link must originate from the remaining backbone
path $P_3$ since the current fragment does not contain any
switches; choose the leftmost node $R_3$ in $P_3$ connected to the
right of $R_2$; this node could potentially be connected to
multiple nodes in $P_2$. Let $R_4$ be the leftmost node in $P_2$
connected to $R_3$. Now set $R_1$, $R_2$ and $R_3$ as the three
nodes for the left partition and $R_1$, $R_4$ and $R_3$ as the
three nodes for the right partition. An example of such a layer is
given in Fig.~\ref{fig:CI_T1_2}.

\eit

A feature of the layers produced by this last step is that no
switches are contained within the layer and we will label such
layers as Type-1 layers (T-$1$). This sequential procedure has
resulted in the decomposition of the entire network into layers of
type T-$3$, T-$2$ or T-$1$.

Once the network is layered, we assign colors to edges, indicating
their activation pattern, as well as delays to nodes.

\subsubsection{Step 3 - Assigning colors and delays}

In this final step, we construct the protocol for the given
network by suitably assigning to each edge, a color that
represents the activation time slot. We also introduce delays at
node in order to make the protocol causal. Note that the given
network does not possess any non-contiguous switches and that the
network has been layered into $N$ layers, $L_1, L_2, \ldots, L_N$,
with each layer being of type T-$1$, T-$2$ or T-$3$.

The conditions corresponding to causal interference
(Prop.~\ref{prop:Causal_Interference}) can be satisfied if at
every node where there is an interference link branching out from
the backbone path, the shortest delay from the node to the
destination through the branch-out link is strictly greater than
the delay on the backbone path. It is easily seen that this
condition does not depend on the network to the ``left'' of this
node. For this reason, in designing the protocol, we can begin at
the rightmost node and make sure that this condition is satisfied
for all nodes.

Hence, to construct the protocol, we start from the rightmost
layer $L_N$ and proceed toward left layers, assigning colors to
edges and delays to nodes in each layer. At the end of this
process, we will obtain a protocol for communication, where every
node will be given a delay that must be added to every input that
it receives and every edge will be given a color (or equivalently,
a time slot) to amplify and forward the last symbol that it
received. We adopt the convention that a delay of zero at a node
corresponds to edges on either side of the node in the backbone
path transmitting consecutively. Under this convention, a delay
which is not equal to $2 \pmod 3$ can be added to any node without
violating the half-duplex constraint at the node.

On each layer $L_i$, we will make sure that the two criteria below
are met. We first provide an outline of the procedure and then
explain the details. \bit \item From any node in the layer, we
will make sure that the delay to the right partition, along the
particular backbone path on which the node is situated, is
strictly less than the delay to the right partition on any other
path that might lead from the same node (In making this
assessment, we will ignore any links connecting nodes within a
partition as noted in Note~\ref{rem:links_on_partitions}). We will
ensure this by adding delays to the three nodes in the right
partition of the layer, which we will denote by $D_i^r$, where
$D_i^r$ is a three-component vector comprising of the three delays
to be added. We refer to this as right-compensation of a layer.
\item We will further add delays to the nodes in the left
partition $D_i^l$ so that the delays incurred in travelling from
left partition to right partition along any of the backbone paths
is the same. We refer to this as left-compensation of the layer.
It is easy to see that adding delays on the left partition does
not change the right compensation of the layer. \eit

For layers of type T-$1$ it can be easily proved that there exists
delays that can lead to right-compensation and left-compensation.
Further, if $D_i$ is a right compensation vector for the T-$1$
layer, there exists another right compensation vector $D_i^{'}$
such that $D_i$ and $D_i^{'}$ differ in only one component and
even in that component the difference in the values is equal to
one. We will utilize this degree of freedom in choosing the right
compensation vector for T-$1$ layers later.

However, as it turns out, just adding delays is not sufficient for
layers containing switches, i.e., for layers of type T-$3$ and
T-$2$. So for these layers, we resort to ``neutralizing'' all
interfering links by carefully designing the protocol. An
interference link is considered neutralized under a protocol, if
the receiving node of the interference link is scheduled to
receive at a different time than the time during which the
transmitting node is active. We will explain this process in
detail in what follows by detailing how the delays, and colors are
assigned within each layer.

The rightmost layer $L_N$ is the destination layer, i.e., it is
comprised of three single-edge paths connecting from the backbone
paths to the destination. We assign three colors $A$,$B$ and $C$
to the three paths, so that the transmissions to the destination
are all orthogonal. We set $D_N^l$ as the zero vector since no
left-compensation is required for this layer. Let $i=N$.

Whenever the layer $L_{i+1}$ is compensated and colors are
assigned to the edges inside it, the following information is
passed on to the layer $L_i$ that is immediately to its left: its
left delays, $D_{i+1}^l$ and the colors on the edges immediately
to the right of layer $L_i$, which we call as right colors and
which can be regarded as a vector comprising of the three colors
on the three respective, ordered, backbone paths.

If layer $L_i$ is a T-$1$ layer, the delays $D_i^r$ for
right-compensation are computed. The delays $D_i^r + D_{i+1}^l$
are added to the right partition. However, if this gives delays
that violate half-duplex constraints (i.e., one of the delays
turns out to be equal to $2$ modulo $3$), then the degree of
freedom in the $T1$ layer is utilized to add more delay to a path
so that the delays do not violate the half-duplex constraint in
any of the paths. Inside the layer, the edges are colored
consecutively so that there is no delay at any node inside the
layer. The left-compensation delays $D_i^l$ are computed so that
the delays on all the backbone paths become equal. We make an
additional modification to this layer in the special case that the
layer $L_{i-1}$ to the left is a T-$3$ layer and all the right
colors output to layer $L_{i-1}$ are the same. In this case, the
degree of freedom in choosing the right compensation delays for
the layer $L_i$ is utilized so that the right colors of layer
$L_{i-1}$ are not the same. The reason for this modification will
become clear later when we consider T-$3$ layers.

\begin{figure}[h]
  \centering
  \subfigure[A KPP(I) Network]{\label{fig:before_color_delay}\includegraphics[width=60mm]{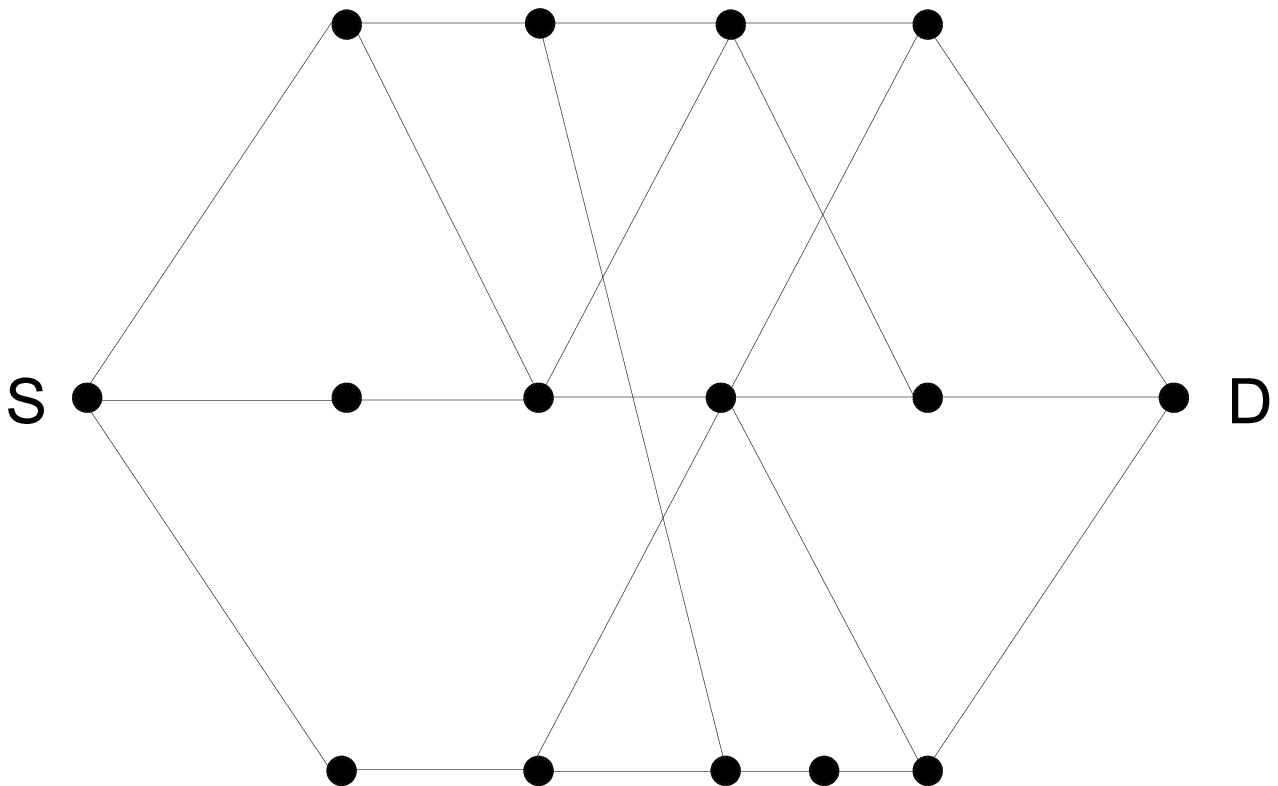}}
  \subfigure[The network after layering, coloring and delaying]{\label{fig:after_color_delay}\includegraphics[width=60mm]{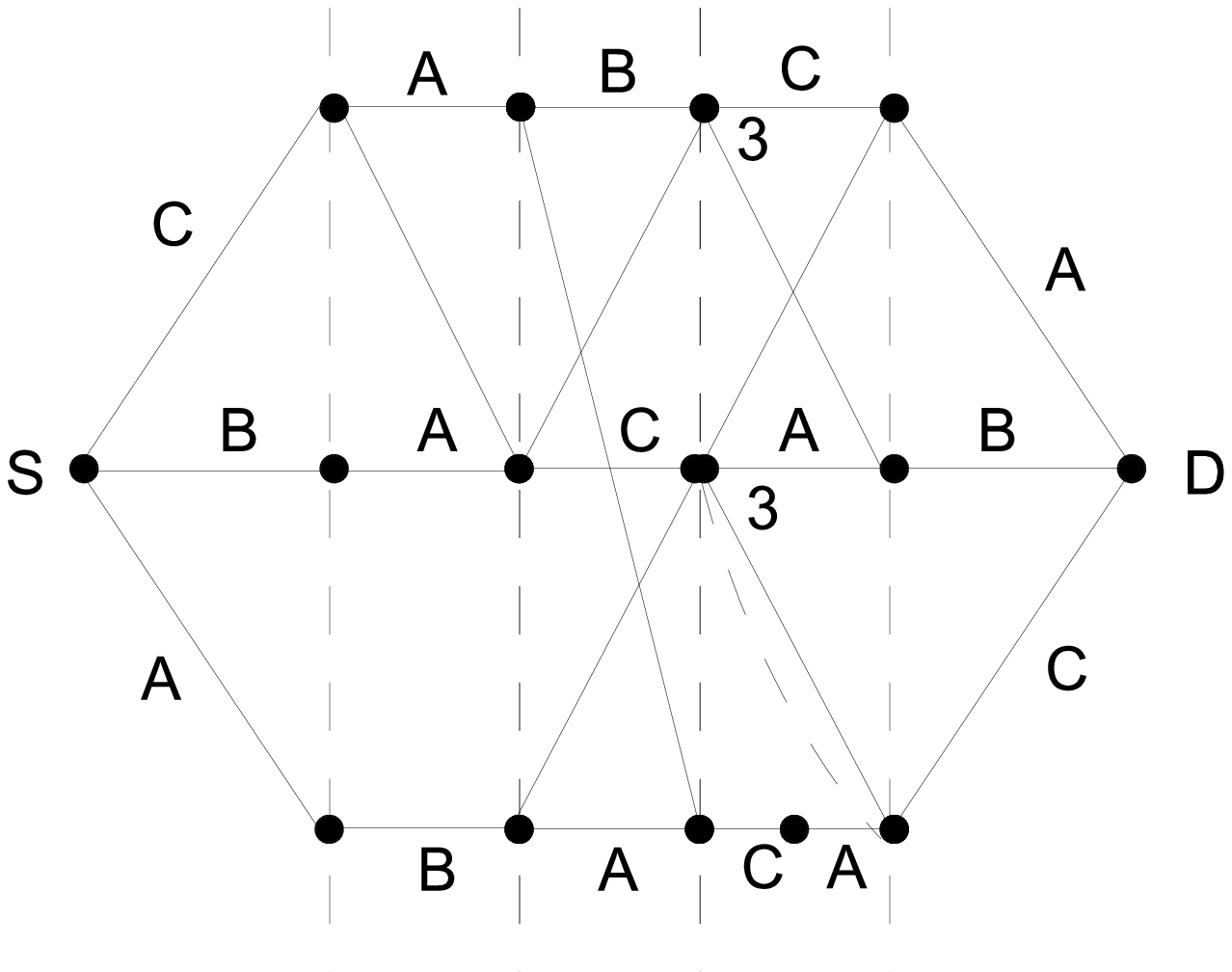}}
  \caption{Adding colors and delay in KPP(I) network}
  \label{fig:color_delay}
\end{figure}

If layer $L_i$ is a T-$2$ layer, we first color the two edges of
the contiguous $2$-switch on the back-bone path with distinct
colors. It can be verified easily that this can always be done
without violating the half-duplex constraint irrespective of the
right colors given to layer $L_i$. This coloring neutralizes the
interference links in the layer and therefore, there is no need
for right compensation of the layer, i.e., we set $D_i^r = 0$.
However delays still need to be added to the right partition of
the layer since this is also the left partition of layer
$L_{i+1}$. Delays are added to the right partition such that the
delay on each node on the right partition is greater than or equal
to the delays $D_{i+1}^l$. We compute the left compensation vector
$D_i^l$ to ensure that all the backbone paths have equal delays.

If layer $L_i$ is a T-$3$ layer, all the interference links can be
effectively neutralized if the three edges on the backbone paths
can be assigned distinct colors. It can be easily shown that this
can be done without violating the half-duplex constraint as long
as the right colors for this layer are not all the same. As an
example, if the right colors are $(A,B,B)$, then we can choose the
colors in layer $L_i$ as $(B,A,C)$. However if the right colors
are $(A,A,A)$, we cannot get distinct colors on the layer $L_i$
without violating the half-duplex constraint, since no edge in
layer $L_i$ is allowed to have color $A$. Therefore, we must
ensure that the right colors are all not the same. The right
colors will not all be the same if the layer $L_{i+1}$ is a T-$3$
or a T-$2$. If the right layer is a T-$1$ layer, then it has
already been modified in such a way that the right colors of $L_i$
are not all same. Thus all the interference links in the T-$3$
layer can be neutralized by choosing distinct colors on the three
paths. Thus, just as in the T-$2$ case, there is no need for right
compensation of the layer, i.e., we set $D_i^r = 0$. Delays are
added to the right partition such that the delay on each node on
the right partition is greater than or equal to the delays
$D_{i+1}^l$. We compute the left compensation vector $D_i^l$ to
ensure all the backbone paths have equal delays.

Now the delay from any node in layer $L_i$ to the right partition
of the layer is strictly greater than that along the corresponding
backbone path to the right partition. Once the right partition has
been reached, both paths incur equal delay (since this would have
been ensured in layer $L_{i+1}$ itself). In this way, we have
satisfied the conditions of Prop.~\ref{prop:Causal_Interference},
hence the interference is causal under the designed protocol. An
example of a KPP(I) network is given in
Fig.~\ref{fig:before_color_delay} and after layering, coloring and
delaying the network is depicted in
Fig.~\ref{fig:after_color_delay}. In this figure, the numbers
denote the amount of delay to be added at that particular node.
For example, the delay of $3$ in Fig.~\ref{fig:after_color_delay},
indicates that the node waits for $3$ time slots (which amounts to
one protocol cycle).

However, as noted in Remark~\ref{rem:links_on_partitions}, we have
not accounted for the interfering links between nodes within a
partition while constructing the protocol. In the proposition
given below, we justify that interference links connecting nodes
inside a partition do not alter the causal nature of the protocol.

\bprop Assume that a protocol is designed for a KPP(I) network by
the procedure detailed above. Then the edges that are present
within any partition of the network do not alter the causal nature
of the protocol. \eprop

\bpf By hypothesis, the protocol is constructed for a KPP(I)
network by the procedure detailed above. Hence, while checking for
the condition that a given interference link causes purely causal
interference, links inside a partition are discounted. We need to
show that notwithstanding this fact, the links inside a partition
do not result in any non-causal interference. Equivalently, it
suffices to show that for any link inside a left partition, the
delay along the backbone path leading from the tail of the
interference link to the right partition is lesser than that along
any other path to the right partition from the same node.

Let the network be decomposed into $N$ layers, $L_1, L_2, \ldots,
L_N$. Without loss of generality, consider an interference link in
the left partition $\mathcal{P}_i$ of an arbitrary layer $L_j, \ 1
\leq j \leq N$ connecting nodes $R_1$ and $R_2$ on the first and
second backbone paths, say $P_1$ and $P_2$. Then focussing on the
path from $R_1$, we will prove that the link under consideration
does not create any non-causal interference. By symmetry the same
will hold even if we choose $R_2$. Assume that delay encountered
by a symbol forwarded via the backbone paths $P_1$ and $P_2$
within the layer $L_j$ be $d_1$ and $d_2$ respectively. As part of
left compensation, let $d_1^{'}$ and $d_2^{'}$  be added to nodes
$R_1$ and $R_2$ respectively.

Any symbol which is received at $R_1$ is delayed for $d_1^{'}$,
and then forwarded. Then the symbol simultaneously starts
following both the backbone path $P_1$ as well as the path via the
interference link. Therefore the delay $d_1^{'}$ does not come
into picture for this relative delay comparison and we have that
the delay through the shortcut path that includes the interference
link is equal to $1 + d_2^{'} + d_2$ and the delay through the
back-bone path from $R_1$ is $d_1$. Therefore in order to show
that the delay on the back-bone path is strictly lesser than the
delay on the short-cut path, we need to show that $d_2^{'} + d_2
\geq d_1$. This is indeed ensured by the algorithm designed above.

Thus the interference remains causal considering the path
beginning from the relay $R_1$. Thus the protocol remains causal
even in the presence of interference link within a partition.

\epf

This completes the proof of the lemma.

\section{Proof of Lemma~\ref{lem:product_channel_DMT} \label{app:product_channel_DMT}}

\bpf Let us assume without loss of generality that $N_1 \geq N_2 \geq ... N_{L+1}$. We also have, \beq {\bf H} = \left[ \begin{array}{cccc} {\bf g}_1 & & & \\
                                                               & \ddots & & \\
                                                               & & & {\bf g}_N \end{array} \right].
                           \eeq
Consider a variable transformation where ${\bm{\alpha}}^{(k)}_{j}$
is defined such that $\rho^{- {\bm{\alpha}}^{(k)}_{j}} = |{\bf
h}^{(k)}_j|^2$. Now the DMT $d(r)$ of the parallel channel is
characterized as,
\beqa \rho^{-d(r)} & \doteq & \Pr\{\log\det(I+\rho \bold{H} \bold{H}^\dagger) \leq r\log\rho\}  \nonumber \\
& = & \Pr\{\det(I+\rho \bold{H} \bold{H}^\dagger) \leq \rho^r \}  \nonumber \\
& = & \Pr\{\prod_{i=1}^{N} (1+\rho{|{\bf g}_i|}^2) \leq \rho^r \}  \nonumber \\
& = & \Pr\{\prod_{i=1}^{N} (1+\rho{ \prod_{k=1}^{L+1} |{\bf
h}^{(k)}_{e(i,k)}|^2 }) \leq \rho^r \}  \nonumber \eeqa

\beqa & = & \Pr\{\prod_{i=1}^{N} (1+\rho{ \prod_{k=1}^{L+1}
\rho^{-{\bm{\alpha}}^{(k)}_{e(i,k)}} }) \leq
\rho^r \}  \nonumber \\
& = & \Pr\{\prod_{i=1}^{N} (1+{  \rho^{1-\sum_{k=1}^{L+1}
{\bm{\alpha}}^{(k)}_{e(i,k)}} }) \leq \rho^r \}  \nonumber  \eeqa

\beqa & \doteq & \Pr\{\prod_{i=1}^{N} {  \rho^{(1-\sum_{k=1}^{L+1}
{\bm{\alpha}}^{(k)}_{e(i,k)})^{+}} }
\leq \rho^r \} \nonumber  \\
& = & \Pr\{\sum_{i=1}^{N} {  {(1-\sum_{k=1}^{L+1} {\bm{\alpha}}^{(k)}_{e(i,k)})^{+}} } \leq r \} \label{eq:exact_layered}  \\
&  \leq & \Pr\{\sum_{i=1}^{N} {  {(1-\sum_{k=1}^{L+1} {\bm{\alpha}}^{(k)}_{e(i,k)})} } \leq r \}  \label{eq:approx_layered} \\
& = & \Pr\{ N - \sum_{k=1}^{L+1} { N_k \sum_{j=1}^{M_k}
{\bm{\alpha}}^{(k)}_j } \leq r \} . \nonumber \eeqa

The last equality follows since each $|\bold{h}_{ij}|^2$ appear in
$N_i$ of the terms in $\mathcal{H}$ irrespective of $j$ and so do
the corresponding ${\bm{\alpha}}_{ij}$. Let $d_1(r)$ be defined as
the SNR exponent of the RHS in the last equation above, i.e.,
\beqa \Pr\{ N - \sum_{k=1}^{L+1} { N_k \sum_{j=1}^{M_k}
{\bm{\alpha}}_{kj} } \leq r \} & \doteq & \rho^{-d_1(r)} \eeqa

Let the set $S$ and $T$ be defined as \beqa S & := &
\{({\alpha}^k_{j}): \sum_{i=1}^{N} { {(1-\sum_{k=1}^{L+1}
{\alpha}^{(k)}_{e(i,k)})^{+}} } \leq r \}, \\
T & := & \{({\alpha}^k_{j}):\sum_{i=1}^{N} { {(1-\sum_{k=1}^{L+1}
\alpha^{(k)}_{e(i,k)})} } \leq r \} . \eeqa

Now, \beqa d(r) & = & \inf_{({\alpha}^k_{j}) \in S} \sum_{k,j}
{\alpha}^k_{j}, \label{eq:exact} \eeqa and \beqa d_1(r) & = &
\inf_{({\alpha}^k_{j}) \in T} \sum_{k,j} {\alpha}^k_{j}.
\label{eq:relaxed}\eeqa Since $S \subseteq T$, we have that $d(r)
\geq d_1(r)$.

\bnote \label{rem:dmt_equal} While the minimizing solution
$({\alpha}^k_{j})$ for \eqref{eq:relaxed} is usually a member of
the set $T$, if the solution happens to be an element of the set
$S$ also, then it follows that $d(r) = d_1(r)$. As will be seen
later, this is the case here.  We proceed to compute $d_1(r)$ and
the optimizing $({\alpha}^k_{j})$ for \eqref{eq:relaxed}. \enote

Now, \beq d_1(r) = \inf_{ \left\{
                    \begin{array}{c} N - \sum_{k=1}^{L+1} { N_k \sum_{j=1}^{M_k} {\alpha}^k_{j} } \ \leq r \ , \\
                    {\alpha}^k_{j} \geq 0 \end{array} \right\} } { \ \ \sum_{k=1}^{L+1}  \sum_{j=1}^{M_k} {\alpha}^k_{j} }. \eeq

Define $ {\alpha}^{(k)}  :=  \sum_{j=1}^{M_k} {\alpha}^k_{j} $ to
obtain,

\beqa d_1(r) & = &  \inf_{ \{ N - \sum_{k=1}^{L+1} { N_k
{\alpha}^{(k)} } \leq r \ , \ {\alpha}^{(k)} \geq 0 \} } { \ \
\sum_{k=1}^{L+1} {\alpha}^{(k)} } . \label{eq:d1_inf}\eeqa

The infimum is attained in the above minimization by $
{\alpha}^{(1)} = \frac{N - r}{N_1}, \ {\alpha}^{(i)} = 0, \forall
i = 2,...,N $ and the value of the infimum is $\frac{N - r}{N_1}$.
Thus $d(r) \geq d_1(r) = \frac{N - r}{N_1}$.

Now we will verify that this lower bound is in fact equal to the
DMT of the channel. An optimizing assignment of ${\alpha}^k_{j}$
for obtaining $d_1(r)$ is given by, $({\alpha}^1_{j})^{*} =
\frac{N - r}{N_1 M_1} = \frac{N-r}{N}, j=1,2,..,M_1$. Clearly
$({\alpha}^1_{j})^{*} \in T$. It can be easily checked that
$({\alpha}^1_{j})^{*} \in S$ also. This implies, by
Remark~\ref{rem:dmt_equal}, that this is the optimizing
${\alpha}^k_{j}$ for \eqref{eq:exact} too. Therefore, we have
\beqa d(r) = d_1(r) = \frac{N - r}{N_1} \eeqa \epf

\section{Achievability of Outage Exponent
\label{app:outage_achievability}}

Consider a compound channel, where a channel, $s$ is chosen from a
set of possible channels $\mathcal{S}$ and the channel once
chosen, remains fixed. Then, the compound channel coding theorem
tells us that,, for a fixed $p_X(x)$, any rate

\beqa R & < &  \ \inf_{s \in \mathcal(S)} I(\bold{X};\bold{Y} |
\bold{S} = s ) \eeqa is achievable on the compound channel, i.e,
irrespective of the particular channel $s$ chosen, the probability
of error at the receiver can be reduced to zero.

We will now consider a specific channel ${\bf Y}={\bf H}{\bf
X}+{\bf W}$, where the channel ${\bf H}$ can be any matrix in
$\mathbb{C}^{m \times n}$. However, if we consider the set of all
possible ${\bf H}$ that are not in outage as a set $\mathcal{H}$,
we can apply a compound channel theorem to this set and whenever a
matrix is chosen from this set, we can drive the probability of
error to zero.

Consider the set of all channels not in outage, $\mathcal{H}$ as
the possible channels in the compound channel. Then $\mathcal{H}$
is defined as \beqa \mathcal{H} = \{ {\bf H}: \log \det(I+{\bf H}
\Sigma_x {\bf H}^{\dagger}
> r \log(\rho) \}. \eeqa Now, any rate lesser than $R$ is achievable on the compound channel by using an optimal
compound channel code. This means that there exists a code for
this compound channel, whose probability of error is less than
$\epsilon$ for any given $\epsilon > 0$.

In this setting, let the optimal covariance matrix for minimizing
outage probability for a given rate be $p_X^{*}$, i.e., $p_X^{*} =
\mathbb{C}\mathcal{N}(0,\Sigma_x)$, where $\Sigma_x$ be optimizing
covariance matrix for the following optimization: \beqn
    P_{\text{out}}(R) = \inf_{\Sigma_x \ \geq \ 0, \
    \text{Tr}(\Sigma_x) \ \leq \ n \rho }
    \Pr(\log \det(I+{\bf H} \Sigma_x {\bf H}^{\dagger} \ \leq \ n R ).
\eeqn

If we use this optimal compound channel code on the slow fading
channel, we know that the probability of error goes to $\epsilon$
whenever, the channel realization $H \in \mathcal{H}$.

The probability of error of this code when used on the slow fading
channel is given by

\beqa P_e & = & P_{\text{out}} P_{\text{e/out}} + P_{\text{out}^{c}} P_{ {e / {out^c}}} \\
& \leq & P_{\text{out}} + P_{ {e / {out^c}}}  \\
& \leq & P_{\text{out}} + \epsilon \\
&  \dot \leq & P_{\text{out}} \eeqa

where $P_{\text{out}}$ is the probability of the channel being in
outage and $P_{\text{out}^{c}}$ is the probability of the channel
not being in outage. $P_{\text{e/out}}$ is the probability of
error of the code given the channel is in outage and $P_{ {e /
{out^c}}}$ is the probability of error of the code given the
channel is not in outage. Thus the outage probability is
achievable.

\section{Different Fractions of Activation in Multi-antenna KPP(I) Networks\label{app:fraction}}

In this appendix, we study the achievable region when different
paths are to be activated for different fractions of time. For
choosing $3$ parallel paths from the KPP network, the total number
of possibilities is $M := K \choose 3$, let us number these
possibilities as $1,2,...,M$. If we use the $3$ paths specified by
the combination for a fraction $\lambda_i$ fraction of time. Let
us construct a matrix of size $K \times M$ with each column being
composed of distinct vectors of weight $3$. Now $\frac{1}{3} A
\lambda$ yields a vector $f = [f_1, f_2,...,f_K]^{t}$ of size $K$
that gives us the fraction of duration $f_i$ for which the
parallel path $P_i$ is activated. Now we have to identify which
are the possible fractions $f$ that can be obtained by choosing
various combinations of $\lambda$. The lemma below proves that all
valid activation fractions $f$ such that each component is lesser
than $\frac{1}{3}$ can be obtained using this scheme.

\blem Let $K \in \mathbb{Z}$, $K \geq 4$, and let $M = { K \choose
3}$. Construct a matrix $A \in \{0,1\}^{K \times M}$ constituting
of distinct columns, each being a vector of weight $3$. Then the
equation $\frac{1}{3}A \ [y_1 \ y_2 \ldots \ y_M]^t \ = \ [f_1 \
f_2 \ldots f_K]^t$ has a solution $\lambda \ = \ [\lambda_1 \
\lambda_2 \ldots \lambda_M]$ satisfying, \ben \item $0 \leq
\lambda_i \leq 1, \ \forall \ 1 \leq i \leq M$ and \item
$\lambda_1 + \lambda_2 + \ldots + \lambda_M = 1$, \een for every
$[f_1 \ f_2 \ldots f_K]$ in the region \beqn \mathcal{F} = \{[f_1
\ f_2 \ \ldots \ f_K]^t: \sum_{i=1}^{K} f_i = 1, \ \ 0 \leq f_i
\leq \frac{1}{3} \}. \eeqn \elem

\bpf The region $\mathcal{F}$ is convex, and thus every point in
$\mathcal{F}$ can be expressed as a linear combination of its
extreme points, where the coefficients appearing in the linear
combination lie between $0$ and $1$, and add up to $1$. Hence, it
is sufficient to prove that the extreme points of $\mathcal{F}$
are $\frac{1}{3}$ times columns of $A$. The claim is that this is
precisely the case, i.e., extreme points of $\mathcal{F}$ are
vectors containing $\frac{1}{3}$ as entries in $3$ positions, and
zero elsewhere.

Suppose it is not the case. Then there exists an extreme point $x
\ = \ [x_1 \ x_2 \ \ldots \ x_K]$, such that at least one entry,
say $x_1$ without loss of generality, is less than $\frac{1}{3}$.
But, due to constraints of the region, this forces one more entry,
say $x_2$ without loss of generality, to be greater than
$\frac{1}{3}$.
Then clearly, $\exists \ \delta > 0$ such that \beqan x_1 + \delta & \leq & \frac{1}{3} \\
x_2 - \delta & \geq & 0 \\
x_1 - \delta & \geq & 0 \\
x_2 + \delta & \leq & \frac{1}{3}, \eeqan so that $x^\prime \ = \
[x_1+\delta \ x_2-\delta \ \ldots \ x_K]$ and $x^{\prime\prime} \
= \ [x_1-\delta \ x_2+\delta \ \ldots \ x_K]$ belong to the region
$\mathcal{F}$. Now, \beqn x = \frac{1}{2}x^{\prime}+
\frac{1}{2}x^{\prime\prime}, \eeqn which contradicts our
hypothesis that $x$ is an extreme point. This completes the proof.
\epf

\section*{Acknowledgment}

Thanks are due to K.~Vinodh and M.~Anand for useful discussions.

\begin{spacing}{1.2}

\end{spacing}


\begin{thebibliography}{1}
\bibitem{Part1} K.~Sreeram, S.~Birenjith, and P.~V.~Kumar, ``DMT of multi-hop cooperative networks - Part I:
Basic results,'' submitted to {\em IEEE Trans. Inform. Theory}.

\bibitem{AzaGamSch} K.~Azarian, H.~El Gamal, and P.~Schniter, ``On
the achievable diversity--multiplexing tradeoff in half--duplex
cooperative channels,'' {\em IEEE Trans. Inform. Theory}, vol. 51,
no. 12, pp. 4152--4172, Dec. 2005.

\bibitem{YukErk} M.~Yuksel and E.~Erkip, ``Multiple--antenna cooperative wireless systems: A diversity-–multiplexing tradeoff perspective'',
{\em IEEE Trans. Inform. Theory }, vol 53, no.10, pp. 3371-3393,
Oct. 2007.

\bibitem{EliVinAnaKum} P.~Elia, K.~Vinodh, M.~Anand, and
P.~V.~Kumar, ``D-MG tradeoff and optimal codes for a class of AF
and DF cooperative communication protocols,'' {\em IEEE Trans.
Inform. Theory,} accepted for publication pending revision.
Available Online: http://arxiv.org/abs/cs/0611156 , Nov. 2006.


\bibitem{TavVis} S.~Tavildar and P.~Viswanath,
``Approximately universal codes over slow-fading channels'' {\em
IEEE Trans. Inform. Theory}, vol. 52, no. 7, pp. 3233--3258, July
2006.

\bibitem{EliRajPawKumLu} P.~Elia, K.~Raj Kumar, S.~A.~Pawar,
P.~V.~Kumar, and H-F.~Lu, ``Explicit, minimum-delay space-time
codes achieving the diversity-multiplexing gain tradeoff,'' {\em
IEEE Trans. Inform. Theory}, vol. 52, no. 9, pp. 3869--3884, Sept.
2006.

\bibitem{EliKum} P.~Elia and P.~V.~Kumar, ``Approximately--universal space--time codes for the parallel, multi--block and cooperative dynamic--decode--and--forward channels,'' {\em Available Online: http://arxiv.org/abs/0706.3502
}, June 2007.

\bibitem{YanBelMimoAf} S.~Yang and J.-C.~Belfiore, ``Optimal space--time codes for the MIMO amplify-and-forward cooperative channel,'' {\em IEEE Trans. Inform. Theory,} vol. 53, Issue 2, pp 647-663,
Feb. 2007.

\bibitem{YanBelNew} S.~Yang and J.-C.~Belfiore, ``Diversity of MIMO multihop relay channels,'' submitted to {\em  IEEE Trans. on Inform. Theory},
Available Online: http://arxiv.org/abs/0708.0386, Aug. 2007.

\bibitem{YanBelSaf} S.~Yang and J.-C.~Belfiore, ``Towards the optimal amplify--and--forward cooperative diversity scheme,'' {\em
IEEE Trans. Inform. Theory}, vol. 53, Issue 9, pp 3114-3126, Sept.
2007.

\bibitem{SetRajSas} B.~A.~Sethuraman, B.~Sundar Rajan, and V.~Shashidhar, ``Full--diversity, high--rate, space--time block codes from division algebras,'' {\em IEEE Trans. Inform. Theory},
vol. 49, no. 10, pp. 2596--2616, Oct. 2003.

\bibitem{ITA} K.~Sreeram, S.~Birenjith, and P.~V.~Kumar, ``Multi--hop cooperative wireless networks: Diversity
multiplexing tradeoff and optimal code design,'' {\em Proceedings
of Information Theory and Applications Workshop, UCSD}, Feb. 2008.

\bibitem{WPMC} K.~Sreeram, S.~Birenjith, K.~Vinod, M.~Anand, and P.~V.~Kumar ``On the throughput, DMT and
optimal code construction of the K-parallel-path cooperative
wireless fading network,'' {\em Proc. 10th Int. Symp. Wireless
Personal Multimedia Commun.,} Dec. 2007.

\bibitem{ISIT1} K.~Sreeram, S.~Birenjith, and P.~V.~Kumar, ``DMT of multi--hop cooperative networks--Part I:
K--parallel--path--networks'' {\em Proc. IEEE Int. Symp. Inform.
Theory}, Toronto, July 6-11, 2008.

\bibitem{ISIT2} K.~Sreeram, S.~Birenjith, and P.~V.~Kumar, ``DMT of multi--hop cooperative networks--Part II:
Layered and multi--antenna networks,'' {\em Proc. IEEE Int. Symp.
Inform. Theory}, Toronto, July 6-11, 2008.

\bibitem{Arxiv}K.~Sreeram, S.~Birenjith, and P.~V.~Kumar,
``Multi--hop cooperative wireless networks: Diversity multiplexing
tradeoff and optimal code design,'' { \em Available Online }:
http://arxiv.org/pdf/0802.1888 , Feb. 2008.

\bibitem{TechReport}K.~Sreeram, S.~Birenjith, and P.~V.~Kumar,
``On the throughput, DMT and optimal code construction of the
K--parallel--path cooperative wireless fading network,'' {\em USC
CSI Technical Report}, CSI-2007-06-07, June 2007.

\bibitem{BorZheGal} S.~Borade, L.~Zheng, and R.~Gallager, ``Amplify and forward in wireless relay networks: Rate, diversity and network size,'' {\em IEEE Trans. Inform. Theory}, vol 53, no.10, pp 3302-3318, Oct.
2007.

\bibitem{GhaBayKha} S.~O.~Gharan, A.~Bayesteh, and A.~K.~Khandani, ``On the diversity-multiplexing tradeoff in multiple-relay network,'' submitted to {\em IEEE Trans. Inform. Theory}, April 2008.

\bibitem{AveDigTse1} A.~S.~Avestimehr, S.~N.~Diggavi, and D.~Tse,  ``Wireless network information flow,'' {\em Proc. Forty-fifth Allerton Conf. Commun. Contr. Comput.}, Illinois, Sep
2007.

\bibitem{VazHea} R.~Vaze and R.~W.~Heath~Jr., ``Maximizing reliability in multi-hop wireless networks with cascaded space-time codes ,'' {\em
Proc. Inform. Theory and Appln. Workshop, UCSD}, Feb. 2008.

\bibitem{Lu} Hsiao-feng Lu, ``Explicit construction of multi-block
space-time codes that acheive the diversity-multiplexing gain
tradeoff,'' {\em Proc. IEEE Int. Symp. Inform. Theory}, Seattle,
USA , 2006 .

\bibitem{YanBelRek} S. Yang, J.-C. Belfiore and G. Rekaya, ``Perfect
space-time block codes for parallel MIMO channels,'' {\em Proc.
IEEE Int. Symp. Inform. Theory}, Seattle, USA , 2006.

\bibitem{KodNan} M.~Kodialam and T.~Nandagopal, ``Characterizing achievable rates in multi-hop wireless mesh networks with orthogonal channels,'' {\em IEEE/ACM Trans. Networking}, Vol 13, No.4, pp 868-880, Aug. 2005.

\bibitem{CovTho} T.~M.~Cover and J.~A.~Thomas, {\em Elements of information theory}, 2nd Edition, John Wiley and Sons, New York, 2006.

\bibitem{RibCaiGia} A.~Ribeiro, X.~Cai, and G.~B.~Giannakis, ``Symbol error probabilities
for general cooperative links,'' {\em IEEE Trans. Wireless Comm.
}, Vol. 4, No. 3, pp 1264-1273, May 2005.


\end{thebibliography}
\end{document}